\documentstyle[12pt,epsf,epsfig,fleqn]{article}
\newcommand{\fips}[1]{\epsffile{#1.eps}}

\newcommand{\mbf}[1]{\mbox{\boldmath $#1$}}

\newcommand{\floadeps}[1]{\epsfig{file=#1.eps,width=20mm}}

\begin{document}

\begin{center}

{\LARGE 
Solving QCD Hamiltonian for bound states
\footnote{Based on talks given at JLAB (August 30, 1999), Duke University 
(October, 1999), BNL (November 3, 2000), MIT (December 11, 2001).} \\} 
\vskip 20pt
{\Large
Elena Gubankova\\}
\bigskip
{\it
Institute of Theoretical and Experimental Physics,
B. Cheremushkinskaya 25,\\ RU-117 218 Moscow, Russia} 

\bigskip
\end{center}


\begin{abstract}

We consider the eigenstate problem for a Hamiltonian operator of the
field theory.  Methods of construction the effective field theoretical
Hamiltonians for which the eigenstate problem may be solved are
discussed.  In particular, we discuss the method of flow equations
from a general perspective as well as in application to the gauge
field theories.  Flow equations transform the Hamiltonian to a
block-diagonal form with the number of particles conserved in each
block and thus reduce the original bound state problem to a set of
coupled eigenstate equations with an effective Hamiltonian in each
sector.

Applications of flow equations to the Hamiltonians of QED and QCD in
the light-front gauge and the QCD Hamiltonian in the Coulomb gauge are
considered.  Using flow equations, we derive the effective
Hamiltonians as well as the renormalized gap equations and the
Bethe-Salpeter equations for the bound states in these theories.  We
show that the obtained equations are finite in both UV and IR regions
and are completely renormalized in UV, i.e. the corresponding
solutions do not depend on the cut-off $\Lambda$.

We calculate positronium spectrum, glueball masses, $\pi-\rho$ mass
splitting, gluon and chiral quark condensates and compare our results
with the covariant calculations and experimental results.  Use of flow
equations to calculate the dynamical terms is critical to achive good
agreement with experimental results.

\end{abstract}

\newpage\tableofcontents\newpage

\section{Flow equations. Idea and technique.}
\label{sec:1}
\subsection{Hamiltonian bound state problem}

{\it Motivation}

Nonabelian gauge field theories, like Quantum Chromodynamics (QCD),
are well understood in the high-momentum region, since due to the asymptotic
freedom coupling constant is small and Feynmann rules of the covariant
perturbation theory, using $L_{QCD}$, provide convincing agreement with experiment. 
However, low energy QCD, where such nonperturbative phenomena as confinement and 
chiral symmetry breaking are taking place, is still understood only 
on a phenomenological level. Except for the lattice studies 
and some phenomenological
models, there is a lack of analytical nonperturbative methods. Here
a systematic analytical method to solve bound state problem, using Hamiltonian, 
\begin{eqnarray}
H_{QCD}|\psi\rangle &=& E|\psi\rangle
\,,\label{eq:1.1}\end{eqnarray} 
is presented. 

{\it Problems}

Two general problems occure when solving the eigenstate equation for a relativistic
Hamiltonian of the field theory: field theoretical Hamiltonian is an infinite
dimensional matrix, first, in the particle number space and, second, 
in the energy space.

First, the number of particles is not fixed in the field theory due to allowed
creation and annihilation processes in vacuum. Therefore, any physical state
contains, in principle, infinite many Fock components
\begin{eqnarray}
|\psi\rangle_{meson}&=&c_{q\bar{q}}|q\bar{q}\rangle 
+ c_{q\bar{q}g}|q\bar{q}g\rangle + ... 
\,.\label{eq:1.2}\end{eqnarray}
Thus, we need a method to construct an effective Hamiltonian, which acts 
in a smaller truncated space and provides the same eigenvalues as the original
Hamiltonian.

Second, there are states with (infinite) large energies/momenta in the Hamiltonian,
which contribute UV divergent matrix elements. 
Hamiltonian matrix in a momentum space (plane wave space) reads
\begin{eqnarray}
 H &=& 
  \begin{array}{c|cc|c}
       0&\phantom{1111}&\phantom{111}&  \\
       &&&     \\
       &&&     \\
       &&&  \Lambda\rightarrow\infty  
  \end{array} 
\,,\label{eq:1.3}\end{eqnarray}
where the top left matrix element correspond to small (zero) momentum,
and the bottom right -- to (infinite) large momentum with the cut-off UV scale 
$\Lambda_{UV}$. Thus, we need a method of UV regularization and subsequent 
renormalization directly for a Hamiltonian operator.

\subsection{Hamiltonian methods}

There are following Hamiltonian methods, which address both problems.
However, there are some disadvantageous in these methods, restriting 
the range of their applicability.

\subsubsection{Tamm-Dancoff truncation and iterated resolvents}

\begin{figure}
$$
\fips{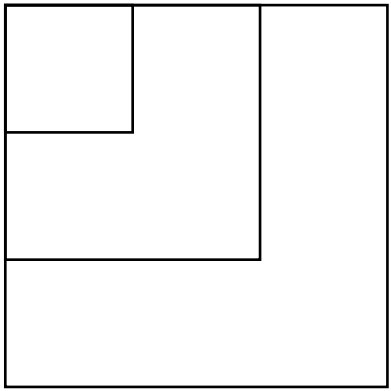}
\setlength{\unitlength}{0.240900pt}
\begin{picture}(0,0)
\end{picture}
$$
\caption{Projection of high Fock components to the low ones.}
\label{fig:1.1}
\end{figure}

The method of Tamm-Dancoff truncation, treating the 'infinities' in Fock space,
has been invented in 50's, and later generalized in the method of iterated
resolvents by Pauli in 80's. The main idea of this method is to project 
high Fock components onto the lower ones in sequence, ending up with an effective
Hamiltonian which acts in a smaller truncated space (Fig. \ref{fig:1.1})
and therefore it can be solved for eigenstates and eigenvalues.

%
%


Technically, by projecting high Fock states
one inverts a resolvent, which is the energy denominator 
containing matrix element
of the original Hamiltonian, $(E-\langle n|H|m\rangle)^{-1}$. 
Inversion is done within some approximation scheme.
It turns out that this scheme breaks down for QCD, convergence cannot be achieved,
and eigenvalues of an effective QCD Hamiltonian do not predict QCD mass spectrum 
correctly. Physically, starting with a canonical QCD Hamiltonian and working
in terms of bare fields, one fails to find a representation 
for the QCD Hamiltonian with fixed number of particles, since QCD vacuum is
unstable and due to a strong QCD interaction it creates and annihilates 
bare particles. In addition, unregulated UV divergences, comming from high energies,
appear in the effective Hamiltonian. A way out to proceed with QCD is suggested
by various phenomenological studies, where instead of bare (current) 
degrees of freedom
one uses the renormalized (constituent) ones. In other words, one needs 
a scheme to regulate and renormalize Hamiltonians.
     
\subsubsection{Similarity renormalization}

\begin{figure}
$$
\fips{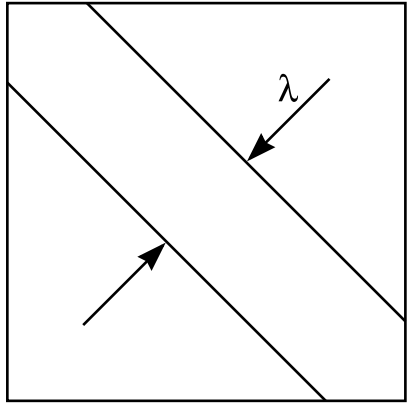}
\setlength{\unitlength}{0.240900pt}
\begin{picture}(0,0)
\end{picture}
$$
\caption{Band-diagonal Hamiltonian in the energy space.}
\label{fig:1.2}
\end{figure}

Several years ago, in 1994, the similarity renormalization scheme, which treats
'infinities' in the energy space, has been suggested by Glazek and Wilson.
Similarity scheme is an analog to the Wilson's renormalization through 
an effective action where high energy modes integrated out in a path integral, 
but formulated directly for a Hamiltonian matrix. The aim is to regulate and 
renormalize away UV divergencies. The main idea of the similarity renormalization 
is to find a unitary transformation, $U$, which brings a Hamiltonian matrix 
to a band-diagonal form (Fig. \ref{fig:1.2})
with the width of the band, $\lambda$, being much smaller than the UV cut-off, 
$\Lambda_{UV}$,
\begin{eqnarray}
U^{\dagger}(\lambda,\Lambda)H(\Lambda)U(\lambda,\Lambda) &=&
H_{eff}(\lambda,\Lambda) \nonumber\\
|E_i-E_j|\l\lambda \ll \Lambda
\,\,.\label{eq:1.4}\end{eqnarray}

%
%


High and low energy states decouple from each other in a band-diagonal effective 
Hamiltonian. Indeed, high order in the perturbation theory
\begin{eqnarray}
g^{2\lambda/\Lambda}\ll 1 
\,,\label{eq:1.5}\end{eqnarray} 
with $\lambda\ll\Lambda$, connects high and low energies. Therefore, a low-energy
effective Hamiltonian, $H_{eff}$, can be taken and solved for few lowest states.
Since transformation is unitary, eigenvalues of $H_{eff}$ should be the same
as low lying states of the original Hamiltonian. Similarity renormalization
is formulated as a perturbation expansion in coupling constant. As one shrinks
the band width, $\lambda$, one breaks perturbation theory at some point, since
diagonalization is a nonperturbative problem. In similarity, higher orders
of perturbation theory correspond to higher Fock states. Therefore, breaking
of the perturbation theory means that mixing of higher Fock components 
becomes important and cannot be neglected. In QED, a perturbative mixing
of electron-positron and electron-positron-photon states provides reliable
approximation for positronium bound states. Also, the window for the band width
can be easily found, $m\alpha^2\ll\lambda\ll m\alpha$ where
$\alpha=e^2/4\pi$. However, for QCD a perturbative mixing does not work.
One needs a nonperturbative scheme to incorporate high Fock states mixing.

\subsection{Flow equations}
\label{subsec:1.3}

\begin{figure}
$$
\fips{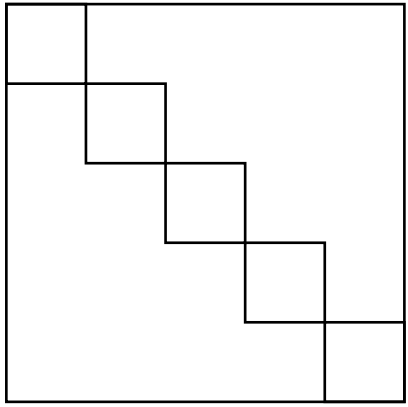}
\setlength{\unitlength}{0.240900pt}
\begin{picture}(0,0)
\end{picture}
$$
\caption{Block-diagonal Hamiltonian in the particle number space.}
\label{fig:1.3}
\end{figure}

Method of flow equations for Hamiltonians was suggested independently
by Wegner in 1994. Applied to the field theory, flow equations try to
solve both problems: 'infinities' in the Fock space as well as in the energy space.
$H_{eff}$ incorporates effects from high Fock components and large energies.
This method can be viewed as a synthesis of the Tamm-Dancoff and 
similarity renormalization approaches. However, instead of band-diagonalization
in the energy space, that is hard to achieve, one block-diagonalizes Hamiltonian
in the particle number space (Fig. \ref{fig:1.3}).

One seeks for a unitary transformation,
which brings a Hamiltonian matrix 
\begin{eqnarray}
H &=&  \left( \begin{array}{cc}
       PHP & PHQ \\
       QHP & QHQ          
 \end{array}
\right)
\,,\label{eq:1.6}\end{eqnarray}
where $P$ and $Q=1-P$ are projection operators in the particle number space,
to a block-diagonal form 
\begin{eqnarray}
H &=&  \left( \begin{array}{cc}
       PH_{eff}P &  \\
        & QH_{eff}Q          
 \end{array}
\right)
\,,\label{eq:1.7}\end{eqnarray}
with the number of particles being conserved in each block.
Since blocks decouple from each other
in a block-diagonal effective Hamiltonian, full bound state problem is reduced
to several eigenstate equations in each block. Usually it is easier to diagonalize
separately $P$ or $Q$ sectors than to diagonalize the original Hamiltonian matrix. 

Unitary transformation is governed by a continuous flow parameter $l$,
\begin{eqnarray}
 U(l)HU^{-1}(l) &=& H(l)
\,,\label{eq:1.8}\end{eqnarray}
with initial condition $U(l=0)=1$. As $l\rightarrow\infty$ an effective Hamiltonian
coincides with its diagonal part, Eq. (\ref{eq:1.7}), 
$H(l\rightarrow\infty)=PH_{eff}P$. Unitary transformation is given by
\begin{eqnarray}
 U(l)=T_l {\rm exp}\left(\int_0^l\eta(l^{\prime})dl^{\prime}\right)  
\,,\label{eq:1.9}\end{eqnarray}
where $\eta(l)$ is the generator of transformation, and $T_l$ is the $l$-ordering.
Wegner's flow equations are written in a diffrential form
\begin{eqnarray}
\frac{dH(l)}{dl} &=& [\eta(l),H(l)] \nonumber\\
\eta(l) &=& [H_d(l),H(l)] 
\,,\label{eq:1.10}\end{eqnarray}
where the generator, chosen by Wegner as a commutator of
a diagonal part,  
\begin{eqnarray}
H_d(l) &=&  \left( \begin{array}{cc}
       PH(l)P &  \\
        & QH(l)Q          
 \end{array}
\right)
\,,\label{eq:1.11}\end{eqnarray}
and an off-diagonal part, 
\begin{eqnarray}
H(l)-H_{d}(l) &=&  \left( \begin{array}{cc}
      & PH(l)Q  \\
   QH(l)P  &          
 \end{array}
\right)
\,,\label{eq:1.12}\end{eqnarray}
insures that an effective Hamiltonian
is diagonal at $l\rightarrow\infty$, 
$H(l\rightarrow\infty)=H_{d}(l\rightarrow\infty)$.
Diagonal -- particle number conserving, Eq. (\ref{eq:1.11}),
and off-diagonal -- particle number changing, Eq. (\ref{eq:1.12}),
parts of a Hamiltonian contain sector Hamiltonians which are fuctions of 
the flow parameter $l$.

Using Eqs. (\ref{eq:1.10}) and (\ref{eq:1.11}), (\ref{eq:1.12}),
flow equations for the diagonal and off-diagonal Hamiltonians,
and the generator of transformation are written in the operator form as 
\begin{eqnarray}
\frac{d}{dl}PHP &=& P\eta QHP - PHQ\eta P \nonumber\\
\frac{d}{dl}PHQ &=& P\eta QHQ - PHP\eta Q \nonumber\\
P\eta Q &=& PHPHQ - PHQHQ 
\,.\label{eq:1.13}\end{eqnarray}
Choosing eigenstates of initial $P$ and $Q$ sector Hamiltonians for a basis
(suppose it can be done), Eq. (\ref{eq:1.13}) is given in the matrix form
\begin{eqnarray}
\frac{d}{dl}h_{pp'}(l) &=&\sum_{q}\left(\eta_{pq}(l)h_{qp'}(l)
-h_{pq}(l)\eta_{qp'}(l) \right) \nonumber\\
\frac{d}{dl}h_{pq}(l) &=& -\left(E_{p}(l)-E_{q}(l)\right)
\eta_{pq}(l) \nonumber\\
\eta_{pq}(l) &=&  \left(E_{p}(l)-E_{q}(l)\right)
h_{pq}(l) 
\,,\label{eq:1.14}\end{eqnarray}
with $PH(l)P\rightarrow E_{p}(l)$, $QH(l)Q\rightarrow E_{q}(l)$, and
$PH(l)Q\rightarrow h_{pq}(l)$; $p,p'$ are energy/momentum indices
in the $P$ space. 
Substituting the generator of transformation,
we get a coupled system of equations for particle number conserving
\begin{eqnarray}
\frac{dh_{pp'}(l)}{dl} &=& -\sum_{q}\left(\frac{dh_{pq}(l)}{dl}
\frac{h_{qp'}(l)}{E_{p}(l)-E_{q}(l)}+
\frac{h_{pq}(l)}{E_{p'}(l)-E_{q}(l)}
\frac{dh_{qp'}(l)}{dl}\right)
\nonumber\\
\frac{dE_{p}(l)}{dl} &=& -\sum_{q}\frac{1}{E_{p}(l)-E_{q}(l)}
\frac{d}{dl}\left(h_{pq}(l)h_{qp}(l)\right)
\,,\label{eq:1.15}\end{eqnarray}
and particle number changing sectors
\begin{eqnarray}
\frac{dh_{pq}(l)}{dl} &=& -(E_{p}(l)-E_{q}(l))^2h_{pq}(l)
\,,\label{eq:1.16}\end{eqnarray} 
respectively. Formal solution of these equations in $P$ space is given
at the flow parameter $l\rightarrow\infty$. Particle number conserving part
\begin{eqnarray}
 h_{pp'}(\infty) &=& h_{pp'}(0)-\int_{0}^{\infty}dl
\sum_{q}\left(\frac{dh_{pq}(l)}{dl}\frac{h_{qp'}(l)}{E_{p}(l)-E_{q}(l)}
+\frac{h_{pq}(l)}{E_{p'}(l)-E_{q}(l)}\frac{dh_{qp'}(l)}{dl}\right)
\nonumber\\
 E_{p}(l) &=& E_{p}(0)-\int_{0}^{l}dl'
\sum_{q}\frac{1}{E_{p}(l)-E_{q}(l)}
\frac{d}{dl}\left(h_{pq}(l)h_{qp}(l)\right)
\,,\label{eq:1.17}\end{eqnarray}
includes the initial $P$ sector Hamiltonian plus terms, generated by flow equations
when eliminating particle number changing sector. If $p$ corresponds to one-particle
degree of freedom, the second equation in (\ref{eq:1.17}) is the energy gap equation,
and the first equation is the Bethe-Salpeter integral equation for a bound state.   
Eq. (\ref{eq:1.17}) resembles
the second order perturbation theory, where $h_{pq}(l)$ is a vertex
and $1/(E_{p}(l)-E_{q}(l))$ is the energy denominator. It is important, that 
energies also flow with $l$, and are given by diagonal matrix elements
in $P$ space, $p=p'$. To find $P$ space effective Hamiltonian, $h_{pp'}$, 
one should know $l$ dependent solutions for the energies, 
$E_{p}(l)$ and $E_{q}(l)$, and for the particle number changing part, $h_{pq}(l)$.   
Solution for the particle number changing part
\begin{eqnarray}
h_{pq}(l) &=& h_{pq}(0){\rm exp}\left(-\int_{0}^{l}dl'
(E_{p}(l')-E_{q}(l'))^2 \right)
\,,\label{eq:1.18}\end{eqnarray}
characterizes by an exponential function. For $l$-independent energies,
$h_{pq}$ decays exponentially to zero with $l\rightarrow\infty$. Renormalization
of energies, i.e. $l$-dependence of energies, brings some additional factor
before the exponential decay. Energy renormalization is crucial for degenerate
matrix elements, when initially $|E_{p}(0)-E_{q}(0)|=0$ for $l=0$.
In this case, $l$-dependence of energies provides power law decay of $h_{pq}$
instead of exponential one. Thus, the Wegner's generator (Eq. (\ref{eq:1.10}))
insures that the off-diagonal part is eliminated always, even for degenerate states.
Eqs. (\ref{eq:1.17}) and (\ref{eq:1.18}) are coupled, therefore they 
should be solved selfconsistently.

Our strategy is to solve first the equation for an effective renormalized energy
(the gap equation), and then using energy solutions to find an effective
Hamiltonian in $P$-sector, $h_{pp'}$, (kernel of the Bethe-Salpeter equation),
which is diagonalized numerically for bound states. Success of the procedure
depends on a particular choice of diagonal and off-diagonal Hamiltonian parts,
that is discussed in application to QED and QCD further.

\begin{figure}
$$
\fips{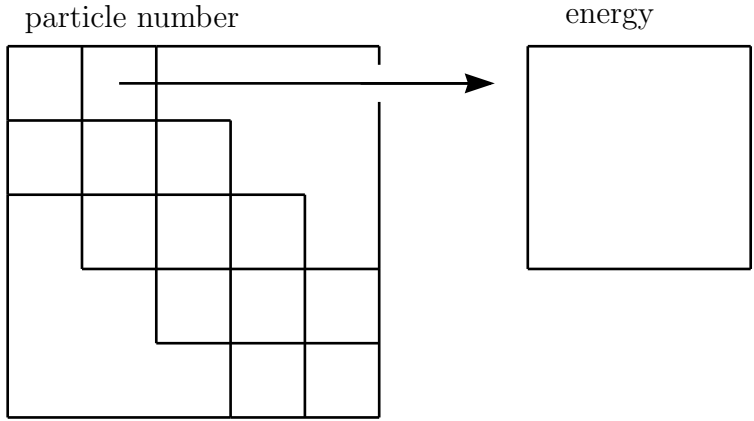}
\setlength{\unitlength}{0.240900pt}
\begin{picture}(0,0)
\put(-1000,650){\makebox(0,0){particle number}}
\put(-250,650){\makebox(0,0){energy}}
\end{picture}
$$
\caption{Pentadiagonal form of Hamiltonian in the 
particle number space; each sector contains matrix elements
with all possible energy changes.}
\label{fig:1.4}
\end{figure}

In a block-diagonal Hamiltonian, there is some freedom left to transform inside 
of each block without changing a block-diagonal structure. This freedom can be used
by choosing different similarity functions, $f(x)$,
\begin{eqnarray}
h_{pq}(l) &=& h_{pq}(0)f(x_{pq}) \nonumber\\
x_{pq}(l) &=& \int_{0}^{l}dl'(E_{p}(l')-E_{q}(l'))^2
\,,\label{eq:1.19}\end{eqnarray}
with general properties to be equal unity at the origin, and to fall down
at large arguments
\begin{eqnarray}
f(0)=1\,,\,f(x\rightarrow\infty)=0 
\,.\label{eq:1.20}\end{eqnarray}    
Similarity function reflects the rate how quickly the particle number 
changing sector is eliminated. The generator is written through the similarity
function as
\begin{eqnarray} 
\eta_{pq}(l) &=& -\frac{h_{pq}(l)}{E_{p}(l)-E_{q}(l)}\frac{d}{dl}
\left( \ln f_{pq}(x) \right)
\,,\label{eq:1.21}\end{eqnarray} 
$f_{pq}(x)=f(x_{pq})$. Flow equation scheme with $P$ and $Q$ sectors
can be easily generalized to the case of many Fock components 
(Fig. \ref{fig:1.4}). It is important,
that the relevant degrees of freedom are renormalized during this procedure.
As a consequence, it is not coupling constant which is used as a small parameter,
but rather some dimensionless combination, say $x_{pq}=|h_{pq}|/|E_{p}-E_{q}|$.

\begin{figure}
$$
\fips{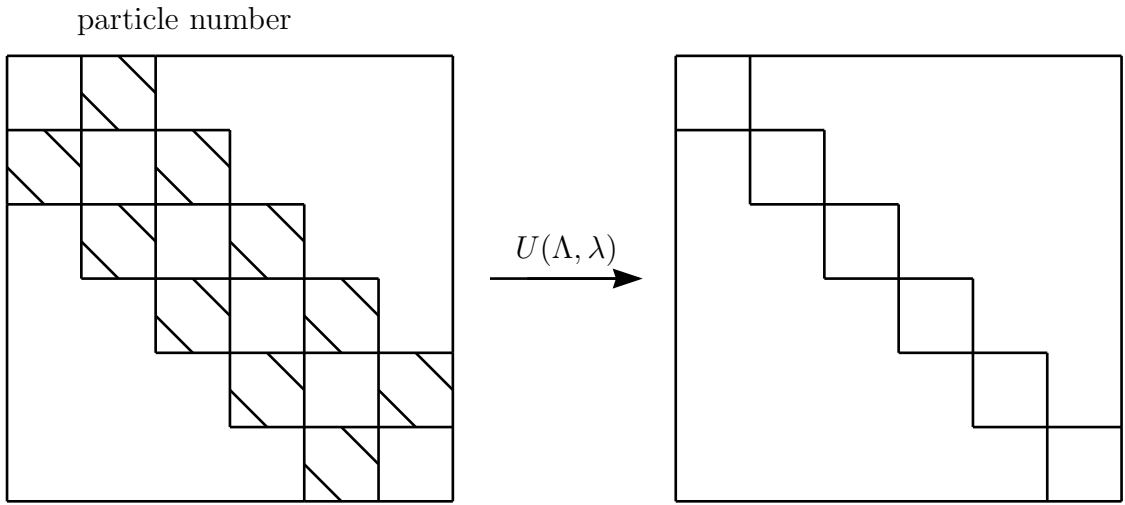}
\setlength{\unitlength}{0.240900pt}
\begin{picture}(0,0)
\put(-1500,780){\makebox(0,0){particle number}}
\put(-900,420){\makebox(0,0){$U(\Lambda,\lambda)$}}
\end{picture}
$$
\caption{Flow equations block-diagonalize
the bare regulated by $\Lambda$ Hamiltonian ($H(\Lambda)$) in the
particle number space.
For the finite $\lambda$ (after the unitary transformation
$U(\Lambda\rightarrow\infty,\lambda)$ is performed) the matrix elements 
of the particle number changing sectors 
are squeezed in the energy band $|E_i-E_j|<\lambda$ on the left hand side 
the picture and are eliminated completely as $\lambda\rightarrow 0$
(that corresponds to $U(\Lambda\rightarrow\infty,\lambda\rightarrow 0)$)
on the right hand side of the picture. Block-diagonal effective 
Hamiltonian is also UV renormalized by flow equations.}
\vspace{0.5cm}
\label{fig:1.5}
\end{figure}

Elimination of the particle number changing sectors is carried out not 
in one step but sequentially for energy differences which exceed the band width
$\lambda$ (flowing cut-off), Eq. (\ref{eq:1.18}), 
\begin{eqnarray} 
\lambda=\frac{1}{\sqrt{l}}\leq |E_{p}(0)-E_{q}(0)|\leq 
\frac{1}{\sqrt{l\rightarrow 0}}= \Lambda\rightarrow\infty 
\,,\label{eq:1.22}\end{eqnarray}
for $l$-independent energies, with $l=1/\lambda^2$ (Fig. \ref{fig:1.5}).

Lowering the cut-off $\lambda$, we effectively scale the theory down 
to low energies, that provides a connection between flow equations and similarity 
renormalization. System of coupled flow equations contain equations of 
the renormalization group, in addition to appearing equations for 
noncanonical operators. In this way, we incorporate renormalization in 
the many-body technique of block-diagonalizing.

{\bf Summary I}

1) Summarizing, flow equations perform Hamiltonian 'renormalization' in the energy 
and particle number space, in the sence that an effective Hamiltonian is finite 
in both spaces and contains effects of high energies and high Fock components.
Technically, flow equations allow to include directly the dynamical perturbative
corrections, responsible for the renormalization and the UV asymptotic region, 
into the many-body calculations. 

2) For the first time, due to the dynamical interactions, 
the physical equations are obtained UV and IR finite.

\section{Application of flow equations to the gauge field theories}
\label{sec:2}

Flow equations have been applied mostly in solid state physics 
by Wegner and his group. Also some toy models have been considered 
in the field theory. Here we consider several examples 
in the gauge field theories, such as the light-front QED \cite{LFQED} 
and light-front QCD \cite{LFQCD} Hamiltonians as well as Hamiltonian 
of gluodynamics \cite{Gluon} and QCD Hamiltonian with dynamical quarks
\cite{Quark} in the Coulomb gauge.
In Hamiltonian dynamics of the gauge field theory, first
a gauge should be fixed and constraint equations should be solved.
Hamiltonian operator is expressed only through 
the physical degrees of freedom.
Generally fewer equations of motion and current algebra relations 
(like Ward identities) occure in the Hamiltonian dynamics 
than for an effective action. Moreover, there are
no ambiguities in these equations since the physical fields have been used. 
However, one should keep track on covariance and gauge invariance manifest 
for physical observables. This serves to check the validity of the 
calculations and approximations.

\subsection{Flow equations in the light-front QED and QCD}
\label{subsec:2.1}

Light-front dynamics is similar
to the equal time one, except that the light-front time is given by 
$x^{+}=t+z$, and the commutation relations and initial conditions are formulated 
at a quantizaton plane $x^{+}=0$, instead of $t=0$ in the equal time framework.
Evolution (propagation) is considered along the light-front, $0 < x^{+}$.
In the collinear limit (small light-front x) which corresponds to the light-front
origin point, the light-front field theory has a singular behavior
associated with nontrivial vacuum effects in the equal time.

The Lagrangian density for QED/QCD 
\begin{eqnarray}
   {\cal L}&=&-\frac{1}{4}F_{\mu \nu} F^{\mu \nu} +
   \overline{\psi}(i \not\! {\partial } +g \not\!\! {A }-
    m )\psi
\,\label{eq:2.1}\end{eqnarray}
is considered here in the light-cone gauge $A^{+}=A^0+A^3=0$. 
Zero modes will be disregarded.
The constrained degrees of freedom, $A^-$ and $\psi_-$
($\Lambda_{\pm}=\frac{1}{2}\gamma^0\gamma^{\pm}$ 
are projection operators, thus $\psi_{\pm}=\Lambda_{\pm}\psi$,
and $\psi=\psi_++\psi_-$)
are removed explicitly and produce the canonical QED/QCD Hamiltonian. 
It is  defined through the independent
physical fields $A_{\perp}$ and $\psi_+$.
To solve the constrained equations for $A^-$ and $\psi_-$  
the auxiliary fields 
\begin{eqnarray}
&& \widetilde A_+ = A_+ - {g\over (i\partial^+)^2}\,J^+ 
\,,\nonumber\\
&& \widetilde\Psi = \Psi _+
   + \left(  m\beta - i\alpha ^i \partial  _{\!\perp i}\right) 
   {1\over 2i\partial _-} \Psi _+
\,,\label{eq:2.2}\end{eqnarray}
are introduced. The fermion current is 
$\widetilde J ^{\mu} (x) 
   =\overline{\widetilde\Psi}\gamma^{\mu} \widetilde\Psi$.  
The resulting canonical Hamiltonian $H=P_+$ is given  
as a sum of the free Hamiltonian and the interaction
\begin{equation}
 H=P_{+}=H_0 + V + W
\,.\label{2.3} \end{equation}
The free Hamiltonian $H_0$, the quark and gluon kinetic energies, is 
\begin{equation}
   H_0 = {1\over2}\int\!dx_+d^2x_{\!\perp} 
   \biggl(\overline{\widetilde\Psi} \gamma^+
   {m^2 +(i\nabla_{\!\!\perp}) ^2 \over i\partial^+}
   \widetilde\Psi   +
   \widetilde A ^\mu (i\nabla_{\!\!\perp}) ^2 
   \widetilde A _\mu     \biggr)
\,.\label{eq:2.4}\end{equation}
In the interaction $V+W$, the vertex interaction $V$ is 
the light-cone analogue of the minimal coupling interaction
in covariant QED/QCD and $W=W_1+W_2$ is the sum of the instantaneous-gluon 
$W_1$ and the instantaneous-quark interactions $W_2$.
The latter arise from the constraint equations
(analog of the light-front Gauss law). 
More explicitly, the interaction is given by
\begin{eqnarray}
   V &=& g  \int\!dx_+d^2x_{\!\perp} 
    \ \widetilde J ^\mu \widetilde A _\mu 
,\nonumber\\
   W_1 &=& { g  ^2 \over 2} \int\!dx_+d^2x_{\!\perp} 
      \ \widetilde J ^+
      {1\over \left(i \partial ^+ \right)^2} \widetilde J ^+ 
,\nonumber\\
   W_2 &=& { g  ^2 \over2} \int\!dx_+d^2x_{\!\perp} 
    \ \overline{\widetilde\Psi}  \gamma ^\mu 
    \widetilde A _\mu \ {\gamma ^+\over i\partial ^+}
    \left( \gamma ^\nu  \widetilde A _\nu 
    \widetilde\Psi  \right)
,\nonumber\\
   V^{\prime} &=& \int\!dx_+d^2x_{\!\perp}
  \widetilde B^{\mu\nu}_{a}\widetilde B_{\mu\nu}^{a}
\,,\label{eq:2.5}\end{eqnarray}
where the current includes the quark and gluon parts
\begin{eqnarray}
\widetilde J ^{\mu} (x) 
   &=& \overline{\widetilde\Psi}\gamma^{\mu} \widetilde\Psi T^{a}
+\frac{1}{i}[\widetilde F^{\mu k},\widetilde A_{k}]
\,,\label{eq:2.6}\end{eqnarray}
with $\widetilde A_{\mu}=A_{\mu}^{a}T^{a}$.
The instantaneous gluon and quark interactions behave as 
$1/q^{+2}$ and $1/q^{+}$, respectively, with the momentum transfer 
$q=(q_{\perp},q^{+})$.

It is convenient to work in a second quantized form, decomposing
the physical fields through creation and annihilation operators.
By definition, the fields 
$\widetilde\Psi=\widetilde\Psi_++\widetilde\Psi_-$ and 
$\widetilde A^\mu = \left( 0,\vec A_{\!\perp },\widetilde A^+ \right)$ 
are the free solutions which in momentum space are parametrized as 
\begin{eqnarray}
   \widetilde\Psi  _{\alpha } (x) &=&
   \sum _\lambda\!\int\!\!  
   {dp^+ d^2p_{\!\bot}\over \sqrt{2p^+(2\pi)^3}}
   \left( b (p) u_\alpha (p ,\lambda ) e^{-ipx} +
   d^\dagger (p)v_\alpha(p,\lambda) 
   e^{+ipx}\right)
,\nonumber \\
   \widetilde A _\mu (x) &=& 
   \sum _\lambda\!\int\!\!  
   {dp^+ d^2p_{\!\bot}\over \sqrt{2p^+(2\pi)^3}}
   \left( a(p)\epsilon_\mu(p,\lambda) e^{-ipx} +
   a^\dagger(p)\epsilon_\mu^\star(p,\lambda)
   e^{+ipx} \right) 
\,.\label{eq:2.7}\end{eqnarray} 
The single particle operators obey the commutation relations 
\begin{equation} 
   \left [ a(p), a^\dagger(p^\prime)\right]  = 
   \left\{ b(p), b^\dagger(p^\prime)\right\} = 
   \left\{ d(p), d^\dagger(p^\prime)\right\} = 
   \delta (p^+-p^{+\,\prime}) 
   \delta ^{(2)}(\vec p_{\!\bot}-\vec p_{\!\bot} ^{\,\prime}) 
   \delta _\lambda ^{\lambda ^\prime}  
\,.\label{eq:2.8}\end{equation}
The free part is given by 
\begin{eqnarray}
   \widehat H_0(l) &=&\int\!{[d^3p]\over\sqrt{2p^+}}  
   \ E(p;l)
   \left(b^\dagger(p)b(p)+d^\dagger(p)d(p)\right)
   + \int\!{[d^3p]\over\sqrt{2p^+}}
   \ \omega(p;l)\,a^\dagger(p)a(p)
\,.\label{eq:2.9}\end{eqnarray}
The single particle energies ($E=p^{-}$) depend on the
3-momentum $p=(p^+,\vec p_\perp)$
\begin{eqnarray}
   E(p;l) = \frac{\vec{p}_{\perp}^{\ 2}+m^2(p;l)} {p^+}\,,\,
   \omega(p;l) = \frac{\vec{p}_{\perp}^{\ 2}+\mu^2(p;l)} {p^+}
\,,\label{eq:2.10}\end{eqnarray}
and potentially on the flow parameter through
the mass $m^2(p;l)$ of the particle in question.
Inserting the free fields into the Hamiltonian yields 
for the quark-gluon vertex interaction
\begin{eqnarray}
   \widehat V(l) &=& {1\over\sqrt{(2\pi)^3}} 
   \int\!{[d^3p_1]\over\sqrt{2p^+_1}} 
   \int\!{[d^3p_2]\over\sqrt{2p^+_2}} 
   \int\!{[d^3p_3]\over\sqrt{2p^+_2}}
   \ \delta^{(3)}(p_{1} - p_{2} - p_{2}) 
  \nonumber\\
   && \left[g(p_1,p_2,p_3;l)
   b^\dagger(p_1) b(p_2) a(p_3) \, 
   \overline u(p_1) \slash\!\!\!\epsilon(p_3) u(p_2)+ ...  
   \right] 
\,,\label{eq:2.11}\end{eqnarray}
and the quark-antiquark interaction in the exchange channel
\begin{eqnarray}
   \widehat  W_{e\bar{e}}(l) &=&{1\over (2\pi)^3} 
   \int\!{[d^3p_1]\over\sqrt{2p^+_1}} 
   \int\!{[d^3p_2]\over\sqrt{2p^+_2}} 
   \int\!{[d^3p_3]\over\sqrt{2p^+_3}}
   \int\!{[d^3p_4]\over\sqrt{2p^+_4}}
   \ \delta^{(3)}(p_1-p_2+p_3-p_4)
\nonumber\\  
   &&  \left[ W_{\mu\nu}(p_1,p_2,p_3,p_4;l)
   \ b^\dagger (p_1) b(p_2)
    d (-p_3) d^{\dagger}(-p_4) T_{12}^{a}T_{34}^{a}\right.
\nonumber\\
   && \left.  \overline u(p_1)\gamma^{\mu}u(p_2)
       \overline v(-p_3)\gamma^{\nu}v(-p_4)+...
   \right] 
\,.\label{eq:2.12}\end{eqnarray}
The integration symbols denote
\begin{equation}
   [d^3p]= dp^+ d^2p_{\perp}\sum _{\lambda}
\,,\label{eq:2.13}\end{equation}
and the abbreviations $u(p)\equiv u(p,\lambda)$ and 
$\slash\!\!\!\epsilon^\star\equiv\gamma^\mu\epsilon\mu^\star(p,\lambda)$
are introduced. 
When unitary transformation with the finite flow parameter $l$ is applied, 
all coupling constants and masses become unknown functions of momenta and $l$
(This is reflected in explicit dependence of $g$ and $m,\,\mu$, and $W_{\mu\nu}$). 
Initial conditions are given at $l=0$.
The effective coupling constant has the initial value
\begin{eqnarray}
   g(l=0) &=& g
\,,\label{eq:2.14}\end{eqnarray}
with the fine structure constant $\alpha=g^2/4\pi\sim 1/137$ in QED,
and the strong coupling $g=g_s$ in QCD. The two-point interaction includes 
the instantaneous and dynamical generated
by flow equations interactions, $W=W^{inst}+W^{gen}$, with initial
conditions
\begin{eqnarray} 
   W_{\mu\nu}^{inst}(l=0) &=& -\frac{\eta_{\mu}\eta_{\nu}}{q^{\,+2}}\,,\,
   W_{\mu\nu}^{gen}(l=0) = 0
\,,\label{eq:2.15}\end{eqnarray}
where the momentum transfer is $q=p_1-p_2$.

\subsubsection{Effective Hamiltonian for the light-front QED}

Representing through creation and annihilation operators, we define explicitly
the particle number conserving and particle number changing parts
which directly define the generator of transformation. In QED,
all canonical interactions are shown in the Table \ref{tab:2.1}; for example, 
the electron-photon minimal coupling is the matrix element $H_{12}$, 
the electron and photon instantaneous interactions are $H_{14}$ and
$H_{22}$, respectively. Particle number changing terms
are off-diagonal in this table and are eliminated using flow equations.

The generator of the unitary transformation is
\begin{eqnarray}
\widehat\eta(l) &=& \widehat\eta_1(l)+\widehat \eta_2(l)
\,.\label{eq:2.16}\end{eqnarray}
where the generator  
$\widehat \eta_1 = [\widehat H_0,\widehat V]$ eliminates
the electron-photon coupling $V$ and is given by 
\begin{eqnarray}
   \widehat \eta (l) = {1\over\sqrt{(2\pi)^3}}
   \int\!{[d^3p_1]\over\sqrt{2p^+_1}} 
   \int\!{[d^3p_2]\over\sqrt{2p^+_2}} 
   \int\!{[d^3p_3]\over\sqrt{2p^+_3}}
   \,\delta^{(3)}(p_{1} - p_{2} - p_{3}) \, 
\nonumber\\
   \times  
   \left[\eta (p_1,p_2,p_3;l) b^\dagger_1 b_2 a_3 \, 
   \overline u(p_1) \slash\!\!\!\epsilon(p_3) u(p_2)+...\right] 
\,,\label{eq:2.17}\end{eqnarray}
and the generator $\widehat \eta_2 = [\widehat H_0,\widehat W]$
eliminates the off-diagonal two-point interactions and its explicit
form depends on the particular sector where is $\widehat W$.

\begin{table}
\vspace{2cm}
\begin{tabular}{|r|c|c|c|c|c|} \hline
 & $|\gamma>$ & $|e\bar{e}>$ & $|\gamma\gamma>$ & $|e\bar{e}\gamma>$ 
& $|e\bar{e}e\bar{e}>$   \\ \hline
$|\gamma>$ &  & \floadeps{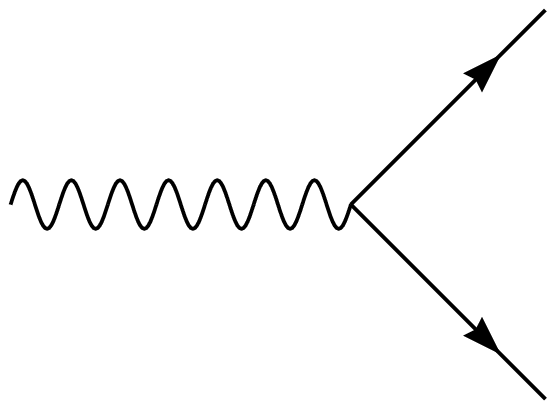} 
&     & \floadeps{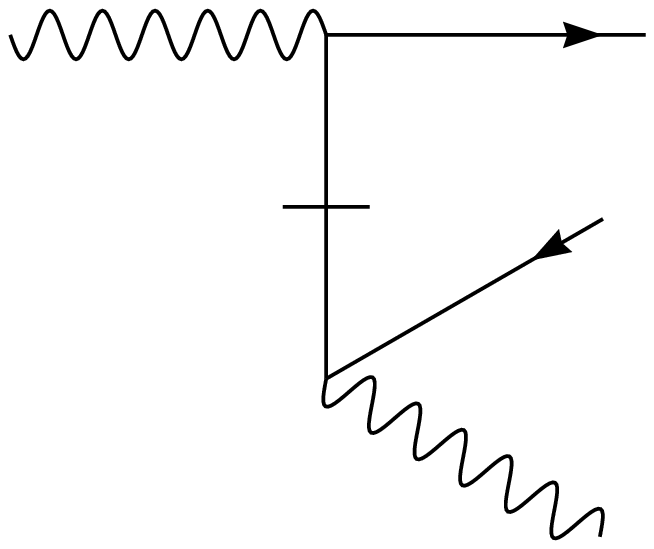} &  \\ \hline
$|e\bar{e}>$ & \floadeps{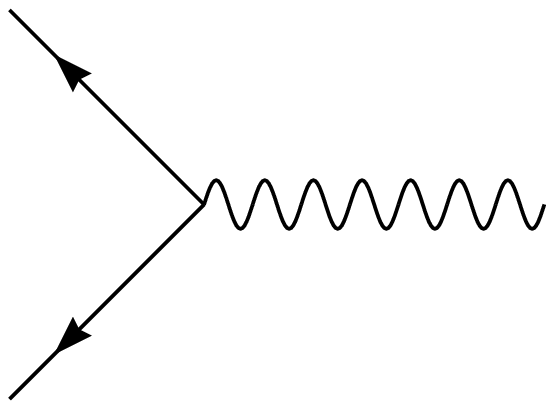} & \floadeps{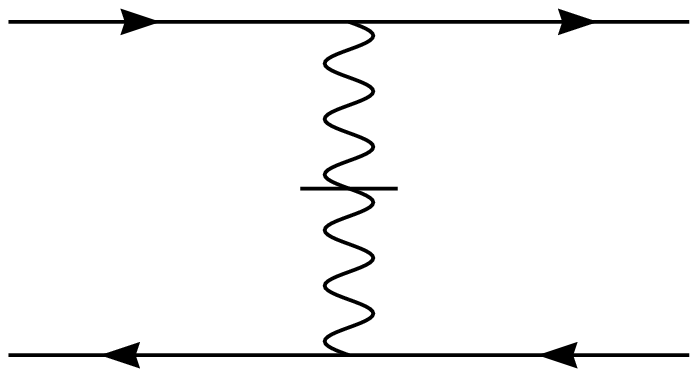} 
& \floadeps{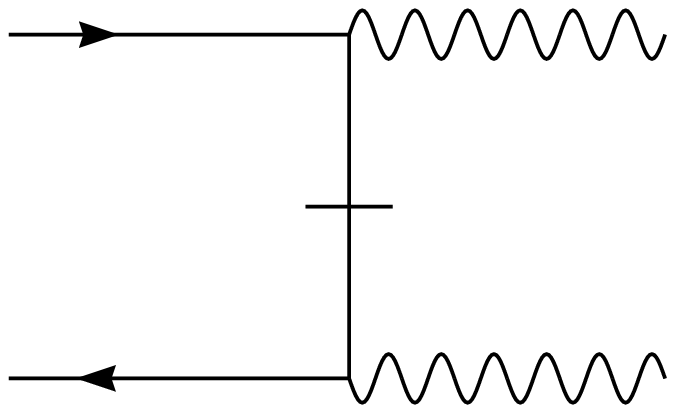} & \floadeps{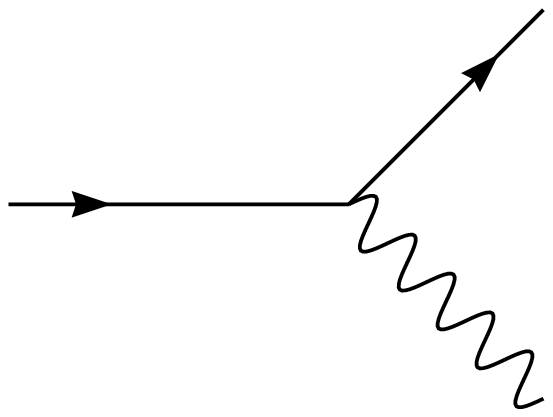} & \floadeps{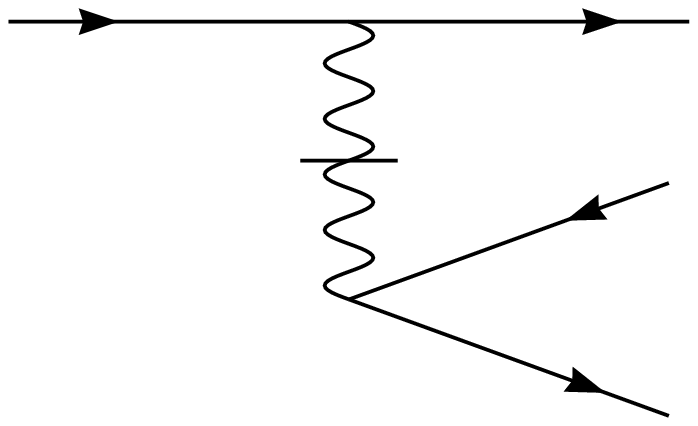} \\ \hline
$|\gamma\gamma>$ &             & \floadeps{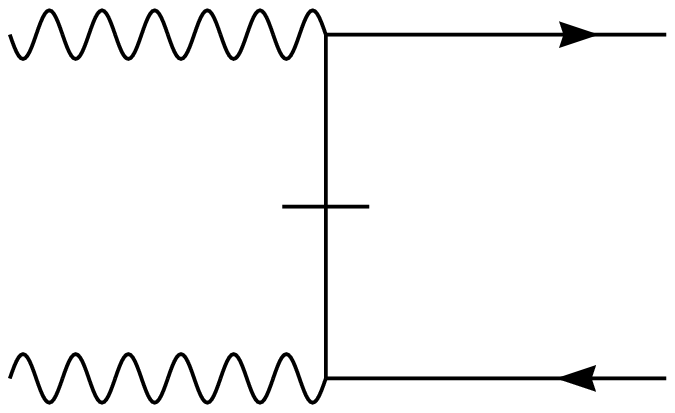} 
&            & \floadeps{tab12} &  \\ \hline
$|e\bar{e}\gamma>$ & \floadeps{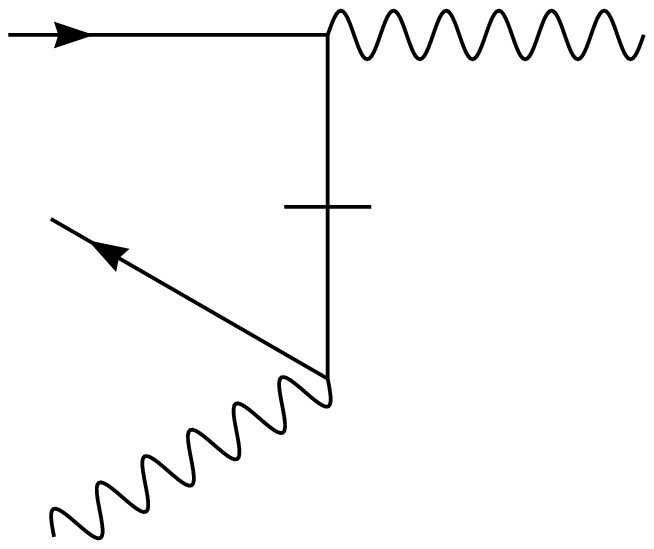} & \floadeps{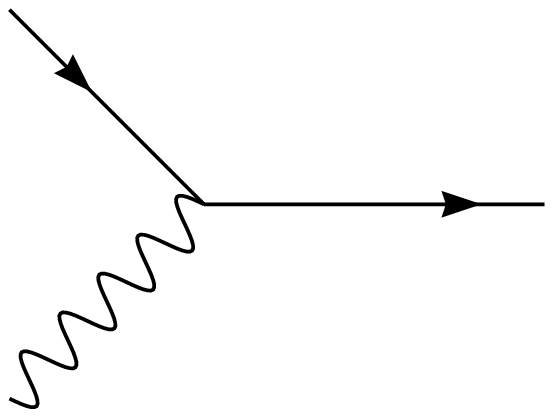} 
& \floadeps{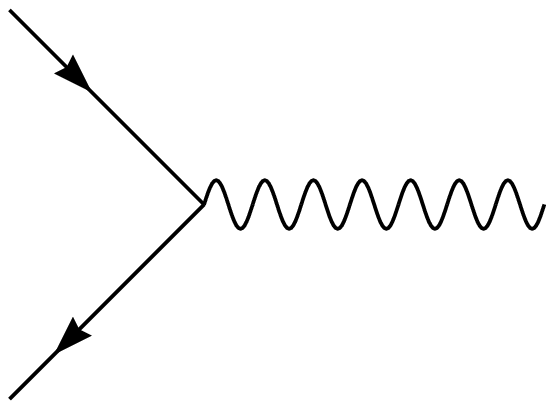} & \floadeps{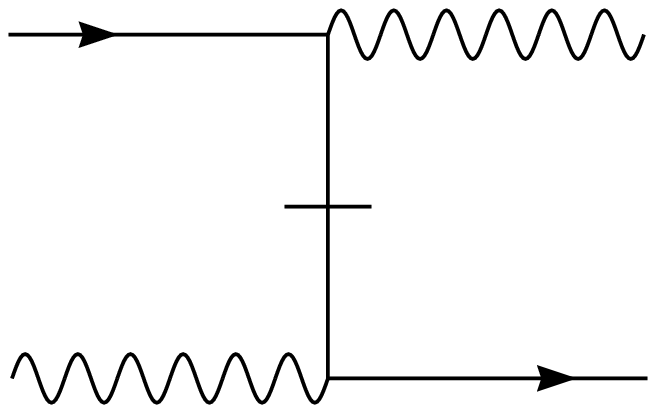} & \floadeps{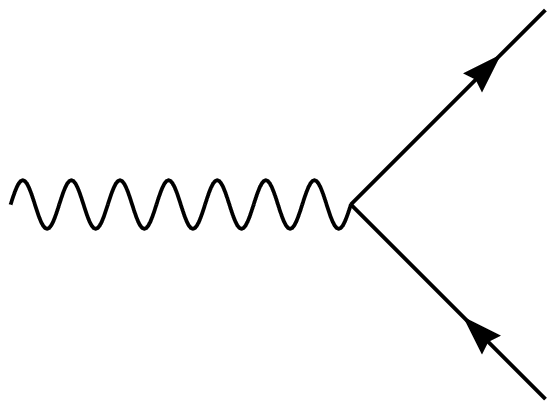} \\ \hline
$|e\bar{e}e\bar{e}>$ &        & \floadeps{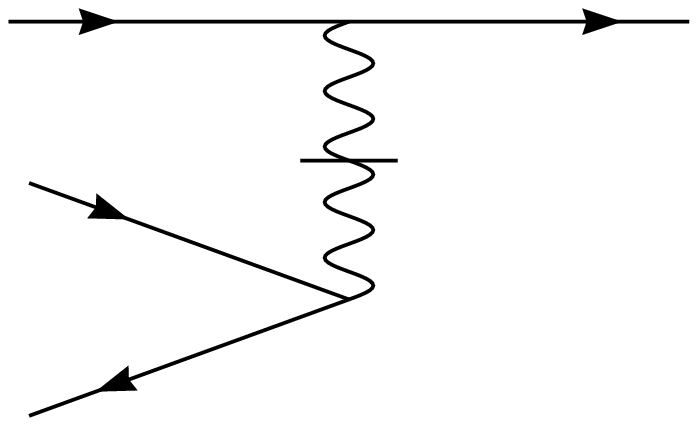} 
&      & \floadeps{tab43} & \floadeps{table22a} \\ \hline
\end{tabular}
\vspace{1cm}
\caption{QED canonical Hamiltonian in the light-front gauge.}
\label{tab:2.1}
\end{table}

\begin{table}
\vspace{2cm}
\begin{tabular}{|r|c|c|c|c|c|} \hline
 & $|\gamma>$ & $|e\bar{e}>$ & $|\gamma\gamma>$ & $|e\bar{e}\gamma>$ 
& $|e\bar{e}e\bar{e}>$   \\ \hline
$|\gamma>$ & \floadeps{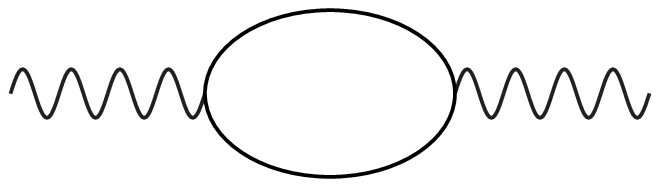} &  &     &   &  \\ \hline
$|e\bar{e}>$ &  & \floadeps{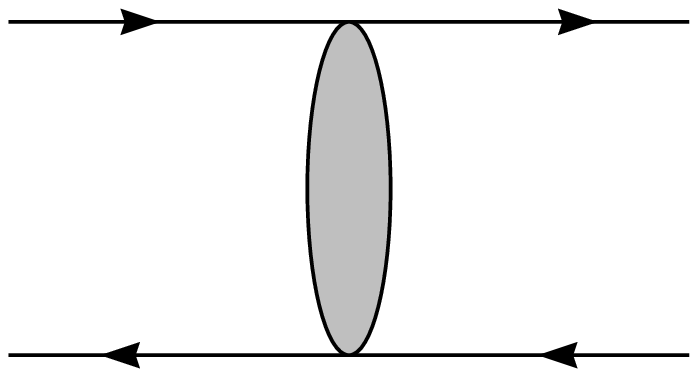} &  &  &  \\ \hline
$|\gamma\gamma>$ &  &  & \floadeps{} & &  \\ \hline
$|e\bar{e}\gamma>$ &  &  
&  & \floadeps{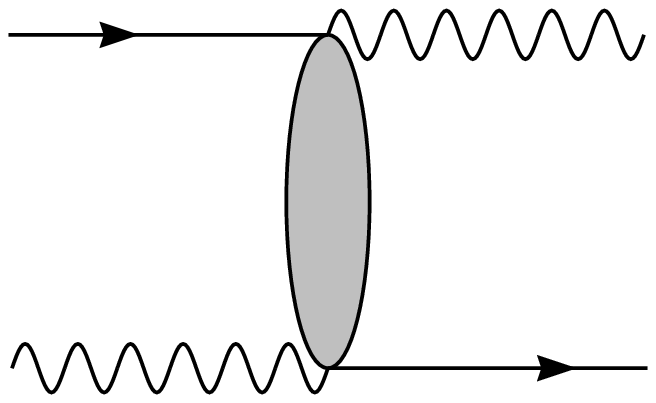} &  \\ \hline
$|e\bar{e}e\bar{e}>$ &        &  &      &  & \floadeps{table22} \\ \hline
\end{tabular}
\vspace{1cm}
\caption{Effective QED Hamiltonian generated through the second order 
by flow equations in the light-front gauge.}
\label{tab:2.2}
\end{table}

QED coupling is a small parameter. Eliminating the particle number changing
sector to the second order in coupling, we generate new terms
which are given in the Table \ref{tab:2.2}. In the second order $O(g^2)$,
new dynamical two-point interactions arise from elimination the electron-photon 
vertex terms, while elimination of the instantaneous terms $W$ generates
new interactions in the third and high orders which are not depicted. 
In the Table \ref{tab:2.2}, black area in the two-point
interaction depicts an effective kernel which depends on all four in- and 
out-going momenta.

Effective electron-positron interaction (matrix element $H_{22}$) includes 
the instantaneous and dynamically generated by flow equations terms. 
In the exchange channel, it is given by 
a product of the current-current term and the interaction kernel,
\begin{eqnarray}
W_{e\bar{e}} &=&-4\pi\alpha_sC_f\langle\gamma^{\mu}\gamma^{\nu}\rangle
B_{\mu\nu}
\,,\label{eq:2.18}\end{eqnarray}
where the photon interaction kernel 
\begin{eqnarray}
B_{\mu\nu} &=& g_{\mu\nu}\left(I_1+I_2\right)
+\eta_{\mu}\eta_{\nu} \frac{\delta Q^2}{q^{+2}}
\left(I_1-I_2\right)
\,,\label{eq:2.19}\end{eqnarray}
contains the integrals $I_1$ and $I_2$ which are defined by
two similarity form factors in each vertex
\begin{eqnarray}
I_1=\int_0^{\infty}d\lambda\frac{1}{Q_1^2}
\frac{df(Q_1^2;\lambda)}{d\lambda}
f(Q_2^2;\lambda)
\,,\label{eq:2.20}\end{eqnarray}
and for $I_2$ the indeces $1$ and $2$ are interchanged.
Here, $Q_1^2$ and $Q_2^2$ are momenta transfers in two verteces.
In the light-front framework, they are given by
\begin{eqnarray}
Q_1^2 &=& \frac{(x'k_{\perp}-xk'_{\perp})^2+m^2(x-x')^2}{xx'}
\nonumber\\
Q_2^2 &=& Q_1^2|_{x\rightarrow (1-x);~x'\rightarrow (1-x')}
\,,\label{eq:2.21}\end{eqnarray} 
On mass shell, these momenta are reduced to the three photon
momentum transfer, 
\begin{eqnarray}
Q_1^2=Q_2^2 \rightarrow \vec{q}^2
\,.\label{eq:2.22}\end{eqnarray}
Using three choices of similarity function; 
exponential $f={\rm exp}\left(-Q^2/\lambda^2\right)$,
gaussian $f={\rm exp}\left(-Q^4/\lambda^4\right)$
and sharp $f=\theta\left(1-Q^2/\lambda^2\right)$;
we obtain the following kernels for the effective 
electron-positron interaction,  
\begin{eqnarray}
B_{\mu\nu} &=& g_{\mu\nu}
\frac{1}{Q^2}
\nonumber\\
B_{\mu\nu} &=& g_{\mu\nu}
\frac{1}{Q^2}
-\left[\frac{g_{\mu\nu}}{Q^2}
-\frac{\eta_{\mu}\eta_{\nu}}{q^{+2}}\right]
\frac{\delta Q^4}{Q^4+\delta Q^4} 
\nonumber\\
B_{\mu\nu} &=& g_{\mu\nu}
\frac{1}{Q^2}
-\left[
\frac{g_{\mu\nu}}{Q^2}
- \frac{\eta_{\mu}\eta_{\nu}}{q^{+2}}
\right]
\frac{\left|\delta Q^2\right|}
{Q^2+\left|\delta Q^2\right|} 
\,,\label{eq:2.23}\end{eqnarray}
respectively. Here, $Q^2$ is the average momentum transfer and 
$\delta Q^2$ shows the off-shellness.
On the energy shell, i.e. $\delta Q^2=0$, the electron-positron 
interaction is reduced in each case to the $3$-d Coloumb potential
with behavior $1/\vec{q}^2$.

Picking the $e\bar{e}$-sector, we solve an effective interaction 
for positronium bound states.
The light front Schr\"odinger equation for the positronium model reads
\begin{eqnarray}
H_{LC}|\psi_n>=M_n^2|\psi_n>
\,,\label{eq:2.24}\end{eqnarray}
where $H_{LC}=P^{\mu}P_{\mu}$ is the invariant mass (squared) operator, 
refered for convenience to as the light front Hamiltonian of positronium 
and $|\psi_n>$ being the corresponding eigenfunction; $n$ labels all
the quantum numbers of the state. Projecting this equation on 
the $e\bar{e}$ state yields  
\begin{eqnarray}
H_{LC}^{eff}|(e\bar{e})_n>=M_n^2|(e\bar{e})_n>.
\,.\label{eq:2.25}\end{eqnarray}
In the $e\bar{e}$ sector, the effective light-front Hamiltonian consists of
the free part and the effective electron-positron interaction
\begin{eqnarray}
H_{LC}^{eff}=H_{LC}^{(0)}+V_{LC}^{eff}
\,,\label{eq:2.26}\end{eqnarray}
The light front equation Eq. (\ref{eq:2.25}) is then expressed 
by the integral equation 
\begin{eqnarray}
&& \hspace{-2cm} \left(\frac{m^2+\vec{k}^{'2}}{x'(1-x')}-M_n^2\right)
\psi_n(x',\vec{k}_{\perp}^{'};s_3,s_4)
\nonumber\\
&+&\sum_{s_1,s_2}\int_D \frac{dx d^2 k_{\perp}}{2(2\pi)^3}
<x',\vec{k}_{\perp}^{'};s_3,s_4|V_{LC}^{eff}|x,\vec{k}_{\perp};s_1,s_2>
\psi_n(x,\vec{k}_{\perp};s_1,s_2)=0
\,,\label{eq:2.27}\end{eqnarray}
where the integration domain $D$ is restricted by the covariant cutoff
condition of Brodsky and Lepage 
\begin{eqnarray}
\frac{m^2+\vec{k}^{2}}{x(1-x)}\leq \Lambda^2+4m^2
\,,\label{eq:2.28}\end{eqnarray}
allowing for states which have a kinetic energy below the 
cutoff $\Lambda$.
The effective $e\bar{e}$ interaction, $V_{LC}^{eff}$, contains
two-point dynamical and instantaneous interactions 
(with the kernel $W_{e\bar{e}}$ given by Eqs. (\ref{eq:2.18}) and 
(\ref{eq:2.23})) as well as the electron self-energy terms in $e\bar{e}$ sector 
which are renormalized by the electron mass counterterm.
We include the exchange and annihilation channels
in the effective interaction, $V_{LC}^{eff}=V_{exch}+V_{ann}$.

\begin{figure}[thbp]
\centerline{\epsfxsize=\textwidth \epsfbox{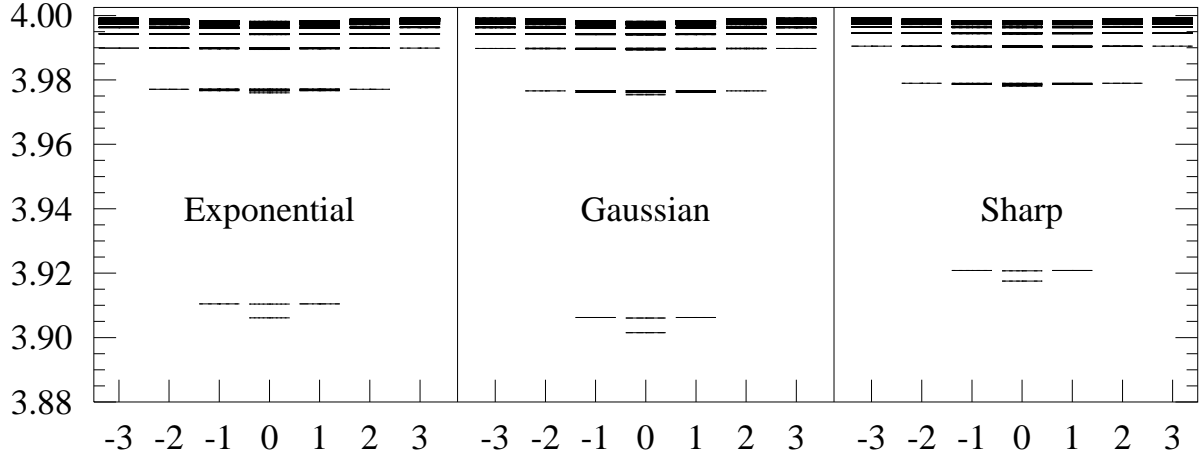}}
\caption{The invariant mass-squared spectrum $M_i^2$ for positronium 
   versus the projection of the total spin, $J_z$, excluding
   annihilation with exponential, Gaussian and sharp cutoffs.
   The number of integration points is $N_1=N_2=21$.}
\label{fig:2.1}
\end{figure}

\begin{figure}[thbp]
\centerline{\epsfxsize=\textwidth \epsfbox{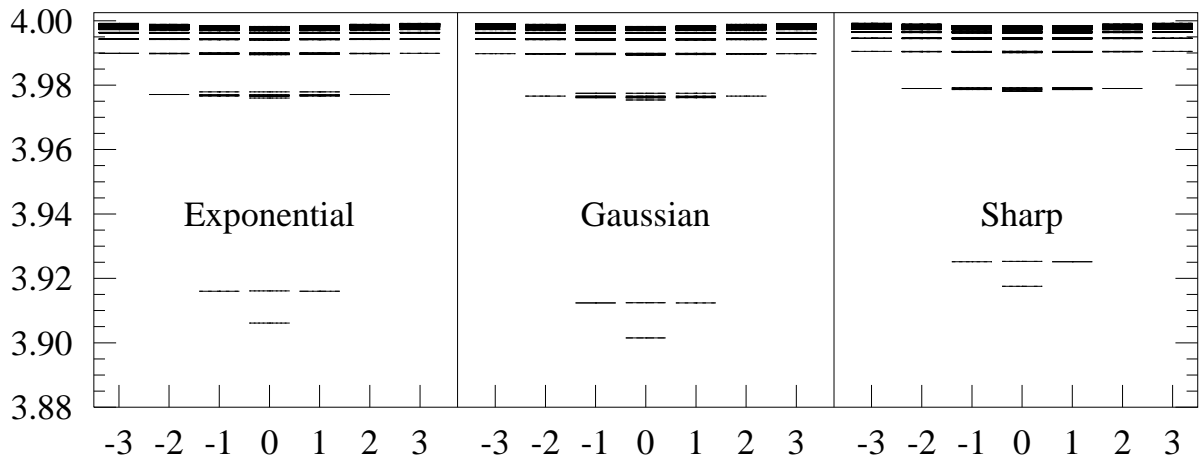}}
\caption{The invariant mass-squared spectrum $M_i^2$ for positronium 
   versus the projection of the total spin, $J_z$, including
   annihilation with exponential, Gaussian and sharp cutoffs.
   The number of integration points is $N_1=N_2=21$.}
\label{fig:2.2}
\end{figure}

Numerical solution of the integral equation (\ref{eq:2.27}) 
produces positronium spectrum, without (Fig. \ref{fig:2.1})
and with (Fig. \ref{fig:2.2}) annihilation channel. 
Including the annihilation channel, we obtain for the singlet-triplet 
splitting a supprising agreement with the equal time calculations.
Degeneracy of triplet state shows that though calculations are done 
in the light-front framework the rotational invariance is manifest.
Next excited states ($n=2$) also show good agreement with the results
of the equal time perturbation theory.

\subsubsection{Effective Hamiltonian for the light-front QCD}

QED calculations have been done in terms of bare unrenormalized parameters
and success of these calculations is based on aplicability of perturbation theory
in terms of small QED charge. The same naive picture does not hold for QCD;
coupling constant is strong. However, as we show below 
(see \cite{LFQCD} for details), renormalization of the gluon energies 
might change the situation and QCD calculations can be still
made using flow equations.

The effective quark-antiquark interaction including instantaneous 
and dynamically generated by flow equations terms is given by 
\begin{eqnarray}
V_{q\bar{q}} &=&-4\pi\alpha_sC_f\langle\gamma^{\mu}\gamma^{\nu}\rangle
B_{\mu\nu}
\,,\label{eq:2.29}\end{eqnarray}
which formally coincides with the equation (\ref{eq:2.18}) for the $e\bar{e}$
interaction in QED. However, the gluon interaction kernel is given by 
\begin{eqnarray}
B_{\mu\nu}=\lim_{\mu_0,\lambda_0\rightarrow 0}\left[
g_{\mu\nu}\left(I_1+I_2\right)
+\eta_{\mu}\eta_{\nu} \frac{\delta Q^2}{q^{+2}}
\left(I_1-I_2\right)\right]
\,,\label{eq:2.30}\end{eqnarray}
where the integrals $I_1$ and $I_2$ are defined by
two similarity form factors containing the cut-off dependent
momenta transfer in each vertex,
\begin{eqnarray}
I_1=\int_0^{\infty}d\lambda\frac{1}{Q_1^2(\lambda)}
\frac{df(Q_1^2(\lambda);\lambda)}{d\lambda}
f(Q_2^2(\lambda);\lambda)
\,,\label{eq:2.31}\end{eqnarray}
and with interchange $1$ and $2$ indeces for $I_2$. The gluon interaction
kernel Eq. (\ref{eq:2.30})) is defined in the limit of zero gluon mass, $\mu_0=0$, 
and zero renormalization point, $\lambda_0=0$, 
that is the renormalization condition
in the gluon gap equation (see below).  
In the light-front framework, momenta transfer are given by
\begin{eqnarray}
Q_1^2(\lambda) &=& \frac{(x'k_{\perp}-xk'_{\perp})^2+m^2(x-x')^2}{xx'}
+\mu^2_{ren}(\lambda)
\nonumber\\
Q_2^2(\lambda) &=& Q_1^2(\lambda)|_{x\rightarrow (1-x);~x'\rightarrow (1-x')}
\,,\label{eq:2.32}\end{eqnarray} 
where $\mu_{ren}(\lambda)$ is the renormalized gluon mass. On mass shell,
momenta transfer are reduced to an effective gluon energy, 
\begin{eqnarray}
Q_1^2(\lambda)=Q_2^2(\lambda)\rightarrow \vec{q}^2+\mu^2_{ren}(\lambda)
\,,\label{eq:2.33}\end{eqnarray}
which contains three gluon momentum transfer and the renormalized gluon mass.
Due to the presence of the renormalized gluon mass flow equations 
eliminate off-diagonal matrix elements even for vanishing gluon momenta,
$\vec{q}=0$, that corresponds to degenerate states. This cannot be achieved
within perturbation theory (PT), since it breaks down at $\vec{q}=0$ and
an effective interaction has zero energy denominators. Nontrivial solution
for the effective gluon mass (gluon gap) is a consequence of a special behavior of
the light-front QCD in the collinear limit, $x^{+}\rightarrow 0$,
in the contrary to QED (there is no mass gap in QED). 
As mentioned before, this behavior of QCD might be attributed to QCD vacuum effects.

The renormalized gluon mass is defined as a solution of the gluon gap equation,
$\mu(\lambda)$, renormalized by the canonical gluon mass counterterm 
of the perturbation theory, $m_{CT}$,
\begin{eqnarray}
\mu^2_{ren}(\lambda)=\mu^2(\lambda)+m_{CT}^2(\lambda)
\,.\label{eq:2.34}\end{eqnarray}
One-point flow equations for the light-front energies provide
the gap equations for the quark ($m^2$) and gluon ($\mu^2$) 
renormalized masses  
\begin{eqnarray}
\frac{dm^2(\lambda)}{d\lambda} &=&
C_f\int_0^1\frac{dx}{x(1-x)}
\int_0^{\infty}\frac{d^2k_{\perp}}{16\pi^3}
g_q^2(\lambda)\frac{1}{Q^2_3(\lambda)}
\frac{df^2(Q^2_3(\lambda);\lambda)}{d\lambda}
\nonumber\\
&\times&
\left[k_{\perp}^2(\frac{2}{1-x}+\frac{4}{x^2})
+2m^2(\lambda)\frac{x^2}{1-x}\right]
\nonumber\\
\frac{d\mu^2(\lambda)}{d\lambda} &=&
2T_fN_f\int_0^1\frac{dx}{x(1-x)}
\int_0^{\infty}\frac{d^2k_{\perp}}{16\pi^3}
g_q^2(\lambda)\frac{1}{Q^2_2(\lambda)}
\frac{df^2(Q^2_2(\lambda);\lambda)}{d\lambda}
\nonumber\\
&\times&
\left[\frac{k_{\perp}^2+m^2}{x(1-x)}
-2k_{\perp}^2\right]
\nonumber\\
&+& 2C_a\int_0^1\frac{dx}{x(1-x)}
\int_0^{\infty}\frac{d^2k_{\perp}}{16\pi^3}
g_g^2(\lambda)\frac{1}{Q_1^2(\lambda)}
\frac{df^2(Q_1^2(\lambda);\lambda)}{d\lambda}
\nonumber\\
&\times&
\left[k_{\perp}^2(1+\frac{1}{x^2}+\frac{1}{(1-x)^2})\right]
\,,\label{eq:2.35}\end{eqnarray}
where energies are given by
\begin{eqnarray}
Q_1^2(\lambda) &=& \frac{k_{\perp}^2+\mu^2(\lambda)}{x(1-x)}
-\mu^2(\lambda)
\nonumber\\
Q_2^2(\lambda) &=&
\frac{k_{\perp}^2+m^2}{x(1-x)}
-\mu^2(\lambda)
\nonumber\\
Q_3^2(\lambda) &=& \frac{k_{\perp}^2+m^2(\lambda)}{x}
+\frac{k_{\perp}^2+\mu^2(\lambda)}{(1-x)}
-m^2(\lambda)
\,.\label{eq:2.36}\end{eqnarray}
These equations (\ref{eq:2.35})
are coupled, since unknown cut-off dependent quark and gluon masses enter
the r.h.s. of both equations. The energy denominators, Eq. (\ref{eq:2.36}), 
show a one-loop structure. When expanded in the coupling constant 
at large cut-offs, $\lambda=\Lambda_{UV}$, these equations are reduced to 
the Dyson-Schwinger equation in the rainbow approximation. 
We solve the gluon gap equation with a constant
current quark mass, using special prescription to regulate collinear small
light-front $x$ in the integral over $dx$. Absorbing trivial perturbative
cut-off behavior by a canonical mass counterterm, Eq. (\ref{eq:2.34}),
we obtain the renormalized cut-off dependent effective gluon mass. 

Taking into account the cut-off dependent renormalized
gluon mass, the quark-antiquark interaction (Eq. (\ref{eq:2.30}))
has the following form with three choices of similarity function,
\begin{eqnarray}
B_{\mu\nu} &=& g_{\mu\nu}\left[
\frac{1}{Q^2}
+\frac{\sigma}{Q^4}\right]
+\left[\frac{g_{\mu\nu}}{Q^2}
-\frac{\eta_{\mu}\eta_{\nu}}{q^{+2}}\right]
\frac{\sigma}{Q^2}
\frac{\delta Q^4}{Q^4-\delta Q^4} 
\nonumber\\
B_{\mu\nu} &=& g_{\mu\nu}\left[
\frac{1}{Q^2}
+\frac{\sigma}{Q^4}\right]
-\left[\frac{g_{\mu\nu}}{Q^2}
(1+\frac{\sigma}{Q^2})
-\frac{\eta_{\mu}\eta_{\nu}}{q^{+2}}\right]
\frac{\delta Q^4}{Q^4+\delta Q^4} 
\nonumber\\
B_{\mu\nu} &=& g_{\mu\nu}\left[
\frac{1}{Q^2}
+\frac{\sigma}{Q^4}\right]
-\left[
\frac{g_{\mu\nu}}{Q^2}
(1+\frac{\sigma}{Q^2}
(1+\frac{Q^2}
{Q^2+\left|\delta Q^2\right|}))\right.
\nonumber\\
&-& \left.\frac{\eta_{\mu}\eta_{\nu}}{q^{+2}}
(1+\frac{\sigma}{Q^2}
\frac{Q^2}
{Q^2+\left|\delta Q^2\right|})
\right]
\frac{\left|\delta Q^2\right|}
{Q^2+\left|\delta Q^2\right|} 
\,,\label{eq:2.37}\end{eqnarray}
that should be compared with the electron-positron interaction in QED,
Eq. (\ref{eq:2.23}). On the energy shell, the $q\bar{q}$ interaction 
is reduced to the sum of the Coloumb, $1/\vec{q}^2$, and linear confining, 
$1/\vec{q}^4$, potentials. Here, $\sigma$ plays the role of the string tension
depending on the cut-off which regulates small light-front $x$. 
This result cannot be obtained in the perturbation 
theory by naive summation over perturbative gluon exchanges in a few low 
orders. There are complex calculations, where infinite orders of given
subclasses of diagramms result in a confining interaction.

{\bf Summary II}

1) Though calculations are performed in the fixed light-front gauge where
the light-front and perpendicular coordinates are treated differently,
we obtain the covariant result using flow equtaions
(triplet states are degenerate; effective interaction depends 
on the covariant four-momenta $Q^2$ and $\delta Q^2$). Dynamical 
terms generated by flow quations are crucial to maintain covariance.

2) Flow equations allow to obtain results beyond the perturbation theory.
Also, a covariant result corresponding to resummation of large number of
perturbative diagrams may be obtained by flow equations straightforward.

3) Flow equations allow to track the covariance and gauge invariance much easier 
than in the standard perturbation theory. (To maintain covariance in the PT, 
all diagrams in the given order including crossed diagrams should be summed.
Except for the planar diagrams summation of high orders of PT is a problem.)

\subsection{Flow equations for QCD in the Coulomb gauge}
\label{subsec:3.1}

\subsubsection{Flow equation scaling}

Flow equations eliminate matrix elements which couple states with large
energy differences, greater than $1/\sqrt{l}=\lambda$, and later more
degenerate states. This is reminisent of the energy scaling separation
underlying perturbative scaling. Flow equations treats different energies 
separately, starting from UV and scaling down towards low energies,
separating physics at each characteristic scale.
 
In strong interactions/QCD, moving from UV to IR, we encounter the following
characteristic scales and corresponding regimes: at $\Lambda_{UV}$
the conformal canonical perturbation theory (PT) without any scale
and with bare quarks and gluons as degrees of freedom; 
at $\Lambda_{QCD}$ PT breaks down due to the dimensional transmutation 
in the renormalization group (RG) ($\Lambda_{QCD}$ is the RG invariant scale); 
at $\sqrt{\sigma}$ confinement steps in and hadron bound states
(instead of quarks and gluons) become relevant degrees of freedom; 
at $m_{\pi}$ chiral symetry dynamically breaks, leading to 
the Goldstone boson-pion; nuclear physics decouples further down the scale
with $N$ and $\pi$ as degrees of freedom. 
Physical ranges decouple from each other due to weak interaction
(say in the chiral PT, interaction between the Goldstone mode and hadrons
is weak). Flow equations decouple these regions, providing dynamics
on all energy scales including crossover between weak and strong coupling.

In what follows, we consider regime where physics is dictated
by phenomena of confinement and chiral symmetry breaking and take into account
using flow equations effects of high energy QCD which influence 
the hadron scale physics.

\subsubsection{Model calculations in QCD}

Calculations made in the Coulomb gauge QCD illustrate clear the scheme 
of flow equations. The question we address is, how starting with canonical
QCD Hamiltonian to obtain using flow equations an effective 
block-diagonal Hamiltonian with fixed number of Fock components
in each block. It seems, that this goal is possible to achieve only for  
a 'confined QCD' with constituent quarks and gluons. Indeed, in the presence
of a strong confining interaction current (bare) quarks and gluons
are renormalized, getting dressed and becoming constituent quarks and gluons.
This QCD motivated Hamiltonian can be written 
in a Fock space of constituent quarks $q$ and gluons $g$ as
\begin{eqnarray}
H_{QCD}(q,g)= \left(
  \begin{array}{c|c|c}
  q\bar{q} & &  \\  \hline 
  & q\bar{q}g & \\    \hline
  & & gg \\ 
  & & q\bar{q}q\bar{q}
  \end{array} \right)
\,,\label{eq:3.1}\end{eqnarray}
where in every next Fock state a $q\bar{q}$ pair or a gluon are added. 
In terms of constituent quasiparticles, there is a natural energy gap of order
of one GeV between different sectors. In the gauge field theories,
mixing between sectors is mediated by minimally coupled gauge fields and 
is strongly suppressed. Small parameter is no longer the coupling constant, 
but rather a ratio of the off-diagonal matrix element which mixes sectors to the
energy gap between diagonal sectors. In the canonical QCD, the off-diagonal sector
is given by the mixing Coulomb interaction which value is of order of an inverse
Bohr radius or a current quark mass -- several MeV, and the mass gap between
sectors is defined by a confining scale with a value of order of a 
constituent mass -- one GeV; i.e. a small parameter is
\begin{eqnarray}
\frac{V_{12}}{M_1-M_2}\sim \frac{10 MeV}{1 GeV}\ll 1
\,,\label{eq:3.1a}\end{eqnarray}
where $V_{12}$ is a matrix element of canonical interaction
between the first and second Fock state with masses $M_1$ and
$M_2$ correspondingly. Thus, in the constituent basis, mixing between
sectors is suppressed and can be eliminated perturbatively by flow equations
with a small parameter given by Eq. (\ref{eq:3.1a}).    
However, perturbative expansion in this parameter
holds when working between sectors but not inside a diagonal sector
where a strong confining interaction dominates.
In the effective block-diagonal Hamiltonian, the diagonal sectors 
should be diagonalized numerically.
 
At low energies one could think  
about introducing hadron states as elementary degrees of freedom,
and constructing a phenomenological Hamiltonian 
for strong interactions of the form 
\begin{eqnarray}
H_{strong}(hadron) = \left( 
  \begin{array}{c|c|c}
  meson & & \\ \hline 
  & hybrid & \\ \hline
  && glueball
  \end{array} \right)  
\,,\label{eq:3.2}\end{eqnarray}
where hadron states are put on diagonal with increasing glue content 
from up-left corner to down-right; off-diagonal blocks (here empty cells) 
contain possible interactions which mix different diagonal blocks 
(Fock sectors). In order to find the physical (pure) states   
one should block-diagonalize this Hamiltonian matrix, 
eliminating "off-diagonal" interactions. Probably, one could do it
perturbatively, since as known from phenomenology
the mixing between hadron Fock states
is suppressed and a small parameter for iterative procedure could be
\begin{eqnarray}
\frac{M(hybrid)-M(meson)}{M(hybrid)}\sim\frac{1}{10}\ll 1
\,,\label{eq:3.3}\end{eqnarray}
where $M(hybrid)$ denotes hybrid mass, etc. This description
is along with effective low-energy models of nuclear physics, 
where nucleons $N$ are degrees of freedom and pions $\pi$
are mediating interactions between them.

Summarizing, it seems that the 'confined' QCD provides an effective 
description which matches degrees of freedom in both Hamiltonians given 
by Egs. (\ref{eq:3.1}) and (\ref{eq:3.2}). In this framework, using 
flow equations perturbatively we reduce bound state problem to an 
eigenstate equation with the lowest Fock component of quasiparticles.

\subsubsection{Duality and BV transformation}

Now that we know how to work with the 'confined' QCD which resembles CQM,
the question is, how to get this effective theory. 
In other words, how to transform 
the original canonical QCD which is strong coupled, with nonfixed number of 
particles and has complex vacuum with confinement and chiral symmetry breaking
to the CQM type effective theory which is weak coupled, where the valence
quarks and gluons are confined in bound states, with the chiral symmetry 
dynamically broken, and has simple vacuum state. We are looking for a dual 
transformation between these two theories, in the sense that the  duality is 
between the strong interacting QCD and weak interacting effective theory.

Many body approach suggests, that the BCS type Bogoliubov-Valatin (BV) 
transformation might fill the gap between QCD and CQM. We adopt that the BV 
transformed QCD has chiral invariant simple vacuum, but instead contains chiral 
noninvariant interactions including strong BCS interactions, which reflect 
confinement and the chiral symmetry breaking explicitly, and residual dynamical 
interactions, which are weak. Applying flow equations to the BV transformed 
'confined' QCD, we eliminate off-diagonal dynamical interactions and obtain 
the block-diagonal effective Hamiltonian containing strong confining chiral 
noninvariant interactions in addition to dynamical interactions in each block. 
Diagonalizing sector Hamiltonians numerically, we obtain eigenstates which
are confined and are not invariant under the chiral transformation, and involve
dynamics. Due to the dynamical interactions, we obtain for the first time 
equations and corresponding solutions which are both UV and IR finite.
Also, for the first time, solutions include explicitly confinement and 
chiral symmetry breaking as well as dynamically propagating gluons.

\subsubsection{QCD motivated Hamiltonian in the Coulomb gauge}

One of the ways to implement confining potential in the QCD Hamiltonian
is to use the Couomb gauge, which is defined as $\nabla\cdot A=0$.
The Coulomb gauge QCD Hamiltonian includes the free Hamiltonian, $H_0$, 
and two types of interactions: instantaneous, W, describing 
static properties, and dynamical, $V$, involving propagating gluons,
\begin{eqnarray}
H &=& H_{0}+W+V
\,.\label{eq:3.4}\end{eqnarray}
The free Hamiltonian (quark and gluon kinetic energies) is given by
\begin{eqnarray}
 H_{0} &=& \int d \mbf{x}\psi^{\dagger}(\mbf{x}) 
\left( -i \mbf{\alpha}\cdot\mbf{\nabla} + \beta m \right)\psi(\mbf{x})
\nonumber\\ 
&+& {\rm Tr}\int d \mbf{x} \left( \mbf{\Pi}^2(\mbf{x})
+ \mbf{B}^2_{A}(\mbf{x}) \right)
\,,\label{eq:3.5}\end{eqnarray}
where the non-abelian magnetic field is
$\mbf{B}=B_i=\nabla_j A_k-\nabla_k A_j+g[A_j,A_k]$, and 
its abelian part is represented by $\mbf{B}_{A}$.
Dynamical interaction includes the minimal quark-gluon coupling,
$V_{qg}$, and the non-abelian three- and four-gluon interactions,
$V_{gg}$, i.e. $V = V_{qg} + V_{gg}$.
Explicitly they are given by 
\begin{eqnarray}
 V_{qg} &=& -g \int d \mbf{x}\psi^{\dagger}(\mbf{x})
\mbf{\alpha}\cdot\mbf{A}(\mbf{x}) \psi(\mbf{x})
\nonumber\\
 V_{gg} &=& {\rm Tr}\int d \mbf{x}
\left(J\mbf{\Pi}(\mbf{x})J^{-1}\mbf{\Pi}(\mbf{x})-
\mbf{\Pi}^2(\mbf{x})\right) 
+{\rm Tr}\int d \mbf{x}
\left(\mbf{B}^2(\mbf{x})-\mbf{B}^2_{A}(\mbf{x})\right)
\,.\label{eq:3.6}\end{eqnarray}
The Coulomb gauge fixing produces the instantaneous quark-quark and 
gluon-gluon interactions,
\begin{eqnarray}
 W &=& \frac{1}{2}g^2\int d\mbf{x}d\mbf{y}
J^{-1}\rho^{a}(\mbf{x})
\langle\mbf{x},a|(\mbf{\nabla}\cdot\mbf{D})^{-1}(-\mbf{\nabla}^2)
(\mbf{\nabla}\cdot\mbf{D})^{-1}|\mbf{y},b\rangle
J\rho^{b}(\mbf{y})
\,,\label{eq:3.7}\end{eqnarray}
with the leading order Coulomb behavior, $1/\vec{q}^{\,2}$. 
Here, the charge density, $\rho$, contains both quark and gluon components,
$\rho^{a}(\mbf{x})=\psi^{\dagger}(\mbf{x})T^a\psi(\mbf{x})
+f^{abc}\mbf{A}^{b}(\mbf{x})\cdot\mbf{\Pi}^{c}(\mbf{x})$. 
A complete solution of the Coulomb gauge constraint
equations is encoded in the Fadeev-Popov determinant, 
$J=det(\mbf{\nabla}\cdot\mbf{D})$ with $\mbf{D}=\mbf{\nabla}-g\mbf{A}$, 
which is unknown. We assume, that a nonperturbative solution of $J$ produces
the linear confining potential. Thus, we consider QCD motivated Hamiltonian,
given by the Couomb gauge canonical Hamiltonian 
with the instantaneous interaction
including the Coulomb and linear confining potentials; i.e. $W\rightarrow W_0$,
\begin{eqnarray}
W_0 &=& -\frac{1}{2}\int d\mbf{x}d\mbf{y}
 \rho^a(\mbf{x})V_{L+C}(|\mbf{x}-\mbf{y}|)\rho^a(\mbf{y})
\nonumber\\
C_fV_{L+C}(r) &=& \sigma r - C_f\frac{\alpha_s}{r}
\,.\label{eq:3.8}\end{eqnarray}
This QCD motivated Hamiltonian defines our model, used for calculations
of glueball and meson bound states and presented below.

\subsubsection{Effective Hamiltonian for gluodynamics in the Coulomb gauge}

In a pure gluodynamics, charge density contains only the gluon component. Following
the same strategy as before, we represent physical fields, which are three gluon 
field, $\vec{A}$, and its conjugate momentum, $\vec{\Pi}$, in a second quantized form,
\begin{eqnarray}
&&A_i^a({\bf x}) = \int\frac{d{\bf k}}{(2\pi)^3}\frac{1}{\sqrt{2\omega_{\bf k}}}
    [a_i^a({\bf k})+a_i^{a\dagger}(-{\bf k})]{\rm e}^{i{\bf k}{\bf x}} \nonumber\\
&&\Pi_i^a({\bf x}) = -i\int\frac{d{\bf k}}{(2\pi)^3}\sqrt{\frac{\omega_{\bf k}}{2}}
    [a_i^a({\bf k})-a_i^{a\dagger}(-{\bf k})]{\rm e}^{i{\bf k}{\bf x}} 
\,.\label{eq:3.9}\end{eqnarray}
Since a confining potential has been introduced, 
a trivial perturbative vacuum does not 
insure minimum for the ground state. Ground state is shifted to some unknown
nonperturbative state $|0\rangle_{NP}$. The annihilation operator is defined
to annihilate in this vacuum, i.e. $a|0\rangle_{NP}=0$; and creation operator 
acting on this vacuum produces quasiparticle (effective gluon) with unknown energy 
$\omega(\mbf{k})$. Gluon energy $\omega(\mbf{k})$ is kept as a trial parameter 
through out the calculations and is found variationally by minimizing the vacuum 
(ground) state energy.
The canonical commutation relation is
\begin{eqnarray}
[a_i^a({\bf k}),a_j^{b\dagger}({\bf k}^{\prime})]=(2\pi)^3\delta^{ab}\delta^{(3)}
({\bf k}-{\bf k}^{\prime})D_{ij}({\bf k})
\,,\label{eq:3.10}\end{eqnarray} 
where the gluon operators 
$a_i^a({\bf k})=\sum_{\lambda=1,2}\epsilon_i({\bf k},\lambda)a^a({\bf k},\lambda)$
are transverse, i.e. 
${\bf k}\cdot{\bf a}^a({\bf k})={\bf k}\cdot{\bf a}^{a\dagger}({\bf k})=0$; 
so that polarization sum is
\begin{eqnarray}
D_{ij}({\bf k})=\sum_{\lambda=1,2}
\epsilon_i({\bf k},\lambda)\epsilon_j({\bf k},\lambda)
=\delta_{ij}-{\hat k}_i{\hat k}_j
\,,\label{eq:3.11}\end{eqnarray}
with the unit vector ${\hat k}_i=k_i/k$; and $k_i\cdot D_{ij}({\bf k})=0$.

In a second quantized form, the Coulomb gauge Hamiltonian of gluodynamics 
contains the following terms;
the gluon kinetic energy is given by
\begin{eqnarray}
 H_0(l) &=& \frac{1}{2}\int\frac{d{\bf k}}{(2\pi)^3}
\left[(\frac{{\bf k}^2}{\omega({\bf k},l)}+\omega({\bf k},l))
a_i^{a\dagger}({\bf k})a_i^a({\bf k})\right.
\nonumber\\
&+&\left. (\frac{{\bf k}^2}{\omega({\bf k},l)}-\omega({\bf k},l))
\frac{1}{2}( a_i^a({\bf k})a_i^a(-{\bf k})+ {\rm h.c.})\right]
\,.\label{eq:3.12}\end{eqnarray}
The instantaneous gluon-gluon interaction (Eq. (\ref{eq:3.8})) is given by
\begin{eqnarray} 
H_{L+C}& & = -{1\over8} f^{abc}f^{ade} 
\int\left(\prod_{n=1}^{4} {d{\bf k}_n\over(2\pi)^3}\right)
(2\pi)^3\delta^{(3)}(\sum_m {\bf k}_m) 
\left({\omega(\mbf{k}_2,l)\omega(\mbf{k}_4,l)\over
\omega(\mbf{k}_1,l)\omega(\mbf{k}_3,l)}\right)^{1/2}\label{eq:3.13}\\
& & \hspace{-2cm} V_{L+C}({\bf k}_1+{\bf k}_2)  
{\bf :}
\left[ a^b_i({\bf k}_1) + {a^b_i}^\dagger(-{\bf k}_1)\right]
\left[ a^c_i({\bf k}_2) - {a^c_i}^\dagger(-{\bf k}_2)\right]
\left[ a^d_j({\bf k}_3) + {a^d_j}^\dagger(-{\bf k}_3)\right]
\left[ a^e_j({\bf k}_4) - {a^e_j}^\dagger(-{\bf k}_4)\right]
{\bf :} \nonumber
\,,\end{eqnarray}
where $V_{L+C}({\bf k})$ is a Fourier transform in the momentum 
space of linear confining plus Coulomb potentials,
\begin{eqnarray}
V_{L+C} &=& 2\pi C_{adj}\frac{\alpha_s}{\mbf{k}^2}
+4\pi\frac{\sigma_{adj}}{\mbf{k}^4}
\,,\label{eq:3.14}\end{eqnarray}
here the adjoint Casimir is $C_{adj}=N_c$.
The nonabelian gluon part (Eq. (\ref{eq:3.6})) includes
in the order $O(g)$ a triple-gluon coupling 
\begin{eqnarray}
H_{3g}(l) &=& \frac{i}{2\sqrt{2}}f^{abc}
\int\left(\prod_{n=1}^{3} {d{\bf k}_n\over(2\pi)^3}\right)
(2\pi)^3\delta^{(3)}(\sum_m {\bf k}_m)
\frac{\Gamma_{ijk}(\mbf{k}_1,\mbf{k}_2,\mbf{k}_3)}
{\sqrt{\omega(\mbf{k}_1,l)\omega(\mbf{k}_2,l)\omega(\mbf{k}_3,l)}} 
\label{eq:3.15} \\
& &  {\bf :}
\left[g_0(\mbf{k}_1,\mbf{k}_2,\mbf{k}_3,l)
a^a_i({\bf k}_1)a^b_j({\bf k}_2)a^c_i({\bf k}_3) + ... \right]
{\bf :}  \nonumber
\,,\end{eqnarray}
with
\begin{eqnarray}
\Gamma_{ijk}(\mbf{k}_1,\mbf{k}_2,\mbf{k}_3,l) &=&
\frac{1}{6}\left((k_1-k_3)_j\delta_{ik}+(k_2-k_1)_k\delta_{ij}
+(k_3-k_2)_i\delta_{jk}\right)
\,.\label{eq:3.16}\end{eqnarray}
In the order $O(g^2)$ the normal-ordered four-gluon vertex reads
\begin{eqnarray}
H_{4g}(l)  &=& {\alpha_s\pi\over4} f^{abc}f^{ade}
\int\left(\prod_{n=1}^{4} {d{\bf k}_n\over(2\pi)^3}\right)
(2\pi)^3\delta^{(3)}(\sum_m {\bf k}_m)
\nonumber\\ 
&& {1\over\sqrt{\omega(\mbf{k}_1,l)\omega(\mbf{k}_2,l)
\omega(\mbf{k}_3,l)\omega(\mbf{k}_4,l)}} \label{eq:3.17} \\
& &\hspace{-2cm}
{\bf :}
\left[ a^b_i({\bf k}_1) + {a^b_i}^\dagger(-{\bf k}_1)\right]
\left[ a^c_j({\bf k}_2) + {a^c_j}^\dagger(-{\bf k}_2)\right]
\left[ a^d_i({\bf k}_3) + {a^d_i}^\dagger(-{\bf k}_3)\right]
\left[ a^e_j({\bf k}_4) + {a^e_j}^\dagger(-{\bf k}_4)\right]
{\bf :} \nonumber
\,.\end{eqnarray}
Coupling constants and masses (relevant and marginal operators, respectively) 
are functions of the flow parameter $l$ and momenta involved.
In addition terms from normal ordering with respect to the NP 
vacuum $|0\rangle_{NP}$ arise; condensates $O_0$, $O_{L+C}$, $O_{4g}$ and 
gluon polarization operators $\Pi_{L+C}$, $\Pi_{4g}$.

\begin{table}
\vspace{2cm}
\begin{tabular}{|r|c|c|c|c|c|c|} \hline
 & $|0>$ & $|g>$ & $gg>$ & $|ggg>$ & $|gggg>$   \\ \hline
$|0>$ &  &  
&     & \floadeps{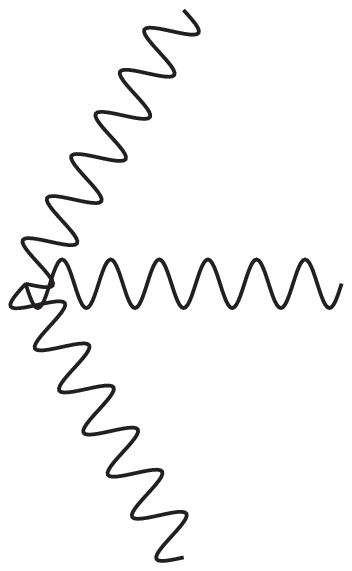} &\floadeps{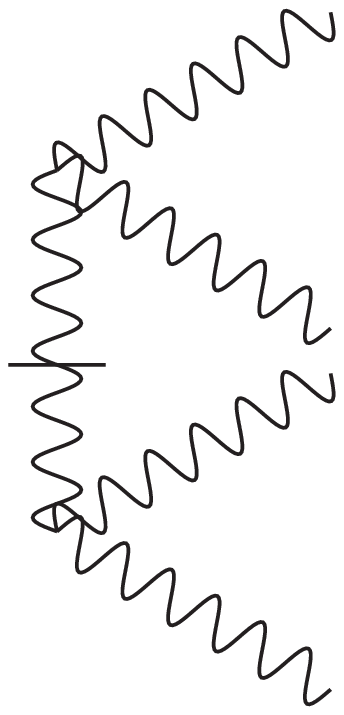}  \\ \hline
$|g>$ &  & & \floadeps{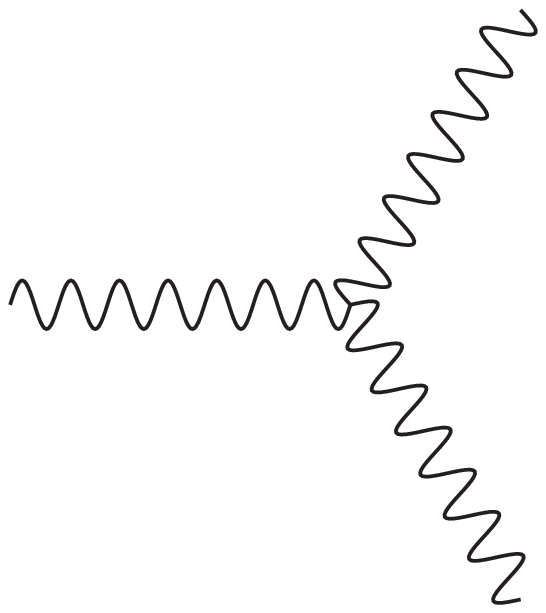} & \floadeps{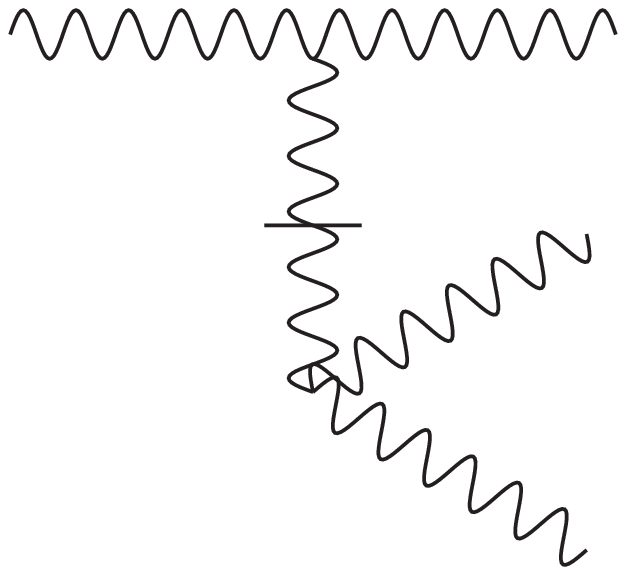} &  \\ \hline
$|gg>$ &  & \floadeps{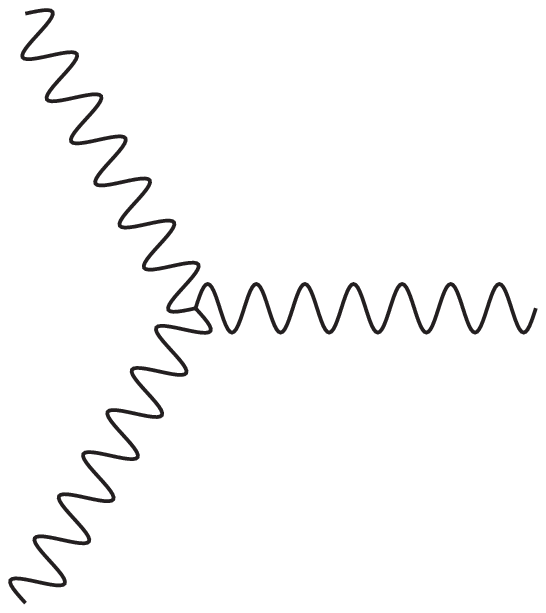} & \floadeps{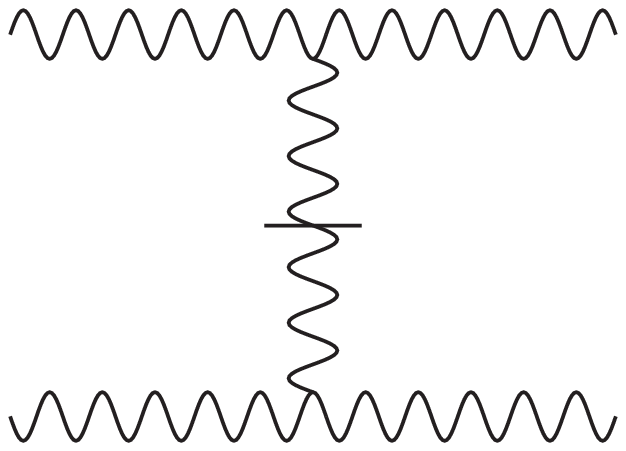}&  &  \\ \hline
$|ggg>$ & \floadeps{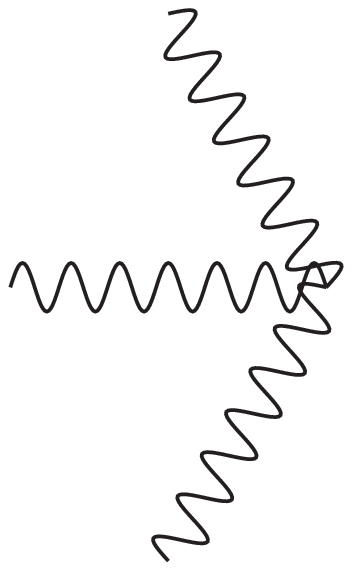} & \floadeps{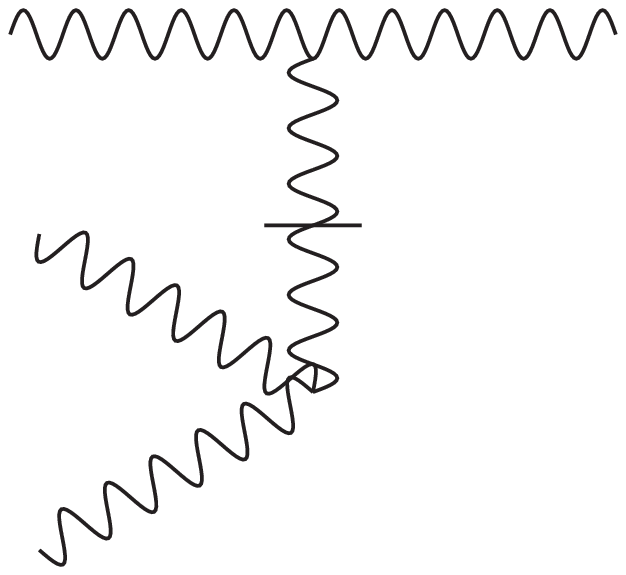} &  &  &  \\ \hline
$|gggg>$ &  \floadeps{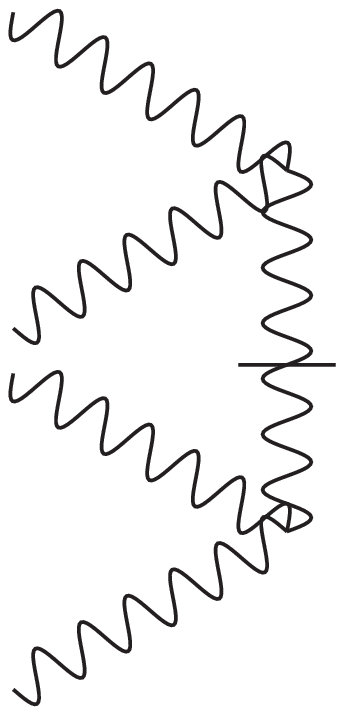}  &  &  &  &  \\ \hline
\end{tabular}
\vspace{1cm}
\caption{Hamiltonian of gluodynamics in the Coulomb gauge.}
\label{tab:3.1}
\end{table}

\begin{table}
\vspace{2cm}
\begin{tabular}{|r|c|c|c|c|c|c|} \hline
 & $|0>$ & $|g>$ & $gg>$ & $|ggg>$ & $|gggg>$   \\ \hline
$|0>$ &\floadeps{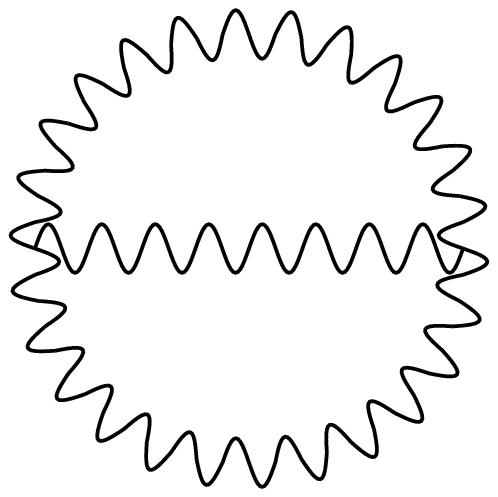}  &  &  &  &  \\ \hline
$|g>$ &  & \floadeps{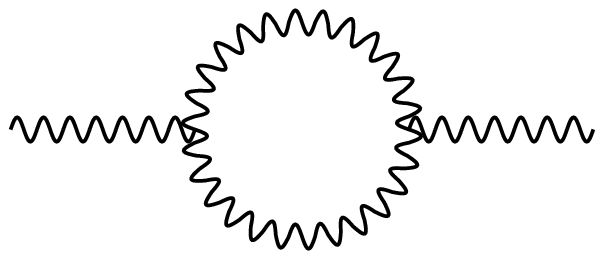} &  &  &  \\ \hline
$|gg>$ &  &  & \floadeps{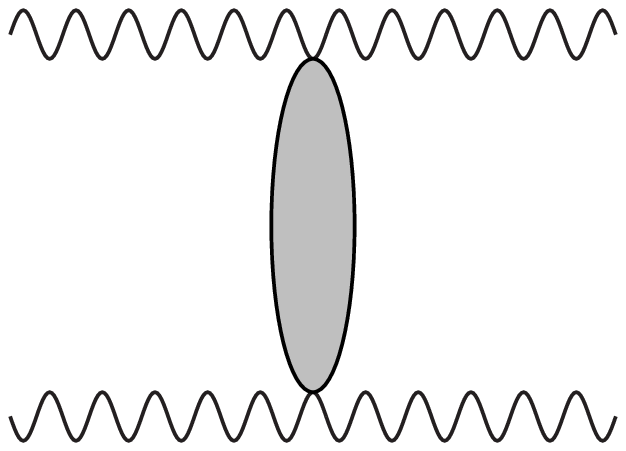}&  &  \\ \hline
$|ggg>$ &  &  &  &\floadeps{}  &  \\ \hline
$|gggg>$ &    &  &  &  &\floadeps{table00c}  \\ \hline
\end{tabular}
\vspace{1cm}
\caption{Effective Hamiltonian of gluodynamics generated through 
the second order by flow equations in the Coulomb gauge.}
\label{tab:3.2}
\end{table}

Hamiltonian matrix of gluodynamics is depicted in the Table \ref{tab:3.1}. 
It includes the triple-gluon vertex (for example, element $H_{14}$)
and the instantaneous gluon-gluon interaction (for example, element $H_{15}$);
(four-gluon vertex is not depicted).
Off-diagonal elements mixing different Fock sectors are
eliminated by flow equations.

Elimination of the off-diagonal elements through the second order
generates the Hamiltonian matrix depicted in the Table \ref{tab:3.2}.
Explicit form of the effective Hamiltonian is not given
here (for details see \cite{Gluon}). Black area in the effective
$gg$ interaction ($H^{\rm eff}_{33}$) depicts the interaction
generated by flow equations.    

We solve the effective Hamiltonian in the two lowest sectors. 
This gives the gap equation and the Bethe-Salpeter bound equation for 
the glueball states.

Minimizing the ground state energy ($H^{\rm eff}_{11}$), 
\begin{eqnarray}  
\frac{\delta \langle 0|H_{\rm eff}|0\rangle}{\delta\omega_{\bf k}}=0
\,,\label{eq:3.18}\end{eqnarray} 
we obtain the gap equation for the effective gluon energy, $\omega(\mbf{k})$, 
which is a function of gluon momentum $\mbf{k}$ and the cut-off $\Lambda$. 
Gap equation can be also obtained by demanding the
off-diagonal one-body operator ($H^{\rm eff}_{13}$), having the structure 
$(aa+a^{\dagger}a^{\dagger})$, to be equal to zero (thus it is not depicted
in the Table \ref{tab:3.2}).
Explicitly the gap equation is given by
\begin{eqnarray}
\omega_{\bf k}^2 &=& k^2+m_{CT}^2(\Lambda)
\nonumber\\
&+& \frac{1}{4} N_c
\int{d{\bf q}\over(2\pi)^3}{1\over\omega_{\bf q}}
V_{L+C}({\bf k}-{\bf q})\left(1+({\hat k}{\hat q})^2 \right)
(\omega_{\bf q}^2-\omega_{\bf k}^2){\rm e}^{-q^2/\Lambda^2}
\nonumber\\
&+& \alpha_s\pi N_c
\int{d{\bf q}\over(2\pi)^3}{1\over\omega_{\bf q}}
\left(3-({\hat k}{\hat q})^2 \right)
{\rm e}^{-q^2/\Lambda^2}
\nonumber\\
&-& 2\alpha_s\pi N_c
\int{d{\bf q}\over(2\pi)^3}{1\over\omega_{\bf q}\omega_{{\bf k}-{\bf q}} }
\frac{G({\bf k},{\bf q})}{\omega_{\bf q}+\omega_{{\bf k}-{\bf q}}}
{\rm e}^{-4q^2/\Lambda^2}
\,,\label{eq:3.19}\end{eqnarray}
where the mass counterterm $m_{CT}(\Lambda)$ is defined by
\begin{eqnarray}                 
m_{CT}^2(\Lambda) &=& -\frac{\alpha_s}{\pi}N_c\frac{11}{6}\Lambda^2
\,,\label{eq:3.20}\end{eqnarray}
which absorbs the leading ultraviolet divergences in ${\rm H}_{eff}$
as $\Lambda\rightarrow \infty$. In the regulating fuction, we assume all 
external momenta to be soft compared to momentum flowing in the loop.
Here, the tensor structure of generated terms is given by
\begin{eqnarray}
& & G({\bf k},{\bf q})=2(1-({\hat k}{\hat q})^2)
\left( k^2+q^2+\frac{k^2q^2}{2({\bf k}-{\bf q})^2}
(1+({\hat k}{\hat q})^2) \right)
\,,\label{eq:3.21}\end{eqnarray}
and $V_{L+C}$ is the Fourier transform of the sum of 
linear and Coulomb potentials.
The obtained gluon gap equation is both UV and IR finite. 
UV divergent behavior arising from the Coulomb interaction, four-gluon, 
and dynamically generated terms is cancelled by the canonical
gluon mass counterterm.

We approximate a glueball bound state to consist of two 
valence constituent gluons.
The glueball wave function in the rest frame is 
\begin{eqnarray}
|\psi_n\rangle=\int{d{\bf q}\over(2\pi)^3}\phi_n^{ij}({\bf q})
a_i^{a\dagger}({\bf q})a_j^{a\dagger}(-{\bf q})|0\rangle_{NP}
\,,\label{eq:3.22}\end{eqnarray}
where $a^{\dagger}({\bf q})$ creats a quasiparticle from
a nontrivial vacuum $|0\rangle_{NP}$
with an effective dispersion relation $\omega({\bf q})$.
Since in this constituent basis an effective Hamiltonian
${\rm H}_{eff}$ is block-diagonal, the mixing with four gluon and 
higher states is suppressed. Therefore Tamm-Dancoff approach 
with ${\rm H}_{eff}$ is a reasonable approximaion for bound states.
To obtain the Tamm-Dancoff bound state equation
we project the Schr{\"o}dinger equation 
${\rm H}_{eff}|\psi_n\rangle={\rm E}_n|\psi_n\rangle$
on the two-body sector; the result reads
\begin{eqnarray} 
\langle\psi_n|[H_{eff},a^{i\dagger}(\mbf{q})a^{j\dagger}(-\mbf{q})]|0\rangle
=(E_n-E_0)\left(X^{ij}_{i'j'}({\bf q})\phi_n^{i'j'}({\bf q}) \right)
\,,\label{eq:3.23}\end{eqnarray}
with the notation
\begin{eqnarray}
\left(X^{ij}_{i'j'}({\bf q})\phi_n^{i'j'}({\bf q}) \right) 
= D_{ii'}({\bf q})D_{jj'}({\bf q})\phi_n^{i'j'}({\bf q})
+ D_{ij'}({\bf q})D_{ji'}({\bf q})\phi_n^{i'j'}(-{\bf q})
\,,\label{eq:3.24}\end{eqnarray}
here color indices are omitted, 
the polarization sum $D_{ij}$ is given in Eq.(\ref{eq:3.11}), 
and $H_{eff}$ is given in the Table \ref{tab:3.2}. 
In Eq. (\ref{eq:3.23}) we subtracted the vacuum energy $E_0$,
defined as $H_{eff}|0\rangle_{NP}=E_0|0\rangle_{NP}$. 
Explicitly Tamm-Dancoff equation for scalar and pseudoscalar glueball states 
with total angular momentum $J$, parity $P$, and charge conjugation $C$
and only the instantaneous interactions included is given by
\begin{eqnarray} 
& & M_n \phi_n(q)
= \left[\left(\frac{q^2+\widetilde{m}_{CT}^2(\Lambda)}
{\omega({\bf q})}+\omega({\bf q}) \right)\right.
\nonumber\\
&+& \left.\frac{1}{4} N_c
\int{d{\bf p}\over(2\pi)^3}
V_{L+C}({\bf p}-{\bf q})\left(1+({\hat p}{\hat q})^2 \right)
\frac{\omega^2({\bf p})+\omega^2({\bf q})}{\omega({\bf p})\omega({\bf q})}
{\rm e}^{-p^2/\Lambda^2}
\right]\phi_n(q)
\nonumber\\
&-& \frac{1}{8} N_c \int{d{\bf p}\over(2\pi)^3}
V_{L+C}({\bf p}-{\bf q})
\frac{(\omega({\bf p})+\omega({\bf q}))^2}{\omega({\bf p})\omega({\bf q})}
F^{JPC}({\bf p},{\bf q})\phi_n(p)
\,,\label{eq:3.25}\end{eqnarray}
with
\begin{eqnarray}
F^{0++}({\bf p},{\bf q}) &=& 1+({\hat p}{\hat q})^2
\nonumber\\
F^{0-+}({\bf p},{\bf q}) &=& 2({\hat p}{\hat q})
\,.\label{eq:3.26}\end{eqnarray}
Here the Coulomb mass counterterm is given by 
\begin{eqnarray}                 
\widetilde{m}_{CT}^2(\Lambda) &=& -\frac{\alpha_s}{\pi}\frac{N_c}{3}\Lambda^2
\,,\label{eq:3.27}\end{eqnarray}
and $V_{L+C}$ is the sum of linear and Coulomb potentials 
in the momentum space.
One-body sector of the effective Hamiltonian contributes to the kinetic
part of the Tamm-Dancoff equation, and two-body effective interaction to the
potential part. UV divergences, comming from the Coulomb potential, are cancelled
by the mass counterterm. IR divergent behavior, due to the confining potential,
cancels in the potential and kinetic parts with each other. Such complete
cancelation happens only for the color-singlet state, and does not occure
for colored objects.

For completeness, we summarize the complete Tamm-Dancoff equation in 
the scalar and pseudoscalar channels when all the terms up to the second order  
(instantaneous, four-gluon and dynamical interactions) are included.
It is give by
\begin{eqnarray} 
& & (E_n-E_0)\phi_n(q)
= \left[\left(\frac{q^2+ m_{CT}^2(\Lambda)}
{\omega_{\bf q}}+\omega_{\bf q} \right)\right.
\nonumber\\
&+& \left.\frac{1}{4} N_c
\int{d{\bf p}\over(2\pi)^3}
V_{L+C}({\bf p}-{\bf q})\left(1+({\hat p}{\hat q})^2 \right)
\frac{\omega_{\bf p}^2+\omega_{\bf q}^2}{\omega_{\bf p}\omega_{\bf q}}
{\rm e}^{-p^2/\Lambda^2}\right.
\nonumber\\
&+&\left.\alpha_s\pi N_c 
\int{d{\bf p}\over(2\pi)^3}
\frac{1}{\omega_{\bf p}\omega_{\bf q}}
\left(3-({\hat p}{\hat q})^2 \right)
{\rm e}^{-p^2/\Lambda^2}\right.
\nonumber\\
&-&\left.\alpha_s\pi N_c 
\int{d{\bf p}\over(2\pi)^3}
\frac{1}{\omega_{\bf p}\omega_{\bf q}\omega_{{\bf p}-{\bf q}} }
\frac{G({\bf p},{\bf q})}{\omega_{\bf p}+\omega_{{\bf p}-{\bf q}} }
{\rm e}^{-4p^2/\Lambda^2}
\right]\phi_n(q)
\nonumber\\
&+&\left[ 
-\frac{1}{8} N_c \int{d{\bf p}\over(2\pi)^3}
V_{L+C}({\bf p}-{\bf q})
\frac{(\omega_{\bf p}+\omega_{\bf q})^2}{\omega_{\bf p}\omega_{\bf q}}
F^{JPC}({\bf p},{\bf q})\phi_n(p) \right.
\nonumber\\
&+&\left.\alpha_s\pi N_c 
\int{d{\bf p}\over(2\pi)^3}
\frac{1}{2\omega_{\bf p}\omega_{\bf q}}
\left(3-({\hat p}{\hat q})^2 \right)
E^{JPC}({\bf p},{\bf q})\phi_n(p) \right.
\nonumber\\
&+&\left.\alpha_s 2\pi N_c 
\int{d{\bf p}\over(2\pi)^3}
\frac{1}{\omega_{\bf p}\omega_{\bf q}}
\frac{D^{JPC}({\bf p},{\bf q})}{\omega^2_{{\bf p}-{\bf q}} }
\left(1-\frac{(\omega_{\bf k}-\omega_{\bf q})^2}
{(\omega_{\bf k}-\omega_{\bf q})^2+\omega_{{\bf k}-{\bf q}}^2}
\right) 
\phi_n(p) \right]
\,,\label{eq:3.28}\end{eqnarray}
with the total angular momentum $J$, parity $P$, and 
charge conjugation $C$.
Here the tensor structures are given by  
\begin{eqnarray}
F^{0++}({\bf p},{\bf q}) &=& 1+({\hat p}{\hat q})^2
\nonumber\\
F^{0-+}({\bf p},{\bf q}) &=& 2({\hat p}{\hat q})
\nonumber\\
E^{0++}({\bf p},{\bf q}) &=& 1
\nonumber\\
E^{0-+}({\bf p},{\bf q}) &=& 0
\nonumber\\
D^{0++}({\bf p},{\bf q}) &=& G({\bf p},{\bf q})
= 2(1-({\hat p}{\hat q})^2)\left( p^2+q^2
+ \frac{p^2q^2}{2({\bf p}-{\bf q})^2}(1+({\hat p}{\hat q})^2)
\right)
\nonumber\\
D^{0-+}({\bf p},{\bf q}) &=& 2p^2q^2 (1-({\hat p}{\hat q})^2)\left(
\frac{2}{pq}+\frac{1}{({\bf p}-{\bf q})^2}({\hat p}{\hat q})
\right)
\,.\label{eq:3.29}\end{eqnarray}
Again, this equation is both UV and IR finite.

\begin{figure}[!htb]
\begin{center}
\input{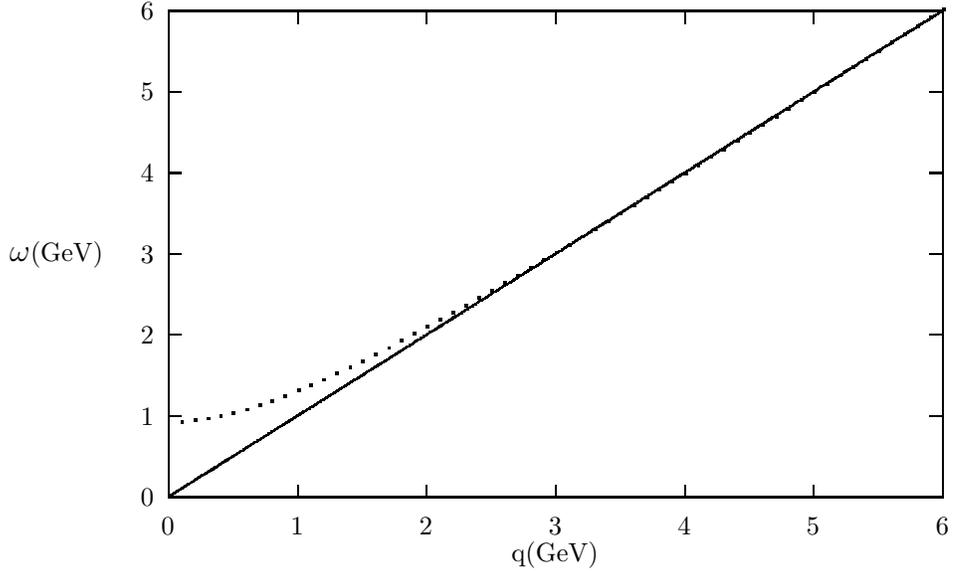}
\end{center}
\caption{One particle dispersion relation.
Dots represent the numerical solution 
of gap equation $\omega({\bf k})$ 
($\alpha_s=0.4, \sigma=0.18 GeV^2, \Lambda=4 GeV, N_c=3$), 
the solid line stays for the free
dispersion relation $\omega({\bf k})=k$.}
\label{fig:3.1}
\end{figure}

\begin{figure}[!htb]
\begin{center}
\input{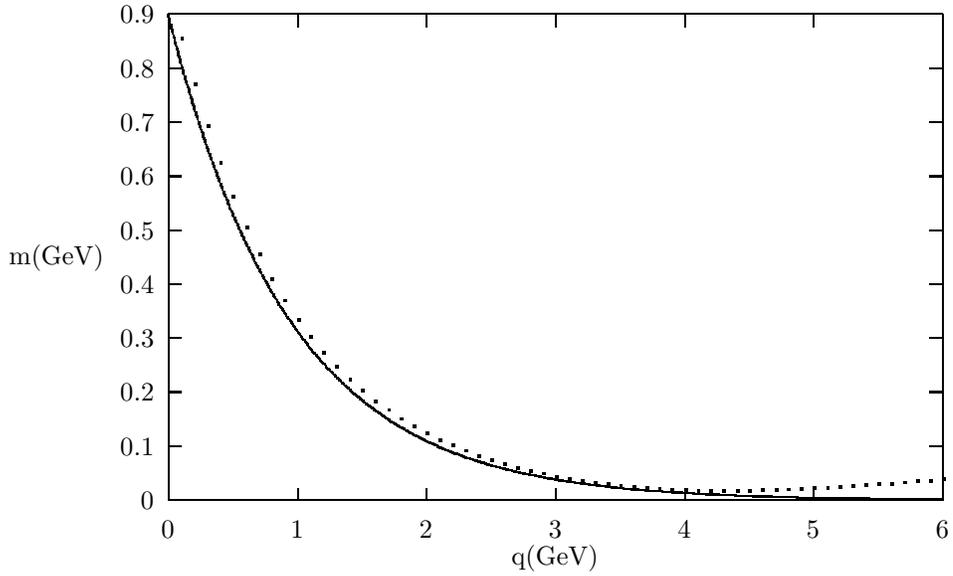}
\end{center}
\caption{Gluon mass. Dots represent the numerical solution 
for $m({\bf k})=\omega({\bf k})-k$
($\alpha_s=0.4, \sigma=0.18 GeV^2, \Lambda=4 GeV, N_c=3$), 
the solid line is a fit
$m({\bf k})=0.9*{\rm exp}(-k/0.95)$ (parameters are in GeV).}
\label{fig:3.2}
\end{figure}

\begin{figure}[!htb]
\begin{center}
\setlength{\unitlength}{0.240900pt}
\ifx\plotpoint\undefined\newsavebox{\plotpoint}\fi
\sbox{\plotpoint}{\rule[-0.200pt]{0.400pt}{0.400pt}}%
\begin{picture}(1500,900)(0,0)
\font\gnuplot=cmr10 at 10pt
\gnuplot
\sbox{\plotpoint}{\rule[-0.200pt]{0.400pt}{0.400pt}}%
\put(220.0,113.0){\rule[-0.200pt]{4.818pt}{0.400pt}}
\put(198,113){\makebox(0,0)[r]{0.7}}
\put(1416.0,113.0){\rule[-0.200pt]{4.818pt}{0.400pt}}
\put(220.0,198.0){\rule[-0.200pt]{4.818pt}{0.400pt}}
\put(198,198){\makebox(0,0)[r]{0.8}}
\put(1416.0,198.0){\rule[-0.200pt]{4.818pt}{0.400pt}}
\put(220.0,283.0){\rule[-0.200pt]{4.818pt}{0.400pt}}
\put(198,283){\makebox(0,0)[r]{0.9}}
\put(1416.0,283.0){\rule[-0.200pt]{4.818pt}{0.400pt}}
\put(220.0,368.0){\rule[-0.200pt]{4.818pt}{0.400pt}}
\put(198,368){\makebox(0,0)[r]{1}}
\put(1416.0,368.0){\rule[-0.200pt]{4.818pt}{0.400pt}}
\put(220.0,453.0){\rule[-0.200pt]{4.818pt}{0.400pt}}
\put(198,453){\makebox(0,0)[r]{1.1}}
\put(1416.0,453.0){\rule[-0.200pt]{4.818pt}{0.400pt}}
\put(220.0,537.0){\rule[-0.200pt]{4.818pt}{0.400pt}}
\put(198,537){\makebox(0,0)[r]{1.2}}
\put(1416.0,537.0){\rule[-0.200pt]{4.818pt}{0.400pt}}
\put(220.0,622.0){\rule[-0.200pt]{4.818pt}{0.400pt}}
\put(198,622){\makebox(0,0)[r]{1.3}}
\put(1416.0,622.0){\rule[-0.200pt]{4.818pt}{0.400pt}}
\put(220.0,707.0){\rule[-0.200pt]{4.818pt}{0.400pt}}
\put(198,707){\makebox(0,0)[r]{1.4}}
\put(1416.0,707.0){\rule[-0.200pt]{4.818pt}{0.400pt}}
\put(220.0,792.0){\rule[-0.200pt]{4.818pt}{0.400pt}}
\put(198,792){\makebox(0,0)[r]{1.5}}
\put(1416.0,792.0){\rule[-0.200pt]{4.818pt}{0.400pt}}
\put(220.0,877.0){\rule[-0.200pt]{4.818pt}{0.400pt}}
\put(198,877){\makebox(0,0)[r]{1.6}}
\put(1416.0,877.0){\rule[-0.200pt]{4.818pt}{0.400pt}}
\put(220.0,113.0){\rule[-0.200pt]{0.400pt}{4.818pt}}
\put(220,68){\makebox(0,0){1}}
\put(220.0,857.0){\rule[-0.200pt]{0.400pt}{4.818pt}}
\put(403.0,113.0){\rule[-0.200pt]{0.400pt}{2.409pt}}
\put(403.0,867.0){\rule[-0.200pt]{0.400pt}{2.409pt}}
\put(510.0,113.0){\rule[-0.200pt]{0.400pt}{2.409pt}}
\put(510.0,867.0){\rule[-0.200pt]{0.400pt}{2.409pt}}
\put(586.0,113.0){\rule[-0.200pt]{0.400pt}{2.409pt}}
\put(586.0,867.0){\rule[-0.200pt]{0.400pt}{2.409pt}}
\put(645.0,113.0){\rule[-0.200pt]{0.400pt}{2.409pt}}
\put(645.0,867.0){\rule[-0.200pt]{0.400pt}{2.409pt}}
\put(693.0,113.0){\rule[-0.200pt]{0.400pt}{2.409pt}}
\put(693.0,867.0){\rule[-0.200pt]{0.400pt}{2.409pt}}
\put(734.0,113.0){\rule[-0.200pt]{0.400pt}{2.409pt}}
\put(734.0,867.0){\rule[-0.200pt]{0.400pt}{2.409pt}}
\put(769.0,113.0){\rule[-0.200pt]{0.400pt}{2.409pt}}
\put(769.0,867.0){\rule[-0.200pt]{0.400pt}{2.409pt}}
\put(800.0,113.0){\rule[-0.200pt]{0.400pt}{2.409pt}}
\put(800.0,867.0){\rule[-0.200pt]{0.400pt}{2.409pt}}
\put(828.0,113.0){\rule[-0.200pt]{0.400pt}{4.818pt}}
\put(828,68){\makebox(0,0){10}}
\put(828.0,857.0){\rule[-0.200pt]{0.400pt}{4.818pt}}
\put(1011.0,113.0){\rule[-0.200pt]{0.400pt}{2.409pt}}
\put(1011.0,867.0){\rule[-0.200pt]{0.400pt}{2.409pt}}
\put(1118.0,113.0){\rule[-0.200pt]{0.400pt}{2.409pt}}
\put(1118.0,867.0){\rule[-0.200pt]{0.400pt}{2.409pt}}
\put(1194.0,113.0){\rule[-0.200pt]{0.400pt}{2.409pt}}
\put(1194.0,867.0){\rule[-0.200pt]{0.400pt}{2.409pt}}
\put(1253.0,113.0){\rule[-0.200pt]{0.400pt}{2.409pt}}
\put(1253.0,867.0){\rule[-0.200pt]{0.400pt}{2.409pt}}
\put(1301.0,113.0){\rule[-0.200pt]{0.400pt}{2.409pt}}
\put(1301.0,867.0){\rule[-0.200pt]{0.400pt}{2.409pt}}
\put(1342.0,113.0){\rule[-0.200pt]{0.400pt}{2.409pt}}
\put(1342.0,867.0){\rule[-0.200pt]{0.400pt}{2.409pt}}
\put(1377.0,113.0){\rule[-0.200pt]{0.400pt}{2.409pt}}
\put(1377.0,867.0){\rule[-0.200pt]{0.400pt}{2.409pt}}
\put(1408.0,113.0){\rule[-0.200pt]{0.400pt}{2.409pt}}
\put(1408.0,867.0){\rule[-0.200pt]{0.400pt}{2.409pt}}
\put(1436.0,113.0){\rule[-0.200pt]{0.400pt}{4.818pt}}
\put(1436,68){\makebox(0,0){100}}
\put(1436.0,857.0){\rule[-0.200pt]{0.400pt}{4.818pt}}
\put(220.0,113.0){\rule[-0.200pt]{292.934pt}{0.400pt}}
\put(1436.0,113.0){\rule[-0.200pt]{0.400pt}{184.048pt}}
\put(220.0,877.0){\rule[-0.200pt]{292.934pt}{0.400pt}}
\put(45,495){\makebox(0,0){\shortstack{$m(0)$\\(GeV)}}}
\put(828,23){\makebox(0,0){$\Lambda$(GeV)}}
\put(220.0,113.0){\rule[-0.200pt]{0.400pt}{184.048pt}}
\put(510,207){\usebox{\plotpoint}}
\multiput(510.58,207.00)(0.499,0.506){149}{\rule{0.120pt}{0.505pt}}
\multiput(509.17,207.00)(76.000,75.951){2}{\rule{0.400pt}{0.253pt}}
\multiput(586.58,284.00)(0.499,0.678){115}{\rule{0.120pt}{0.642pt}}
\multiput(585.17,284.00)(59.000,78.667){2}{\rule{0.400pt}{0.321pt}}
\multiput(645.58,364.00)(0.500,0.645){363}{\rule{0.120pt}{0.616pt}}
\multiput(644.17,364.00)(183.000,234.722){2}{\rule{0.400pt}{0.308pt}}
\multiput(828.00,600.58)(1.066,0.499){169}{\rule{0.951pt}{0.120pt}}
\multiput(828.00,599.17)(181.026,86.000){2}{\rule{0.476pt}{0.400pt}}
\multiput(1011.00,686.58)(5.336,0.496){43}{\rule{4.309pt}{0.120pt}}
\multiput(1011.00,685.17)(233.057,23.000){2}{\rule{2.154pt}{0.400pt}}
\multiput(1253.00,709.60)(26.655,0.468){5}{\rule{18.400pt}{0.113pt}}
\multiput(1253.00,708.17)(144.810,4.000){2}{\rule{9.200pt}{0.400pt}}
\put(510,207){\raisebox{-.8pt}{\makebox(0,0){$\Diamond$}}}
\put(586,284){\raisebox{-.8pt}{\makebox(0,0){$\Diamond$}}}
\put(645,364){\raisebox{-.8pt}{\makebox(0,0){$\Diamond$}}}
\put(828,600){\raisebox{-.8pt}{\makebox(0,0){$\Diamond$}}}
\put(1011,686){\raisebox{-.8pt}{\makebox(0,0){$\Diamond$}}}
\put(1253,709){\raisebox{-.8pt}{\makebox(0,0){$\Diamond$}}}
\put(1436,713){\raisebox{-.8pt}{\makebox(0,0){$\Diamond$}}}
\end{picture}
\end{center}
\caption{Cut-off dependence of the effective gluon mass
($\alpha_s=0.4,\hspace{0.4cm} \sigma=0.18 GeV^2, \hspace{0.5cm} N_c=3$).}
\label{fig:3.3}
\end{figure}
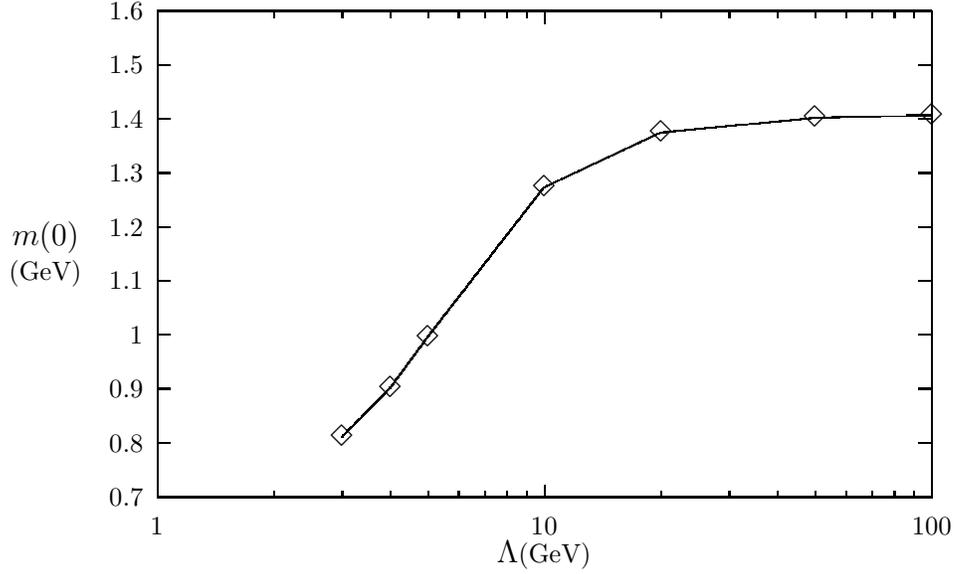

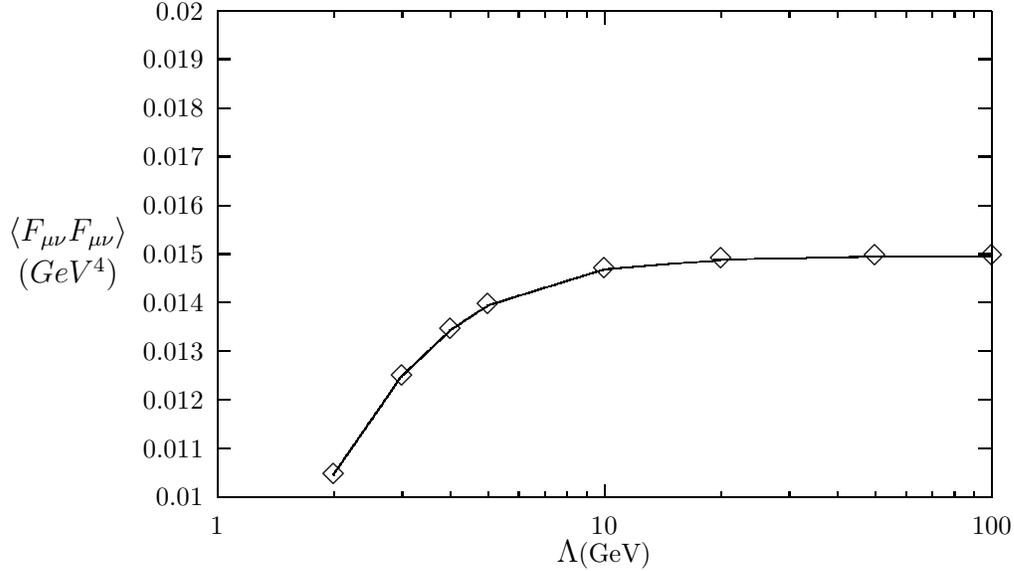
\begin{figure}[!htb]
\begin{center}
\setlength{\unitlength}{0.240900pt}
\ifx\plotpoint\undefined\newsavebox{\plotpoint}\fi
\begin{picture}(1500,900)(0,0)
\font\gnuplot=cmr10 at 10pt
\gnuplot
\sbox{\plotpoint}{\rule[-0.200pt]{0.400pt}{0.400pt}}%
\put(220.0,113.0){\rule[-0.200pt]{4.818pt}{0.400pt}}
\put(198,113){\makebox(0,0)[r]{0.01}}
\put(1416.0,113.0){\rule[-0.200pt]{4.818pt}{0.400pt}}
\put(220.0,189.0){\rule[-0.200pt]{4.818pt}{0.400pt}}
\put(198,189){\makebox(0,0)[r]{0.011}}
\put(1416.0,189.0){\rule[-0.200pt]{4.818pt}{0.400pt}}
\put(220.0,266.0){\rule[-0.200pt]{4.818pt}{0.400pt}}
\put(198,266){\makebox(0,0)[r]{0.012}}
\put(1416.0,266.0){\rule[-0.200pt]{4.818pt}{0.400pt}}
\put(220.0,342.0){\rule[-0.200pt]{4.818pt}{0.400pt}}
\put(198,342){\makebox(0,0)[r]{0.013}}
\put(1416.0,342.0){\rule[-0.200pt]{4.818pt}{0.400pt}}
\put(220.0,419.0){\rule[-0.200pt]{4.818pt}{0.400pt}}
\put(198,419){\makebox(0,0)[r]{0.014}}
\put(1416.0,419.0){\rule[-0.200pt]{4.818pt}{0.400pt}}
\put(220.0,495.0){\rule[-0.200pt]{4.818pt}{0.400pt}}
\put(198,495){\makebox(0,0)[r]{0.015}}
\put(1416.0,495.0){\rule[-0.200pt]{4.818pt}{0.400pt}}
\put(220.0,571.0){\rule[-0.200pt]{4.818pt}{0.400pt}}
\put(198,571){\makebox(0,0)[r]{0.016}}
\put(1416.0,571.0){\rule[-0.200pt]{4.818pt}{0.400pt}}
\put(220.0,648.0){\rule[-0.200pt]{4.818pt}{0.400pt}}
\put(198,648){\makebox(0,0)[r]{0.017}}
\put(1416.0,648.0){\rule[-0.200pt]{4.818pt}{0.400pt}}
\put(220.0,724.0){\rule[-0.200pt]{4.818pt}{0.400pt}}
\put(198,724){\makebox(0,0)[r]{0.018}}
\put(1416.0,724.0){\rule[-0.200pt]{4.818pt}{0.400pt}}
\put(220.0,801.0){\rule[-0.200pt]{4.818pt}{0.400pt}}
\put(198,801){\makebox(0,0)[r]{0.019}}
\put(1416.0,801.0){\rule[-0.200pt]{4.818pt}{0.400pt}}
\put(220.0,877.0){\rule[-0.200pt]{4.818pt}{0.400pt}}
\put(198,877){\makebox(0,0)[r]{0.02}}
\put(1416.0,877.0){\rule[-0.200pt]{4.818pt}{0.400pt}}
\put(220.0,113.0){\rule[-0.200pt]{0.400pt}{4.818pt}}
\put(220,68){\makebox(0,0){1}}
\put(220.0,857.0){\rule[-0.200pt]{0.400pt}{4.818pt}}
\put(403.0,113.0){\rule[-0.200pt]{0.400pt}{2.409pt}}
\put(403.0,867.0){\rule[-0.200pt]{0.400pt}{2.409pt}}
\put(510.0,113.0){\rule[-0.200pt]{0.400pt}{2.409pt}}
\put(510.0,867.0){\rule[-0.200pt]{0.400pt}{2.409pt}}
\put(586.0,113.0){\rule[-0.200pt]{0.400pt}{2.409pt}}
\put(586.0,867.0){\rule[-0.200pt]{0.400pt}{2.409pt}}
\put(645.0,113.0){\rule[-0.200pt]{0.400pt}{2.409pt}}
\put(645.0,867.0){\rule[-0.200pt]{0.400pt}{2.409pt}}
\put(693.0,113.0){\rule[-0.200pt]{0.400pt}{2.409pt}}
\put(693.0,867.0){\rule[-0.200pt]{0.400pt}{2.409pt}}
\put(734.0,113.0){\rule[-0.200pt]{0.400pt}{2.409pt}}
\put(734.0,867.0){\rule[-0.200pt]{0.400pt}{2.409pt}}
\put(769.0,113.0){\rule[-0.200pt]{0.400pt}{2.409pt}}
\put(769.0,867.0){\rule[-0.200pt]{0.400pt}{2.409pt}}
\put(800.0,113.0){\rule[-0.200pt]{0.400pt}{2.409pt}}
\put(800.0,867.0){\rule[-0.200pt]{0.400pt}{2.409pt}}
\put(828.0,113.0){\rule[-0.200pt]{0.400pt}{4.818pt}}
\put(828,68){\makebox(0,0){10}}
\put(828.0,857.0){\rule[-0.200pt]{0.400pt}{4.818pt}}
\put(1011.0,113.0){\rule[-0.200pt]{0.400pt}{2.409pt}}
\put(1011.0,867.0){\rule[-0.200pt]{0.400pt}{2.409pt}}
\put(1118.0,113.0){\rule[-0.200pt]{0.400pt}{2.409pt}}
\put(1118.0,867.0){\rule[-0.200pt]{0.400pt}{2.409pt}}
\put(1194.0,113.0){\rule[-0.200pt]{0.400pt}{2.409pt}}
\put(1194.0,867.0){\rule[-0.200pt]{0.400pt}{2.409pt}}
\put(1253.0,113.0){\rule[-0.200pt]{0.400pt}{2.409pt}}
\put(1253.0,867.0){\rule[-0.200pt]{0.400pt}{2.409pt}}
\put(1301.0,113.0){\rule[-0.200pt]{0.400pt}{2.409pt}}
\put(1301.0,867.0){\rule[-0.200pt]{0.400pt}{2.409pt}}
\put(1342.0,113.0){\rule[-0.200pt]{0.400pt}{2.409pt}}
\put(1342.0,867.0){\rule[-0.200pt]{0.400pt}{2.409pt}}
\put(1377.0,113.0){\rule[-0.200pt]{0.400pt}{2.409pt}}
\put(1377.0,867.0){\rule[-0.200pt]{0.400pt}{2.409pt}}
\put(1408.0,113.0){\rule[-0.200pt]{0.400pt}{2.409pt}}
\put(1408.0,867.0){\rule[-0.200pt]{0.400pt}{2.409pt}}
\put(1436.0,113.0){\rule[-0.200pt]{0.400pt}{4.818pt}}
\put(1436,68){\makebox(0,0){100}}
\put(1436.0,857.0){\rule[-0.200pt]{0.400pt}{4.818pt}}
\put(220.0,113.0){\rule[-0.200pt]{292.934pt}{0.400pt}}
\put(1436.0,113.0){\rule[-0.200pt]{0.400pt}{184.048pt}}
\put(220.0,877.0){\rule[-0.200pt]{292.934pt}{0.400pt}}
\put(45,495){\makebox(0,0){\shortstack{\hspace{-1cm}$\langle F_{\mu\nu}F_{\mu\nu} \rangle$\hspace{-0.4cm}\\ \hspace{-1cm}$(GeV^4)$}}}
\put(828,23){\makebox(0,0){$\Lambda$(GeV)}}
\put(220.0,113.0){\rule[-0.200pt]{0.400pt}{184.048pt}}
\put(403,148){\usebox{\plotpoint}}
\multiput(403.58,148.00)(0.499,0.725){211}{\rule{0.120pt}{0.679pt}}
\multiput(402.17,148.00)(107.000,153.590){2}{\rule{0.400pt}{0.340pt}}
\multiput(510.00,303.58)(0.527,0.499){141}{\rule{0.522pt}{0.120pt}}
\multiput(510.00,302.17)(74.916,72.000){2}{\rule{0.261pt}{0.400pt}}
\multiput(586.00,375.58)(0.758,0.498){75}{\rule{0.705pt}{0.120pt}}
\multiput(586.00,374.17)(57.536,39.000){2}{\rule{0.353pt}{0.400pt}}
\multiput(645.00,414.58)(1.611,0.499){111}{\rule{1.384pt}{0.120pt}}
\multiput(645.00,413.17)(180.127,57.000){2}{\rule{0.692pt}{0.400pt}}
\multiput(828.00,471.58)(6.243,0.494){27}{\rule{4.980pt}{0.119pt}}
\multiput(828.00,470.17)(172.664,15.000){2}{\rule{2.490pt}{0.400pt}}
\multiput(1011.00,486.59)(26.870,0.477){7}{\rule{19.460pt}{0.115pt}}
\multiput(1011.00,485.17)(201.610,5.000){2}{\rule{9.730pt}{0.400pt}}
\put(403,148){\raisebox{-.8pt}{\makebox(0,0){$\Diamond$}}}
\put(510,303){\raisebox{-.8pt}{\makebox(0,0){$\Diamond$}}}
\put(586,375){\raisebox{-.8pt}{\makebox(0,0){$\Diamond$}}}
\put(645,414){\raisebox{-.8pt}{\makebox(0,0){$\Diamond$}}}
\put(828,471){\raisebox{-.8pt}{\makebox(0,0){$\Diamond$}}}
\put(1011,486){\raisebox{-.8pt}{\makebox(0,0){$\Diamond$}}}
\put(1253,491){\raisebox{-.8pt}{\makebox(0,0){$\Diamond$}}}
\put(1436,491){\raisebox{-.8pt}{\makebox(0,0){$\Diamond$}}}
\put(1253.0,491.0){\rule[-0.200pt]{44.085pt}{0.400pt}}
\end{picture}
\end{center}
\caption{Gluon condensate
($\alpha_s=0.4, \sigma=0.18 GeV^2, N_c=3$).}
\label{fig:3.4}
\end{figure}

{\it Numerical results}

Numerical solution of the gap equation is shown on Figure \ref{fig:3.1}, with dots
for numerics, solid line is a perturbative free behavior, $\omega=k$. At large
momenta an effective gluon energy covers the perturbative behavior, while
at small momenta $\omega$ tends to a constant, an effective gluon mass at zero
momentum. We obtain $m(0)=0.9\, GeV$ with the string tension $\sigma=0.2\, GeV^2$.
We define an effective gluon mass as the difference of the effective gluon energy
and the perturbative value, Figure \ref{fig:3.2} (dots is numerics, solid line is a 
parametrization with an exponential fall off). At large momenta, there
is zero mass and at $q\sim 0$ we have $m(0)\sim 1\, GeV$.

Effective gluon mass at zero gluon momentum as a function of the cut-off is given
at Figure \ref{fig:3.3} (in logarithmic scale). Leading $\Lambda^2$ behavior is absorbed 
by the mass counterterm. The logarithmic dependence is left, which is relatively 
slow.

Gluon condensate, calculated using the numerically obtained gluon energy, is 
another nonperturbative chracteristic. It grows as a function of $\Lambda$
logarithmically (Fig. \ref{fig:3.4}). 
We attribute this slow dependence to a numerical artefact.
We regulate the gluon condensate by subtracting 
the perturbative contribution
\begin{eqnarray}   
\langle \frac{\alpha_s}{\pi}F^a_{\mu\nu}F^a_{\mu\nu} \rangle =
\frac{N_c^2-1}{\pi^3}\int_0^{\infty}dk k^2 \alpha_s
\frac{(\omega({\bf k})-k)^2}{2\omega({\bf k})}
\,.\label{eq:3.30}\end{eqnarray}
Using the dispersion relation obtained above $\omega({\bf k})$
the gluon condensate is obtained $1.3\cdot 10^{-2} GeV^4 $ 
(for the cut-off $\Lambda=4$ GeV), that agrees with the sum rules.
(Therefore $\Lambda=4\,GeV$ has been used for all our numerical calculations.)

Numerical calculations of the Tamm-Dancoff equation, Eq.(\ref{eq:3.25}),
are performed variationally with a set of gaussian test functions.
Results of calculations for the lowest glueball states are presented
in the Table (\ref{tab:3.3}) and compared with the available lattice data.

\begin{table}[ht]
$$
\begin{tabular}{|c|c|c|c|c|} \hline 
  $J^{PC}$ & $0^{++}$ & $0^{*++}$ & $0^{-+}$ & $0^{*-+}$ \\ \hline
 Tamm-Dancoff, (MeV) & $1760$ & $2697$ 
& $2142$ & $2895$ \\ \hline
lattice data, (MeV) & $1730(80)$ & $2670(130)$   
& $2590(130)$ & $3640(180)$ \\ \hline 
\end{tabular}
$$
\caption{Glueball spectrum for the lowest scalar and pseudoscalar states
($\alpha_s=0.4, \sigma=0.18 GeV^2, \Lambda=4 GeV, N_c=3$).
Lattice data are from C. Morningstar and M. Peardon, hep-lat/9901004.}
\label{tab:3.3}
\end{table}

Lattice calculations are done for $SU(3)$ pure gluodynamics, 
using anisotropic lattice and improved SII action.
Better agreement with the lattice data is achieved for the scalar channel.
Remarkable, the mass of the lowest scalar glueball $0^{++}$
is roughly twice of the effective gluon mass $m(0)$ obtained before.
This confirms the constituent picture. The better agreement is achieved for
the scalar masses, indicating that dynamical terms are important for the excited
states and the complete Tamm-Dancoff equation with all terms should be solved.

\subsubsection{Effective Hamiltonian for QCD in the Coulomb gauge}

This model is also suitable to study the chiral symmetry breaking. Now, quarks
are dynamical and the charge density has only the quark component. 
The instantaneous interaction contains the Coulomb plus linear confining 
potentials. Gluons are perturbative and quarks have unknown dispersion relation, 
dynamical energy $E(\mbf{k})$ which is parametrized through the BV angle 
$\Phi(\mbf{k})$. Vacuum $|\Omega\rangle$ contains condensates of quark pairs 
(i.e. it is the BCS vacuum). 
The Fock space is constructed from this vacuum using quasiparticle
operators $b^{\dagger}$ and $d^{\dagger}$ 
which appear in the field expansions
\begin{eqnarray}
&& \psi(\mbf{x}) = \sum_{s}\int\frac{d\mbf{k}}{(2\pi)^3}
    [u(\mbf{k},s)b(\mbf{k},s)
    +v(-\mbf{k},s)d^{\dagger}(-\mbf{k},s)] {\rm e}^{i\mbf{k}\mbf{x}} 
    \nonumber\\
&& \mbf{A}(\mbf{x}) = \sum_{a}\int\frac{d\mbf{k}}{(2\pi)^3}
\frac{1}{\sqrt{2\omega(\mbf{k})}}
    [a(\mbf{k},a)+a^{\dagger}(-\mbf{k},a)] {\rm e}^{i\mbf{k}\mbf{x}} 
    \nonumber\\
&& \mbf{\Pi}(\mbf{x}) = -i\sum_{a}\int\frac{d\mbf{k}}{(2\pi)^3}
\sqrt{ \frac{\omega(\mbf{k})}{2} }
    [a(\mbf{k},a)-a^{\dagger}(-\mbf{k},a)]{\rm e}^{i\mbf{k}\mbf{x}} 
\,,\label{eq:3.31}\end{eqnarray}
with $b|\Omega\rangle = d|\Omega\rangle =0$.
Here the quark operators are given in the helicity basis and all descrete numbers
(helicity, color, and flavor for the quarks and color for the gluons)
are collectively denoted by $s$ and $a$, respectively. We use spinors in 
the massive basis, where nonzero effective quark mass explicitly is included
in the spinor
\begin{eqnarray}
 u(\mbf{k},s) &=& \sqrt{E(\mbf{k})+M(\mbf{k})}\left(
  \begin{array}{c}
         1 \\
  \mbf{\sigma}\cdot\mbf{k}/(E(\mbf{k})+M(\mbf{k})) 
  \end{array} \right)\chi_{s}
\nonumber\\
&=& \frac{1}{\sqrt{2}}\left(
  \begin{array}{c}
  \sqrt{1+s(\mbf{k})} \chi_{s}\\
  \sqrt{1-s(\mbf{k})}(\mbf{\sigma}\cdot\mbf{\hat{k}})\chi_{s}
  \end{array} \right)
\nonumber\\
 v(-\mbf{k},s) &=& \sqrt{E(\mbf{k})+M(\mbf{k})}\left(
  \begin{array}{c}
  -\mbf{\sigma}\cdot\mbf{k}/(E(\mbf{k})+M(\mbf{k})) \\
         1 
  \end{array} \right)(-i\sigma_2\chi_{s})
\nonumber\\
&=& \frac{1}{\sqrt{2}}\left(
  \begin{array}{c}
  -\sqrt{1-s(\mbf{k})}(\mbf{\sigma}\cdot\mbf{\hat{k}})
  (-i\sigma_2\chi_{s}) \\
  \sqrt{1+s(\mbf{k})}(-i\sigma_2\chi_{s})
  \end{array} \right)  
\,,\label{eq:3.32}\end{eqnarray}
with sine and cosine of the Bogoliubov angle $\Phi(\mbf{k})$ given by
\begin{eqnarray}
\sin(\Phi(\mbf{k})) &=& s(\mbf{k})
=\frac{M(\mbf{k})}{\sqrt{\mbf{k}^2+M^2(\mbf{k})}}\,,\,
\cos(\Phi(\mbf{k})=c(\mbf{k})
=\frac{k}{\sqrt{\mbf{k}^2+M^2(\mbf{k})}}\nonumber\\
E(\mbf{k})&=& \sqrt{\mbf{k}^2+M^2(\mbf{k})}
\,,\label{eq:3.33}\end{eqnarray}   
$E(\mbf{k})$ is a single-quark energy which we refer as a gap energy below. 
Effective quark mass $M(\mbf{k})$ is kept as an unknown variational parameter 
through out the calculations, and is found from the gap equation by minimizing
the ground state (vacuum) energy. Gluon energy is $\omega(\mbf{k})=k$.

With the definition Eq. (\ref{eq:3.32}), the spinors satisfy the
nonrelativistic normalization and orthogonality relations  
$u^{\dagger}(\mbf{k},s)u(\mbf{k},s)
=v^{\dagger}(-\mbf{k},s)v(-\mbf{k},s)=1$
and $u^{\dagger}(\mbf{k},s)v(-\mbf{k},s)
=v^{\dagger}(-\mbf{k},s)u(\mbf{k},s)=0$.
Canonical (anti)commutation relations are
\begin{eqnarray}
&& \{b(\mbf{k},s),b^{\dagger}(\mbf{k}^{\prime},s^{\prime})\}
=\{d(-\mbf{k},s),d^{\dagger}(-\mbf{k}^{\prime},s^{\prime})\}
=(2\pi)^3\delta(\mbf{k}-\mbf{k}^{\prime})\delta_{s,s^{\prime}}
\nonumber\\
&& [a_i(\mbf{k},a),a^{\dagger}_j(\mbf{k}^{\prime},a^{\prime})]
=(2\pi)^3\delta(\mbf{k}-\mbf{k}^{\prime})D_{ij}(\mbf{k})\delta_{a,a^{\prime}}
\,,\label{eq:3.34}\end{eqnarray}    
where the gluon operators $\mbf{a}=a_i(\mbf{k})^a=
\sum_{\lambda=1,2}\epsilon_i(\mbf{k},\lambda)a^a(\mbf{k},\lambda)$ 
are transverse, i.e. $\mbf{k}\cdot\mbf{a}^a(\mbf{k})=
\mbf{k}\cdot\mbf{a}^{a\dagger}(\mbf{k})=0$, and $D_{ij}(\mbf{k})$ is a
polarization sum
\begin{eqnarray}
D_{ij}(\mbf{k})=\sum_{\lambda=1,2}\epsilon_i(\mbf{k},\lambda)
\epsilon_j(\mbf{k},\lambda)=\delta_{ij}-\hat{k}_i\hat{k}_j
\,,\label{eq:3.35}\end{eqnarray} 
with unit vector component $\hat{k}_i=k_i/|\mbf{k}|$ and
$\hat{k}_i\cdot D_{ij}(\mbf{k})=0$.
 
Following the same scheme, we represent the Hamiltonian in 
a second quantized form.
The free Hamiltonian (kinetic quark and gluon energies) is given by
\begin{eqnarray}
H_0(l) &=& \sum_{s}\int\frac{d\mbf{k}}{(2\pi)^3}\left(
(kc(\mbf{k},l) + ms(\mbf{k},l))
[b^{\dagger}_s(\mbf{k})b_s(\mbf{k})+d^{\dagger}_s(\mbf{k})d_s(\mbf{k})]
\right.\nonumber\\
&+&\left. (ks(\mbf{k},l) - mc(\mbf{k},l)) 
[b^{\dagger}_s(\mbf{k})d^{\dagger}_s({-\bf k})+d_s({-\bf k})b_s(\mbf{k})]
\right)
\nonumber\\
&+& \sum_a \int\frac{d\mbf{k}}{(2\pi)^3}
 \omega(\mbf{k}) a_i^{a\dagger}(\mbf{k})a_i^a(\mbf{k})\,.
\nonumber\\ 
\label{eq:3.36} 
\end{eqnarray}
The quark-gluon vertex is given by
\begin{eqnarray}
&& H_{qg}(l) = -\sum_{s_1,s_2,a}\int\left(\prod_{n=1}^{3}
\frac{d\mbf{k}_n}{(2\pi)^3} \right)\nonumber\\
&& \left[\,
g_0(\mbf{k}_1,\mbf{k}_2,\mbf{k}_3,l)
d_{s_1}(-\mbf{k}_1)T^{a}b_{s_2}(\mbf{k}_2)
\frac{a_i^a(\mbf{k}_3)}{\sqrt{2\omega(\mbf{k}_3)}}\right.\\
&&\left.
v^{\dagger}_{s_1}(-\mbf{k}_1)\alpha_i u_{s_2}(\mbf{k}_2)
(2\pi)^3\delta^{(3)}(\mbf{k}_1-\mbf{k}_2-\mbf{k}_3)+ ...\,\right]
\,,\label{eq:3.37}\end{eqnarray}
where $E(\mbf{k})=\sqrt{\mbf{k}^2+M^2(\mbf{k})}$
and $\omega(\mbf{k})=k$. 
By implementing flow equations effective coupling constants are generated 
($g_0(\mbf{k}_1,\mbf{k}_2,\mbf{k}_3;l)$ ...) 
which are functions of all three momenta corresponding 
to a given Fock sector and depend upon the flow parameter $l$. 
The instantaneous interaction includes the linear confining and Coulomb 
potentials and is given in the quark sector by
\begin{eqnarray}
H_{L+C}(l) &=& \sum_{s_1...s_4}\int 
\left(\prod_{n=1}^{4}\frac{d\mbf{k}_n}{(2\pi)^3}\right)
(2\pi)^3\delta^{(3)}(\mbf{k}_1+\mbf{k}_3-\mbf{k}_2-\mbf{k}_4)
V_{L+C}(\mbf{k}_1,\mbf{k}_2)
\label{eq:3.38} \\
&& \vspace{-2cm}\colon [u^{\dagger}_{s_1}(\mbf{k}_1)b^{\dagger}_{s_1}(\mbf{k}_1)
+ v^{\dagger}_{s_1}(-\mbf{k}_1)d_{s_1}(-\mbf{k}_1)]T^a
[u_{s_2}(\mbf{k}_2)b_{s_2}(\mbf{k}_2)
+ v_{s_2}(-\mbf{k}_2)d^{\dagger}_{s_2}(-\mbf{k}_2)]
\nonumber\\
&& \vspace{-2cm} [u^{\dagger}_{s_3}(\mbf{k}_3)b^{\dagger}_{s_3}(\mbf{k}_3)
+ v^{\dagger}_{s_3}(-\mbf{k}_3)d_{s_3}(-\mbf{k}_3)]T^a
[u_{s_4}(\mbf{k}_4)b_{s_4}(\mbf{k}_4)
+ v_{s_4}(-\mbf{k}_4)d^{\dagger}_{s_4}(-\mbf{k}_4)]\colon\nonumber
\,,\end{eqnarray}
where $V_{L+C}(\mbf{k},\mbf{q})\rightarrow
V_{L+C}(\mbf{k}-\mbf{q})$,
\begin{eqnarray}
C_f V_{L+C}(\mbf{k}) &=& 2\pi C_f \frac{\alpha_s}{\mbf{k}^2}
+4\pi\frac{\sigma}{\mbf{k}^4}
\,,\label{eq:3.39}\end{eqnarray}
with the fundamental Casimir operator $C_f=T^aT^a=(N_c^2-1)/2N_c=4/3$.
The spinors are also functions of the flow parameter, i.e. everywhere
$u_s(\mbf{k},l)$ and $v_s(\mbf{k},l)$.
In addition, the condensate $O_0$, $O_{L+C}$ and self-energy terms
$\Sigma_{L+C}$ arise due to the normal-ordering of the canonical QCD Hamiltonian 
with respect to the BCS vacuum.
Dynamical quark energy is expressed through the BV angle, which is kept as a trial
parameter through calculations.

Matrix elements of the Coulomb gauge normal-ordered Hamiltonian are shown in 
the Table \ref{tab:3.4}. For example the quark-gluon coupling is given
by the element $H_{14}$ and the instantaneous interaction
is given by $H_{33}$. Off-diagonal elements are eliminated by flow equations.

\begin{table}
\vspace{2cm}
\begin{tabular}{|r|c|c|c|c|c|} \hline
 & $|0>$ & $|q>$ & $|q\bar{q}>$ & $|q\bar{q}g>$ 
& $|q\bar{q}q\bar{q}>$   \\ \hline
$|0>$ &  &  &  & \floadeps{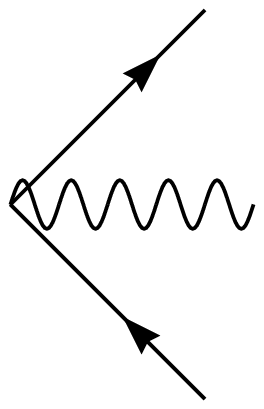} &  \\ \hline
$|q>$ &  &  & \floadeps{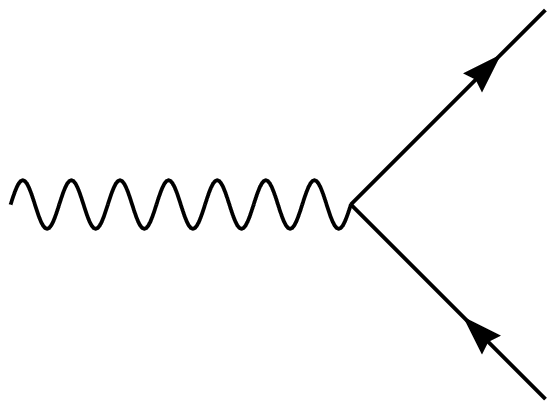} &  &  \\ \hline
$|q\bar{q}>$ &   & \floadeps{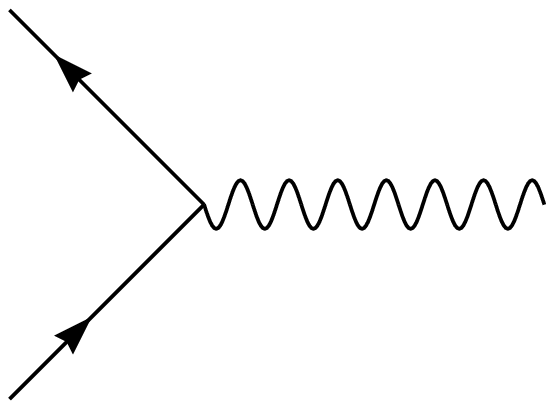} &\floadeps{table22a} 
& \floadeps{tab24} & \floadeps{table25a}  \\ \hline
$|q\bar{q}g>$ & \floadeps{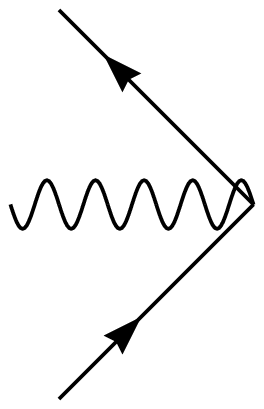} &  & \floadeps{tab42} 
&  & \floadeps{tab45} \\ \hline
$|q\bar{q}q\bar{q}>$ &  & & \floadeps{table52a} 
&\floadeps{tab21} & \floadeps{table22a}   \\ \hline
\end{tabular}
\vspace{1cm}
\caption{QCD Hamiltonian in the Coulomb gauge.}
\label{tab:3.4}
\end{table}

\begin{table}
\vspace{2cm}
\begin{tabular}{|r|c|c|c|c|c|} \hline
 & $|0>$ & $|q>$ & $|q\bar{q}>$ & $|q\bar{q}g>$ 
& $|q\bar{q}q\bar{q}>$   \\ \hline
$|0>$ & \floadeps{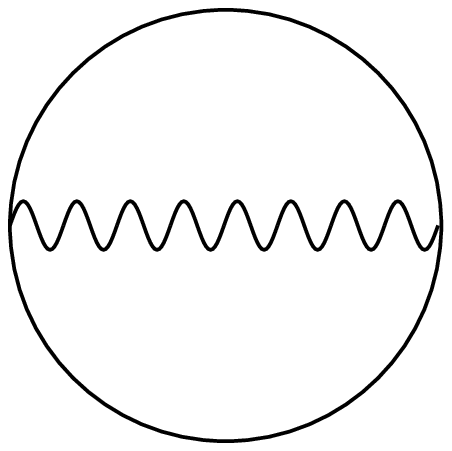} &  &  &  &  \\ \hline
$|q>$ &  & \floadeps{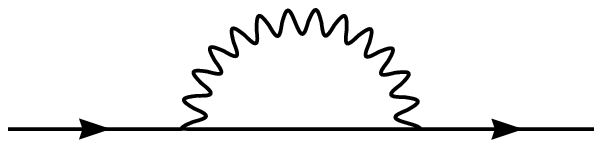} &  &  &  \\ \hline
$|q\bar{q}>$ &   &  & \floadeps{table22} &  &  \\ \hline
$|q\bar{q}g>$ &  &  &  & \floadeps{table44} &  \\ \hline
$|q\bar{q}q\bar{q}>$ &  &  &  &  & \floadeps{table22} \\ \hline
\end{tabular}
\vspace{1cm}
\caption{Effective QCD Hamiltonian generated through the second order
by flow equations in the Coulomb gauge.}
\label{tab:3.5}
\end{table}

As a result of this elimination new terms arise in the diagonal sector,
depicted in the Table \ref{tab:3.5}.

The gap equation allows determination of a nontrivial vacuum
with quark condensates and propagating quasiparticles,
here quarks with a dynamical mass. There are several ways
to obtain this equation, the most common based upon a variational
principle to minimize the vacuum (ground state) energy. The 
variational parameter is the angle of transformation
from undressed to dressed particle (quasiparticle) operators,
$\Phi(\mbf{k})$, which defines a quasiparticle basis
(Eqs. (\ref{eq:3.31}) and (\ref{eq:3.32})) 
with a dynamical quark mass $M(\mbf{k})$.
Therefore, analogosly to the gluon sector,  
minimizing the vacuum energy of the effective Hamiltonian,
\begin{eqnarray}
\frac{\delta\langle\Omega|H_{eff}|\Omega\rangle}{\delta\Phi(\mbf{k})} &=& 0
\,,\label{eq:3.40}\end{eqnarray}
we obtain the gap equation for the unknown $\Phi({\mbf{k}})$
or $M(\mbf{k})$.
Using the condensate terms given by $H_{11}^{eff}$, 
the gap equation is obtained as
\begin{eqnarray}
 ks(\mbf{k})-m(\Lambda)c(\mbf{k})
 &=& \int\frac{d\mbf{q}}{(2\pi)^3} C_f V_{L+C}(\mbf{k},\mbf{q})
\left[\, c(\mbf{k})s(\mbf{q}) - s(\mbf{k})c(\mbf{q})
\mbf{\hat{k}}\cdot\mbf{\hat{q}} \,\right]
{\rm e}^{-q^2/\Lambda^2}
\nonumber\\
&+& \vspace{-2cm}\int\frac{d\mbf{q}}{(2\pi)^3} C_f W(\mbf{k},\mbf{q})
\left[\, c(\mbf{k})s(\mbf{q}) - s(\mbf{k})c(\mbf{q})
\mbf{\hat{k}}\cdot\mbf{\hat{l}}\mbf{\hat{q}}\cdot\mbf{\hat{l}} \,\right]
{\rm e}^{-4q^2/\Lambda^2}
\,,\label{eq:3.41}\end{eqnarray}
where $\mbf{l}=\mbf{k}-\mbf{q}$, the potential functions
are given by 
\begin{eqnarray}
C_f V_{L+C}(\mbf{k},\mbf{q}) &=&\frac{1}{2}\frac{C_fg^2}{(\mbf{k}-\mbf{q})^2}
+\frac{4\pi\sigma}{(\mbf{k}-\mbf{q})^4}
\nonumber\\
C_f W(\mbf{k},\mbf{q}) &=& \frac{C_fg^2}
{\omega(\mbf{k}-\mbf{q})(E(\mbf{q})+\omega(\mbf{k}-\mbf{q}))} 
\,,\label{eq:3.42}\end{eqnarray}
and the running mass $m(\Lambda)$ includes the mass counterterm
and is defined by 
\begin{eqnarray}
m(\Lambda) &=& m+M_{CT}(\Lambda) = 
m\left(1-\frac{C_fg^2}{(4\pi)^2}6\ln\Lambda\right)
\nonumber\\
M_{CT}(\Lambda) &=& -\delta m =  
-\frac{C_fg^2}{(4\pi)^2}6m \ln\Lambda 
\,.\label{eq:3.43}\end{eqnarray}
The mass counterterm is proportional to the bare quark mass, $m$, 
and thus vanishes in the chiral limit $m\rightarrow 0$.
Quark mass counterterm cancels UV divergencies. 
In the chiral limit, no counterterms are required. The gap equation in UV
finite due to the dynamical interactions generated by flow equations.

\begin{figure}[!htb]
\begin{center}
\input{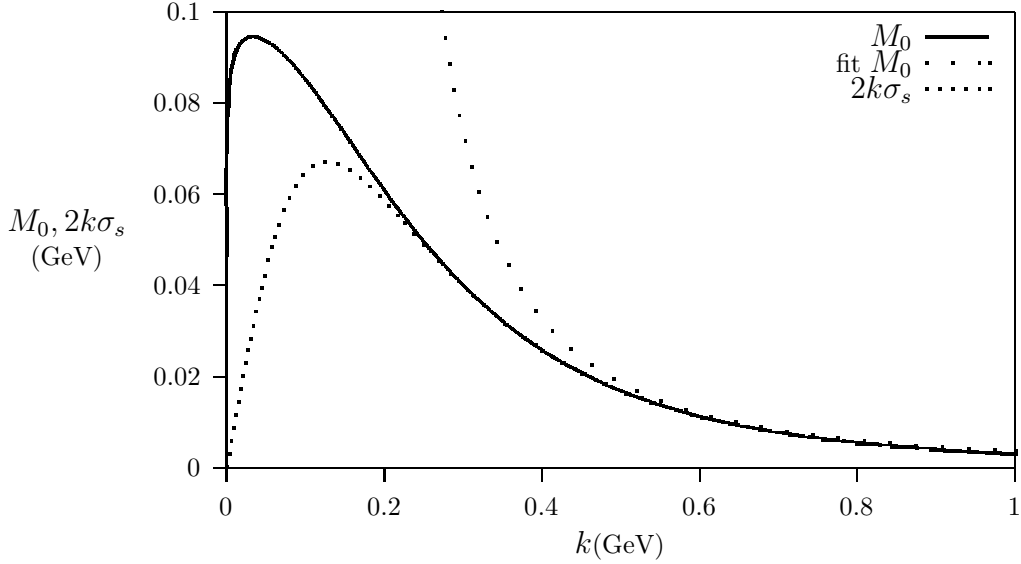}
\end{center}
\caption{The numerical solution of dynamical quark mass, $M_0(k)$, 
and the scalar part of the propagator, $2k\sigma_s$, in the chiral limit
when Coulomb and generated potentials with the running coupling
$\alpha_s(k^2)$ are added to confinement.
The parameters for the numerical solution of 
the gap equation are $\sigma=0.18GeV^2$, $\Lambda=1GeV$.
The results are compared with the fit function given by 
$M_0(k)=0.0060/(k^2[\ln(k^2/0.04)]^{0.43})$.}
\label{fig:3.5}
\end{figure}

\begin{figure}[!htb]
\begin{center}
\input{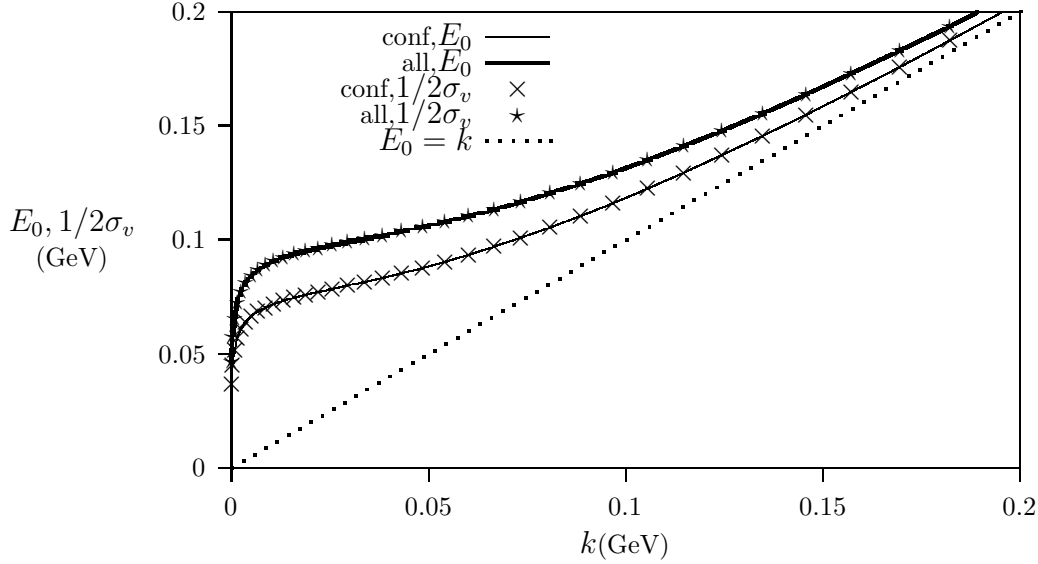}
\end{center}
\caption{One particle dispersion relation, $E_0(k)=\sqrt{k^2+M_0(k)^2}$,
free dispersion, $E_0(k)=k$, 
and the vector part of the propagator, $1/2\sigma_v$, in the chiral limit
with confinement and with confinement plus perturbative potentials (i.e.
when Coulomb and generated potentials are added).
The parameters are same as in Fig. \ref{fig:3.5}.}
\label{fig:3.6}
\end{figure}

Next we consider the quark condensate 
$\langle\Omega|\bar{\psi}\psi|\Omega\rangle$ for a single quark flavor.
We regulate the quark condensate by subtracting
the perturbative contribution 
\begin{eqnarray}
\langle\Omega|\bar{\psi}\psi|\Omega\rangle 
- \langle 0|\bar{\psi}\psi|0\rangle 
&=& -N_c \int\frac{d\mbf{k}}{(2\pi)^3} \left( {\rm Tr}S^{(3)}(\mbf{k})
-{\rm Tr}S_0^{(3)}(\mbf{k}) \right)
\nonumber\\
&=& -2N_c \int\frac{d\mbf{k}}{(2\pi)^3}\left( s(\mbf{k})
-\frac{m}{\sqrt{k^2+m^2}} \right)
\,,\label{eq:3.44}\end{eqnarray}
where $m$ is the bare quark mass.
As $|\mbf{k}|\rightarrow\infty$ the mass gap $M(\mbf{k})\rightarrow m$
and the nonperturbative sine behaves as 
$s(\mbf{k})\rightarrow m/\sqrt{k^2+m^2}$. Thus this subtraction improves 
the convergence of the quark condensate integral in the UV.
Here the equal-time propagator is given by 
\mbox{$S^{(3)}(\mbf{k})=1/2\Omega
\left[mA-\mbf{\gamma}\cdot\mbf{k}(1+B)\right]$}. 
When the scalar part of the propagator is nonzero ($mA\neq 0$), 
the nonzero mass gap ($M_0=mA/(1+B)$) and the chiral condensate 
(\mbox{$\langle\bar{\psi}\psi\rangle_0\sim\int dk mA/\Omega$}) 
are generated.

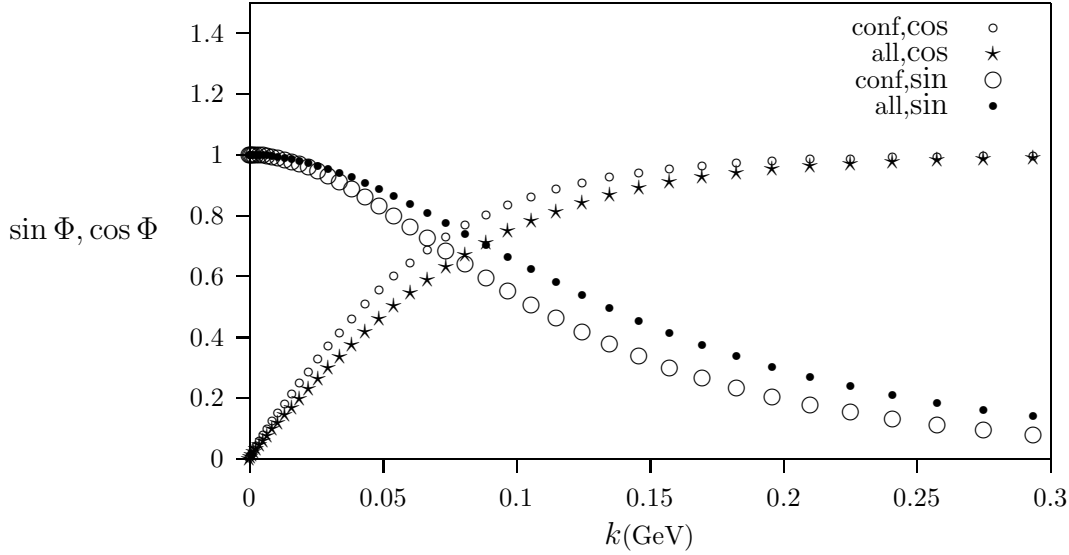
\begin{figure}[!htb]
\begin{center}
\setlength{\unitlength}{0.240900pt}
\ifx\plotpoint\undefined\newsavebox{\plotpoint}\fi
\sbox{\plotpoint}{\rule[-0.200pt]{0.400pt}{0.400pt}}%
\begin{picture}(1500,900)(0,0)
\font\gnuplot=cmr10 at 10pt
\gnuplot
\sbox{\plotpoint}{\rule[-0.200pt]{0.400pt}{0.400pt}}%
\put(141,143){\makebox(0,0)[r]{0}}
\put(161.0,143.0){\rule[-0.200pt]{4.818pt}{0.400pt}}
\put(141,239){\makebox(0,0)[r]{0.2}}
\put(161.0,239.0){\rule[-0.200pt]{4.818pt}{0.400pt}}
\put(141,334){\makebox(0,0)[r]{0.4}}
\put(161.0,334.0){\rule[-0.200pt]{4.818pt}{0.400pt}}
\put(141,430){\makebox(0,0)[r]{0.6}}
\put(161.0,430.0){\rule[-0.200pt]{4.818pt}{0.400pt}}
\put(141,525){\makebox(0,0)[r]{0.8}}
\put(161.0,525.0){\rule[-0.200pt]{4.818pt}{0.400pt}}
\put(141,621){\makebox(0,0)[r]{1}}
\put(161.0,621.0){\rule[-0.200pt]{4.818pt}{0.400pt}}
\put(141,717){\makebox(0,0)[r]{1.2}}
\put(161.0,717.0){\rule[-0.200pt]{4.818pt}{0.400pt}}
\put(141,812){\makebox(0,0)[r]{1.4}}
\put(161.0,812.0){\rule[-0.200pt]{4.818pt}{0.400pt}}
\put(181,82){\makebox(0,0){0}}
\put(181.0,123.0){\rule[-0.200pt]{0.400pt}{4.818pt}}
\put(391,82){\makebox(0,0){0.05}}
\put(391.0,123.0){\rule[-0.200pt]{0.400pt}{4.818pt}}
\put(600,82){\makebox(0,0){0.1}}
\put(600.0,123.0){\rule[-0.200pt]{0.400pt}{4.818pt}}
\put(810,82){\makebox(0,0){0.15}}
\put(810.0,123.0){\rule[-0.200pt]{0.400pt}{4.818pt}}
\put(1020,82){\makebox(0,0){0.2}}
\put(1020.0,123.0){\rule[-0.200pt]{0.400pt}{4.818pt}}
\put(1229,82){\makebox(0,0){0.25}}
\put(1229.0,123.0){\rule[-0.200pt]{0.400pt}{4.818pt}}
\put(1439,82){\makebox(0,0){0.3}}
\put(1439.0,123.0){\rule[-0.200pt]{0.400pt}{4.818pt}}
\put(181.0,143.0){\rule[-0.200pt]{303.052pt}{0.400pt}}
\put(1439.0,143.0){\rule[-0.200pt]{0.400pt}{172.725pt}}
\put(181.0,860.0){\rule[-0.200pt]{303.052pt}{0.400pt}}
\put(40,501){\makebox(0,0){\hspace{-2cm}$\sin\Phi,\cos\Phi$}}
\put(810,21){\makebox(0,0){$k$(GeV)}}
\put(181.0,143.0){\rule[-0.200pt]{0.400pt}{172.725pt}}
\put(1279,820){\makebox(0,0)[r]{conf,$\cos$}}
\put(181,144){\circle{12}}
\put(183,147){\circle{12}}
\put(185,151){\circle{12}}
\put(188,157){\circle{12}}
\put(192,163){\circle{12}}
\put(197,171){\circle{12}}
\put(203,180){\circle{12}}
\put(209,190){\circle{12}}
\put(217,202){\circle{12}}
\put(226,215){\circle{12}}
\put(237,229){\circle{12}}
\put(248,245){\circle{12}}
\put(260,262){\circle{12}}
\put(274,280){\circle{12}}
\put(289,300){\circle{12}}
\put(305,320){\circle{12}}
\put(323,341){\circle{12}}
\put(342,363){\circle{12}}
\put(363,386){\circle{12}}
\put(385,408){\circle{12}}
\put(408,430){\circle{12}}
\put(434,451){\circle{12}}
\put(461,472){\circle{12}}
\put(490,492){\circle{12}}
\put(520,510){\circle{12}}
\put(553,527){\circle{12}}
\put(587,542){\circle{12}}
\put(624,555){\circle{12}}
\put(663,567){\circle{12}}
\put(704,577){\circle{12}}
\put(747,586){\circle{12}}
\put(793,593){\circle{12}}
\put(841,599){\circle{12}}
\put(892,604){\circle{12}}
\put(946,608){\circle{12}}
\put(1002,611){\circle{12}}
\put(1062,614){\circle{12}}
\put(1125,615){\circle{12}}
\put(1191,617){\circle{12}}
\put(1261,618){\circle{12}}
\put(1334,619){\circle{12}}
\put(1412,619){\circle{12}}
\put(1349,820){\circle{12}}
\put(1279,779){\makebox(0,0)[r]{all,$\cos$}}
\put(181,144){\makebox(0,0){$\star$}}
\put(183,146){\makebox(0,0){$\star$}}
\put(185,150){\makebox(0,0){$\star$}}
\put(188,154){\makebox(0,0){$\star$}}
\put(192,159){\makebox(0,0){$\star$}}
\put(197,165){\makebox(0,0){$\star$}}
\put(203,172){\makebox(0,0){$\star$}}
\put(209,180){\makebox(0,0){$\star$}}
\put(217,190){\makebox(0,0){$\star$}}
\put(226,200){\makebox(0,0){$\star$}}
\put(237,212){\makebox(0,0){$\star$}}
\put(248,224){\makebox(0,0){$\star$}}
\put(260,238){\makebox(0,0){$\star$}}
\put(274,253){\makebox(0,0){$\star$}}
\put(289,269){\makebox(0,0){$\star$}}
\put(305,286){\makebox(0,0){$\star$}}
\put(323,304){\makebox(0,0){$\star$}}
\put(342,323){\makebox(0,0){$\star$}}
\put(363,343){\makebox(0,0){$\star$}}
\put(385,363){\makebox(0,0){$\star$}}
\put(408,383){\makebox(0,0){$\star$}}
\put(434,404){\makebox(0,0){$\star$}}
\put(461,424){\makebox(0,0){$\star$}}
\put(490,445){\makebox(0,0){$\star$}}
\put(520,464){\makebox(0,0){$\star$}}
\put(553,483){\makebox(0,0){$\star$}}
\put(587,501){\makebox(0,0){$\star$}}
\put(624,517){\makebox(0,0){$\star$}}
\put(663,532){\makebox(0,0){$\star$}}
\put(704,546){\makebox(0,0){$\star$}}
\put(747,558){\makebox(0,0){$\star$}}
\put(793,569){\makebox(0,0){$\star$}}
\put(841,578){\makebox(0,0){$\star$}}
\put(892,586){\makebox(0,0){$\star$}}
\put(946,593){\makebox(0,0){$\star$}}
\put(1002,599){\makebox(0,0){$\star$}}
\put(1062,603){\makebox(0,0){$\star$}}
\put(1125,607){\makebox(0,0){$\star$}}
\put(1191,610){\makebox(0,0){$\star$}}
\put(1261,613){\makebox(0,0){$\star$}}
\put(1334,615){\makebox(0,0){$\star$}}
\put(1412,616){\makebox(0,0){$\star$}}
\put(1349,779){\makebox(0,0){$\star$}}
\sbox{\plotpoint}{\rule[-0.400pt]{0.800pt}{0.800pt}}%
\put(1279,738){\makebox(0,0)[r]{conf,$\sin$}}
\put(181,621){\circle{24}}
\put(183,621){\circle{24}}
\put(185,621){\circle{24}}
\put(188,621){\circle{24}}
\put(192,621){\circle{24}}
\put(197,620){\circle{24}}
\put(203,620){\circle{24}}
\put(209,619){\circle{24}}
\put(217,617){\circle{24}}
\put(226,616){\circle{24}}
\put(237,613){\circle{24}}
\put(248,610){\circle{24}}
\put(260,606){\circle{24}}
\put(274,601){\circle{24}}
\put(289,595){\circle{24}}
\put(305,587){\circle{24}}
\put(323,578){\circle{24}}
\put(342,567){\circle{24}}
\put(363,555){\circle{24}}
\put(385,541){\circle{24}}
\put(408,525){\circle{24}}
\put(434,508){\circle{24}}
\put(461,490){\circle{24}}
\put(490,470){\circle{24}}
\put(520,450){\circle{24}}
\put(553,428){\circle{24}}
\put(587,407){\circle{24}}
\put(624,385){\circle{24}}
\put(663,364){\circle{24}}
\put(704,343){\circle{24}}
\put(747,323){\circle{24}}
\put(793,304){\circle{24}}
\put(841,286){\circle{24}}
\put(892,270){\circle{24}}
\put(946,254){\circle{24}}
\put(1002,240){\circle{24}}
\put(1062,227){\circle{24}}
\put(1125,216){\circle{24}}
\put(1191,205){\circle{24}}
\put(1261,196){\circle{24}}
\put(1334,188){\circle{24}}
\put(1412,181){\circle{24}}
\put(1349,738){\circle{24}}
\sbox{\plotpoint}{\rule[-0.500pt]{1.000pt}{1.000pt}}%
\put(1279,697){\makebox(0,0)[r]{all,$\sin$}}
\put(181,621){\circle*{12}}
\put(183,621){\circle*{12}}
\put(185,621){\circle*{12}}
\put(188,621){\circle*{12}}
\put(192,621){\circle*{12}}
\put(197,620){\circle*{12}}
\put(203,620){\circle*{12}}
\put(209,620){\circle*{12}}
\put(217,619){\circle*{12}}
\put(226,618){\circle*{12}}
\put(237,616){\circle*{12}}
\put(248,614){\circle*{12}}
\put(260,611){\circle*{12}}
\put(274,608){\circle*{12}}
\put(289,604){\circle*{12}}
\put(305,599){\circle*{12}}
\put(323,593){\circle*{12}}
\put(342,586){\circle*{12}}
\put(363,577){\circle*{12}}
\put(385,567){\circle*{12}}
\put(408,556){\circle*{12}}
\put(434,543){\circle*{12}}
\put(461,529){\circle*{12}}
\put(490,514){\circle*{12}}
\put(520,497){\circle*{12}}
\put(553,479){\circle*{12}}
\put(587,460){\circle*{12}}
\put(624,441){\circle*{12}}
\put(663,421){\circle*{12}}
\put(704,400){\circle*{12}}
\put(747,380){\circle*{12}}
\put(793,360){\circle*{12}}
\put(841,341){\circle*{12}}
\put(892,322){\circle*{12}}
\put(946,304){\circle*{12}}
\put(1002,287){\circle*{12}}
\put(1062,271){\circle*{12}}
\put(1125,257){\circle*{12}}
\put(1191,243){\circle*{12}}
\put(1261,231){\circle*{12}}
\put(1334,220){\circle*{12}}
\put(1412,210){\circle*{12}}
\put(1349,697){\circle*{12}}
\end{picture}
\end{center}
\caption{Sine and cosine of the Bogoliubov-Valatin angle
in the chiral limit with confinement and 
with confinement and with confinement plus perturbative potentials (i.e.
when Coulomb and generated potentials are added). 
The parameters are same as in Fig. \ref{fig:3.5}.}
\label{fig:3.7}
\end{figure}

We approximate the meson bound state as a valence pair of quark and antiquark,
that corresponds to the Tamm-Dancoff approximation.
In terms of the quasiparticle operators the TDA meson creation operator reads
\begin{eqnarray}  
R_n^{\dagger} &=& \int\frac{d\mbf{q}}{(2\pi)^3}\sum_{\delta\gamma}
b_{\delta}^{\dagger}(\mbf{q})d_{\gamma}^{\dagger}(-\mbf{q})
\psi_n^{\delta\gamma}(\mbf{q})
\,,\label{eq:3.45}\end{eqnarray}   
acting on the vacuum, it creates a meson with a wavefunction $|\psi_n\rangle$
with quantum number $n$; annihilation operator defines 
the nonperturbative vacuum, i.e.
\begin{eqnarray}
R_n^{\dagger}|\Omega\rangle &=& |\psi_n\rangle
\nonumber\\
R_n|\Omega\rangle &=& 0 
\,.\label{eq:3.46}\end{eqnarray} 
Projecting Schr{\"o}dinger equation $H_{eff}|\psi_{n}\rangle=E_n|\psi_{n}\rangle$
onto one particle-one hole truncated Fock sector, we get the TDA equation
\begin{eqnarray} 
\langle\Omega|[R_n,[H_{eff},b_{\alpha}^{\dagger}(\mbf{k})
d_{\beta}^{\dagger}(-\mbf{k})]]|\Omega\rangle &=& M_n \psi_n^{\alpha\beta}(\mbf{k})
\,,\label{eq:3.47}\end{eqnarray}  
where the binding energy is defined as $M_n=E_n-E_0$.
Commutation relation of the meson operators 
\begin{eqnarray} 
\langle\Omega|[R_{n^{\prime}},R_{n}^{\dagger}]|\Omega\rangle
=N\delta_{nn^{\prime}}
\,,\label{eq:3.48}\end{eqnarray}  
leads to a normalization condition for the wave functions
\begin{eqnarray} 
\int\frac{d\mbf{q}}{(2\pi)^3} \sum_{\delta\gamma}
\psi_{n^{\prime}}^{\delta\gamma\ast}(\mbf{q})\psi_{n}^{\delta\gamma}(\mbf{q}) 
= N\delta_{nn^{\prime}}
\,,\label{eq:3.49}\end{eqnarray}
with the normalization constant $N$. 

In the Random Phase Approximation, we allow in addition to the $q\bar{q}$ 
creation the $q\bar{q}$ annihilation in vacuum, with corresponding 
wave functions $X$ and $Y$ for each correlation.
Corresponding generalization of the operator given by Eq. (\ref{eq:3.45}) 
containes meson creation and annihilation terms
\begin{eqnarray}  
Q_n^{\dagger} &=& \int\frac{d\mbf{q}}{(2\pi)^3}\sum_{\delta\gamma}
\left[\, b_{\delta}^{\dagger}(\mbf{q})d_{\gamma}^{\dagger}(-\mbf{q})
X_n^{\delta\gamma}(\mbf{q})
-b_{\delta}(\mbf{q})d_{\gamma}(-\mbf{q})
Y_n^{\delta\gamma}(\mbf{q}) \,\right]
\,.\label{eq:3.50}\end{eqnarray} 
The RPA wavefunction and the RPA vacuum are defined by
\begin{eqnarray}
Q_n^{\dagger}|\Omega\rangle &=& |\psi_n\rangle
\nonumber\\
Q_n|\Omega\rangle &=& 0 
\,,\label{eq:3.51}\end{eqnarray} 
where, though the same notations were used above,
they should not be confused with the TDA wave function and TDA
vacuum.
The RPA bound state equation is given as a system of equations for 
both components of the wave function, $X$ and $Y$,  
\begin{eqnarray} 
\langle\Omega|[Q_n,[H_{eff},b_{\alpha}^{\dagger}(\mbf{k})
d_{\beta}^{\dagger}(-\mbf{k})]]|\Omega\rangle &=& 
M_n X_n^{\alpha\beta}(\mbf{k})
\nonumber\\
\langle\Omega|[Q_n,[H_{eff},b_{\alpha}(\mbf{k})
d_{\beta}(-\mbf{k})]]|\Omega\rangle &=& M_n 
Y_n^{\alpha\beta}(\mbf{k})
\,,\label{eq:3.52}\end{eqnarray}  
From the meson commutation relation  
\begin{eqnarray} 
\langle\Omega|[Q_{n^{\prime}},Q_{n}^{\dagger}]|\Omega\rangle
=N\delta_{nn^{\prime}}
\,,\label{eq:3.53}\end{eqnarray}  
the following normalization condition for the wave function
components, $X$  and $Y$, is obtained
\begin{eqnarray} 
\int\frac{d\mbf{q}}{(2\pi)^3} \sum_{\delta\gamma}\left[\,
X_{n^{\prime}}^{\delta\gamma\ast}(\mbf{q})X_{n}^{\delta\gamma}(\mbf{q})
-Y_{n^{\prime}}^{\delta\gamma\ast}(\mbf{q})Y_{n}^{\delta\gamma}(\mbf{q})
\,\right]
= N\delta_{nn^{\prime}}
\,,\label{eq:3.54}\end{eqnarray}
with the normalization constant $N$.

We considered psedoscalar $\pi$-meson ($J^{PC}=0^{++}$ $L=S=1,J=0$) 
and vector $\rho$-meson ($J^{PC}=0^{-+}$ $L=S=J=0$) channels
to illustrate effects of the chiral symmetry breaking.
Based on quantum numbers of these states, 
the tensor structure of $\pi$- and $\rho$-wave functions can be identified as 
\begin{eqnarray} 
X^{\alpha\beta}_{\pi}(\mbf{k}) &=& (i\sigma_2)^{\alpha\beta}X_{\pi}(\mbf{k})
\,,\,
Y^{\alpha\beta}_{\pi}(\mbf{k}) = (-i\sigma_2)^{\alpha\beta}Y_{\pi}(\mbf{k})
\nonumber\\
X^{\alpha\beta}_{\rho}(\mbf{k}) &=& (\mbf{\sigma}i\sigma_2)^{\alpha\beta}
X_{\rho}(\mbf{k})
\,,\,
Y^{\alpha\beta}_{\rho}(\mbf{k}) = 
(-i\sigma_2\mbf{\sigma})^{\alpha\beta}Y_{\rho}(\mbf{k})
\,.\label{eq:3.55}\end{eqnarray}
Normalization is chosen
\begin{eqnarray} 
\int\frac{d\mbf{k}}{(2\pi)^3}\left(X^*(\mbf{k})X(\mbf{k})
-Y^*(\mbf{k})Y(\mbf{k})\right)=1
\,,\label{eq:3.56}\end{eqnarray}
reducing the normalization costant of the full wave function to
\begin{eqnarray}
N=\langle\psi_n|\psi_n\rangle=2N_c
\,,\label{eq:3.57}\end{eqnarray}
where factor $2$ comes from trace in the spinor space, and $N_c=3$.
The RPA equations for the momentum wave function components 
$X(\mbf{k}), Y(\mbf{k})$ have the same form for $\pi$ and $\rho$ states
\begin{eqnarray} 
M_nX(\mbf{k}) &=& 2\varepsilon(\mbf{k})X(\mbf{k})
-\int\frac{q^2dqdx}{4\pi^2}I_{xx}(\mbf{k},\mbf{q})X(\mbf{q})
-\int\frac{q^2dqdx}{4\pi^2}I_{xy}(\mbf{k},\mbf{q})Y(\mbf{q})
\nonumber\\
&-&\int\frac{q^2dqdx}{4\pi^2}G_{xx}(\mbf{k},\mbf{q})X(\mbf{q})
-\int\frac{q^2dqdx}{4\pi^2}G_{xy}(\mbf{k},\mbf{q})Y(\mbf{q})
\nonumber\\
-M_nY(\mbf{k}) &=& 2\varepsilon(\mbf{k})Y(\mbf{k})
-\int\frac{q^2dqdx}{4\pi^2}I_{yy}(\mbf{k},\mbf{q})Y(\mbf{q})
-\int\frac{q^2dqdx}{4\pi^2}I_{yx}(\mbf{k},\mbf{q})X(\mbf{q})
\nonumber\\
&-&\int\frac{q^2dqdx}{4\pi^2}G_{yy}(\mbf{k},\mbf{q})Y(\mbf{q})
-\int\frac{q^2dqdx}{4\pi^2}G_{yx}(\mbf{k},\mbf{q})X(\mbf{q})
\,,\label{eq:3.58}\end{eqnarray}
where the kernels $I$ and $G$ for $\pi$ are given by
\begin{eqnarray} 
I_{xx}^{\pi}(\mbf{k},\mbf{q}) &=& I_{yy}^{\pi}(\mbf{k},\mbf{q})
= C_fV_{L+C}(\mbf{k},\mbf{q})
\frac{1}{2}\left[\,
(1+s(\mbf{k}))(1+s(\mbf{q}))\right.\nonumber\\
&+&\left.(1-s(\mbf{k}))(1-s(\mbf{q}))
+2c(\mbf{k})c(\mbf{q})x\, \right]
\nonumber\\
I_{xy}^{\pi}(\mbf{k},\mbf{q}) &=& I_{yx}^{\pi}(\mbf{k},\mbf{q})
= C_fV_{L+C}(\mbf{k},\mbf{q})
\frac{1}{2}\left[\,
-(1+s(\mbf{k}))(1-s(\mbf{q}))\right.\nonumber\\
&-&\left.(1-s(\mbf{k}))(1+s(\mbf{q}))
+2c(\mbf{k})c(\mbf{q})x\, \right]
\nonumber\\
G_{xx}^{\pi}(\mbf{k},\mbf{q}) &=& G_{yy}^{\pi}(\mbf{k},\mbf{q})=
2C_fW_1(\mbf{k},\mbf{q})
\frac{1}{2}\left[\,\phantom{\frac{x^2}{x^2}}\hspace{-0.5cm}
-(1+s(\mbf{k}))(1-s(\mbf{q}))\right.\nonumber\\
&-&\left.(1-s(\mbf{k}))(1+s(\mbf{q}))
-2c(\mbf{k})c(\mbf{q})
\frac{(1+x^2)kq-x(k^2+q^2)}{(\mbf{k}-\mbf{q})^2}\,\right]
\nonumber\\
G_{xy}^{\pi}(\mbf{k},\mbf{q}) &=& G_{yx}^{\pi}(\mbf{k},\mbf{q})=
2C_fW_2(\mbf{k},\mbf{q})
\frac{1}{2}\left[\,\phantom{\frac{x^2}{x^2}}\hspace{-0.5cm}
(1+s(\mbf{k}))(1+s(\mbf{q}))\right.\nonumber\\
&+&\left.(1-s(\mbf{k}))(1-s(\mbf{q}))
- 2c(\mbf{k})c(\mbf{q})
\frac{(1+x^2)kq-x(k^2+q^2)}{(\mbf{k}-\mbf{q})^2}\, \right]
\,,\label{eq:3.59}\end{eqnarray}
and for $\rho$ are given by
\begin{eqnarray} 
I_{xx}^{\rho}(\mbf{k},\mbf{q}) &=& I_{yy}^{\rho}(\mbf{k},\mbf{q})=
C_fV_{L+C}(\mbf{k},\mbf{q})\frac{1}{2}\left[\,
\phantom{\frac{x^2}{x^2}}\hspace{-0.5cm}
(1+s(\mbf{k}))(1+s(\mbf{q}))\right.\nonumber\\
&+&\left.\frac{1}{3}(1-s(\mbf{k}))(1-s(\mbf{q}))
(4x^2-1)+2c(\mbf{k})c(\mbf{q})x\, \right]
\nonumber\\
I_{xy}^{\rho}(\mbf{k},\mbf{q}) &=& I_{yx}^{\rho}(\mbf{k},\mbf{q})=
C_fV_{L+C}(\mbf{k},\mbf{q})\frac{1}{2}\left[\,
-\frac{1}{3}(1+s(\mbf{k}))(1-s(\mbf{q}))\right.\nonumber\\
&-&\left.\frac{1}{3}(1-s(\mbf{k}))(1+s(\mbf{q})) 
+\frac{2}{3}c(\mbf{k})c(\mbf{q})x\, \right]
\nonumber\\
G_{xx}^{\rho}(\mbf{k},\mbf{q}) &=& G_{yy}^{\rho}(\mbf{k},\mbf{q})=
C_fW_1(\mbf{k},\mbf{q})\frac{1}{2}\left[\,
\frac{1}{3}(1+s(\mbf{k}))(1-s(\mbf{q}))
\left(1-\frac{2(1-x^2)k^2}{(\mbf{k}-\mbf{q})^2}\right)\right.\nonumber\\
&+&\left.\frac{1}{3}(1-s(\mbf{k}))(1+s(\mbf{q}))
\left(1-\frac{2(1-x^2)q^2}{(\mbf{k}-\mbf{q})^2}\right)
-\frac{2}{3}c(\mbf{k})c(\mbf{q})
\left(x+\frac{(1-x^2)kq}{(\mbf{k}-\mbf{q})^2}\right)\, \right]
\nonumber\\
G_{xy}^{\rho}(\mbf{k},\mbf{q}) &=& G_{yx}^{\rho}(\mbf{k},\mbf{q})=
C_fW_2(\mbf{k},\mbf{q})
\frac{1}{2}\left[\,
\frac{1}{3}(1+s(\mbf{k}))(1+s(\mbf{q}))\right.\nonumber\\
&+&\left.\frac{1}{3}(1-s(\mbf{k}))(1-s(\mbf{q}))(2x^2-1)
+\frac{2}{3}c(\mbf{k})c(\mbf{q})
\left(x-\frac{(1-x^2)kq}{(\mbf{k}-\mbf{q})^2}\right)\, \right]
\,,\label{eq:3.60}\end{eqnarray}
where $x=\mbf{\hat{k}}\cdot\mbf{\hat{q}}$; and we have used
\begin{eqnarray}
1-(\mbf{\hat{k}}\cdot\mbf{\hat{l}})^2 &=&\frac{(1-x^2)q^2}{(\mbf{k}-\mbf{q})^2}
\,,\,
1-(\mbf{\hat{q}}\cdot\mbf{\hat{l}})^2 =\frac{(1-x^2)k^2}{(\mbf{k}-\mbf{q})^2}
\nonumber\\
\mbf{\hat{k}}\cdot\mbf{\hat{l}}\mbf{\hat{q}}\cdot\mbf{\hat{l}} &=&
\frac{x(k^2+q^2)-(1+x^2)kq}{(\mbf{k}-\mbf{q})^2}
\,,\label{eq:3.61}\end{eqnarray}
with $\mbf{l}=\mbf{k}-\mbf{q}$.

\begin{figure}[!htb]
\begin{center}
\input{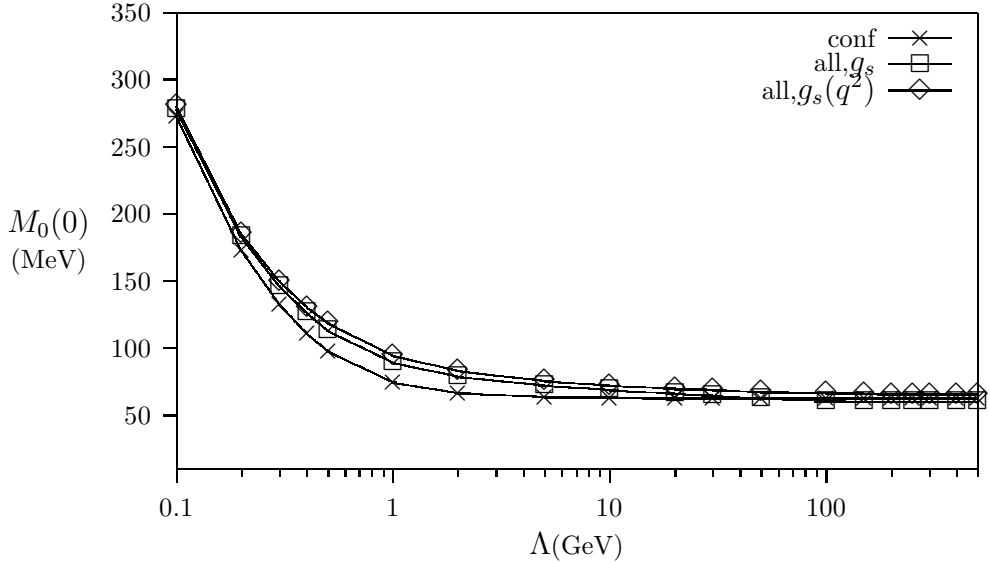}
\end{center}
\caption{Cut-off dependence of the constituent quark mass
in the chiral limit (same parameters as in Fig. \ref{fig:3.5}).
Crosses represent solution with confinement.
Boxes [diamonds] represent solution when Coulomb and generated
potentials are added with the constant value of coupling $g_s$
[with the running coupling $g_s(q^2)$].}
\label{fig:3.8}
\end{figure}

\begin{figure}[!htb]
\begin{center}
\input{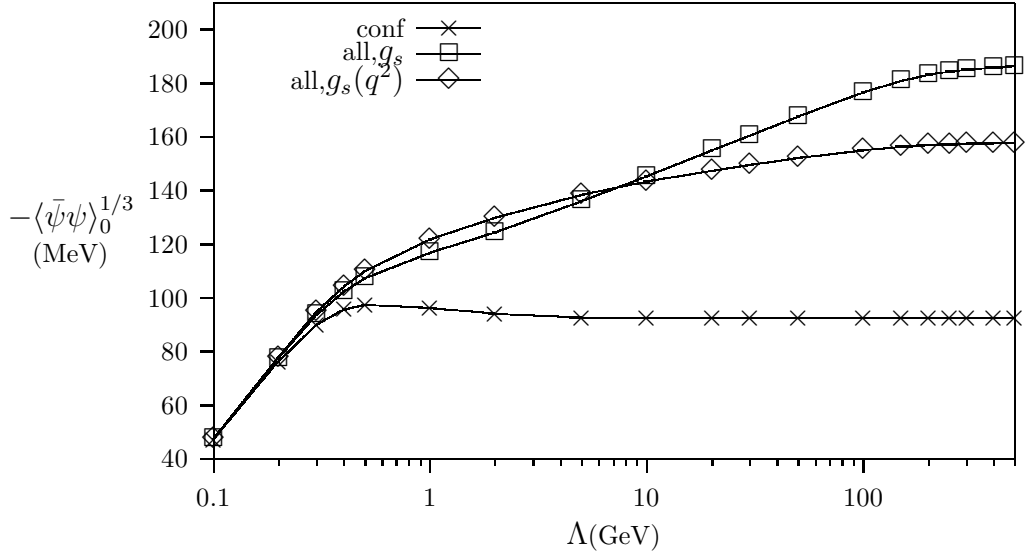}
\end{center}
\caption{Quark condensate cut-off dependence in the chiral limit 
(same parameters as in Fig. \ref{fig:3.5}).
Crosses represent solution with confinement.
Boxes [diamonds] represent solution when Coulomb and generated
potentials are added with the constant value of coupling $g_s$
[with the running coupling $g_s(q^2)$].}
\label{fig:3.9}
\end{figure}

\begin{table}
$$
\begin{tabular}{|c|ccc|ccc|} \hline 
   & &TDA, (MeV)& &&RPA, (MeV)&   \\ \hline
 conf. & $504$&$1364$&$2115$ & $222$&$1416$&$2298$ \\ \hline
 conf.$+$Coul. & $608$&$1514$&$2249$ & $427$&$1521$&$2309$ \\ \hline
 conf.$+$Coul.$+$gen.& $513$&$1411$&$2161$ & $180$&$1413$&$2218$   \\ \hline 
\end{tabular}
$$
\caption{Pion spectrum for the ground, first and second exited states
in the TDA and RPA approaches with confining, confining$+$Coulomb and 
confining$+$Coulomb$+$generated potentials taken. Chiral limit
$m=0$ ($\alpha_s=0.4, \sigma=0.18 GeV^2, \Lambda=10 GeV$).}
\label{tab:3.6}
\end{table}

\begin{table}[!htb]
$$
\begin{tabular}{|c|ccc|ccc|} \hline 
 m, (MeV)  && TDA, (MeV)  & && RPA, (MeV)& \\ \hline
 $150$ & $1038$&$1926$&$2936$ & $1037$&$1986$&$3077$ \\ \hline
 $100$ & $885$&$1762$&$2697$ & $868$&$1811$&$2826$   \\ \hline 
  $50$ & $716$&$1590$&$2431$ & $660$&$1626$&$2537$   \\ \hline 
  $10$ & $553$&$1446$&$2212$ & $366$&$1460$&$2283$   \\ \hline 
   $5$ & $532$&$1428$&$2186$ & $293$&$1437$&$2250$   \\ \hline 
   $0$ & $513$&$1411$&$2161$ & $180$&$1413$&$2218$   \\ \hline 
\end{tabular}
$$
\caption{Pion spectrum for the ground, first and second exited states
in the TDA and RPA approaches for different current masses of constituents.
Confining$+$Coulomb$+$generated potentials are taken, the same parameters
as in the Table \ref{tab:3.6}.}
\label{tab:3.7}
\end{table}

{\it Numerical results}

For the numerical calculations we have used the routines from the SLATEC 
linear algebra archive, part of the Netlib database maintained by 
UTK and ORNL. The routines are found at
$www.netlib.org/slatec/lin/$ and a description of the entire SLATEC archive
can be found at $www.netlib.org/slatec/toc$. 

Effective quark mass and energy are shown in Figures \ref{fig:3.5} and 
\ref{fig:3.6}, respectively.
At large momenta effective mass vanishes and the energy tends to 
the perturbative value, $E(\mbf{k})=k$, while for zero mometum, the energy 
tends to the constant, which we refer as the constituent quark mass.
No counterterms are required in the chiral limit. Due to the dynamical
interactions we recover the large momentum behavior correctly, corresponding
to the UV asymptotic region, which behavior is obtained in the course
of explicit numerical solutions of the QCD renormaliztion group and
operator product expansion and QCD sum rules.

At low momenta cosine behaves lineraly (Fig. \ref{fig:3.7}), 
with the slope $1/M(0)$ (inverse of quark constituent mass), 
$c(\mbf{k})=\mbf{k}/M(0)$.

\begin{table}[!htb]
$$
\begin{tabular}{|c|ccc|ccc|} \hline 
   & &TDA, (MeV)& &&RPA, (MeV)&   \\ \hline
 conf. & $659$&$1484$&$2258$ & $642$&$1482$&$2256$ \\ \hline
 conf.$+$Coul. & $750$&$1678$&$2515$ & $732$&$1676$&$2514$ \\ \hline
 conf.$+$Coul.$+$gen.& $718$&$1592$&$2377$ & $700$&$1590$&$2376$ \\ \hline 
\end{tabular}
$$
\caption{Spectrum of the $\rho$ meson for the ground, first and second exited 
states in the TDA and RPA approaches with confining, confining$+$Coulomb and 
confining$+$Coulomb$+$generated potentials taken. Chiral limit
$m=0$, the same parameters as in the Table \ref{tab:3.6}.} 
\label{tab:3.8}
\end{table}

\begin{table}[!htb]
$$
\begin{tabular}{|c|ccc|ccc|} \hline 
 m, (MeV)  && TDA, (MeV)  & && RPA, (MeV)& \\ \hline
 $150$ & $1130$&$2086$&$3247$ & $1128$&$2086$&$3247$ \\ \hline
 $100$ & $986$&$1916$&$2990$ & $983$&$1915$&$2990$   \\ \hline 
  $50$ & $839$&$1744$&$2692$ & $833$&$1744$&$2692$   \\ \hline 
  $10$ & $727$&$1616$&$2436$ & $714$&$1615$&$2435$   \\ \hline 
   $5$ & $719$&$1603$&$2406$ & $704$&$1601$&$2405$   \\ \hline 
   $0$ & $718$&$1592$&$2377$ & $700$&$1590$&$2376$   \\ \hline 
\end{tabular}
$$
\caption{Spectrum of the $\rho$ meson for the ground, first and second exited 
states in the TDA and RPA approaches for different current masses of constituents.
Confining$+$Coulomb$+$generated potentials are taken, the same parameters
as in the \mbox{Table \ref{tab:3.6}}.}
\label{tab:3.9}
\end{table}

The sensitivity of the constituent quark mass, $M(0)$, and the chiral 
condensate in the chiral limit to the cut-off $\Lambda$ is displayed
in Figures \ref{fig:3.8} and \ref{fig:3.9}. 
Only when all terms, instantaneous and dynamical 
interactions are added, we obtain the stable result. This proves that the
gap equation is renormalized completely. We need to add the renormalization 
group running of the coupling to renormalize the condensate.
For the first time, completely renormalized stable mass and condensate 
are obtained in the chiral limit, with values $M_0(0)=70\,MeV$
and $\langle\bar{\psi}\psi\rangle_0=-(155\, MeV)^3$. Flow equations improve
the chiral condensate by $68\%$.

Relative scales of constituent masses are obtained: 
for the gluon $<\,1 GeV$, for the $s$-quark $> 200\, MeV$,
for the $u,d$-quarks $< 100\,MeV$. 

\begin{figure}[!htb]
\begin{center}
\input{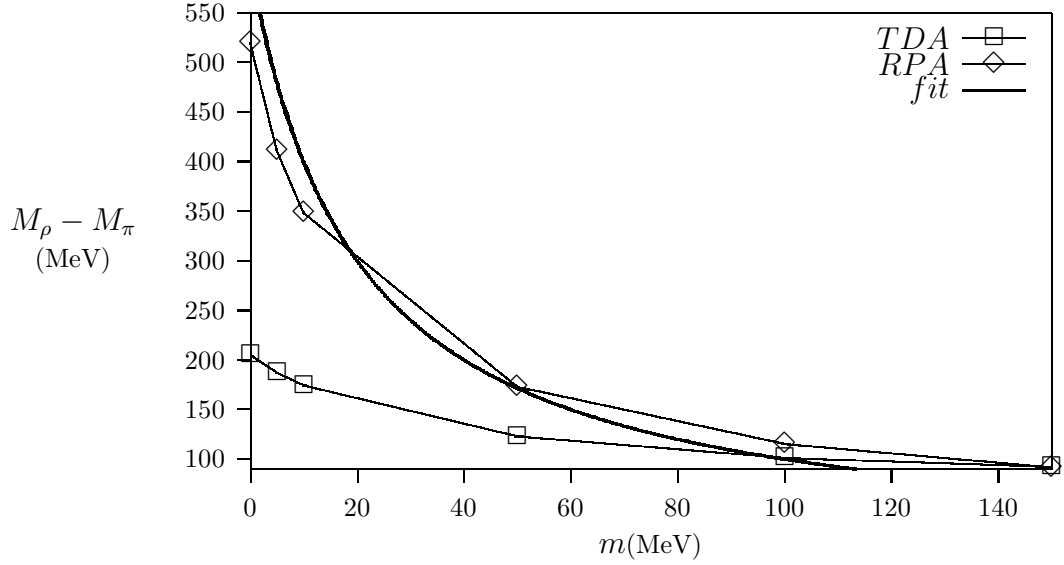}
\end{center}
\caption{$\pi-\rho$ mass splitting for the ground state 
in TDA and RPA approaches.
Chiral limit, $m=0$, and confining$+$Coulomb$+$generated interactions
are taken (same parameters as in the Table \ref{tab:3.6}).
Fit function is 12000/(m+20).}
\label{fig:3.10}
\end{figure}

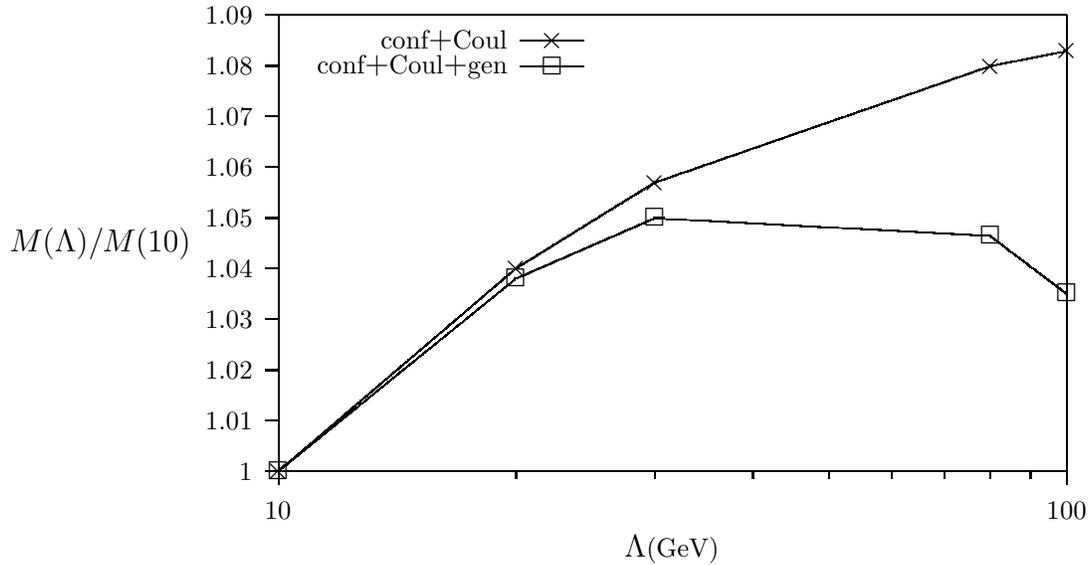
\begin{figure}[!htb]
\begin{center}
\setlength{\unitlength}{0.240900pt}
\ifx\plotpoint\undefined\newsavebox{\plotpoint}\fi
\sbox{\plotpoint}{\rule[-0.200pt]{0.400pt}{0.400pt}}%
\begin{picture}(1500,900)(0,0)
\font\gnuplot=cmr10 at 10pt
\gnuplot
\sbox{\plotpoint}{\rule[-0.200pt]{0.400pt}{0.400pt}}%
\put(161,143){\makebox(0,0)[r]{1}}
\put(181.0,143.0){\rule[-0.200pt]{4.818pt}{0.400pt}}
\put(161,223){\makebox(0,0)[r]{1.01}}
\put(181.0,223.0){\rule[-0.200pt]{4.818pt}{0.400pt}}
\put(161,302){\makebox(0,0)[r]{1.02}}
\put(181.0,302.0){\rule[-0.200pt]{4.818pt}{0.400pt}}
\put(161,382){\makebox(0,0)[r]{1.03}}
\put(181.0,382.0){\rule[-0.200pt]{4.818pt}{0.400pt}}
\put(161,462){\makebox(0,0)[r]{1.04}}
\put(181.0,462.0){\rule[-0.200pt]{4.818pt}{0.400pt}}
\put(161,541){\makebox(0,0)[r]{1.05}}
\put(181.0,541.0){\rule[-0.200pt]{4.818pt}{0.400pt}}
\put(161,621){\makebox(0,0)[r]{1.06}}
\put(181.0,621.0){\rule[-0.200pt]{4.818pt}{0.400pt}}
\put(161,701){\makebox(0,0)[r]{1.07}}
\put(181.0,701.0){\rule[-0.200pt]{4.818pt}{0.400pt}}
\put(161,780){\makebox(0,0)[r]{1.08}}
\put(181.0,780.0){\rule[-0.200pt]{4.818pt}{0.400pt}}
\put(161,860){\makebox(0,0)[r]{1.09}}
\put(181.0,860.0){\rule[-0.200pt]{4.818pt}{0.400pt}}
\put(201,82){\makebox(0,0){10}}
\put(201.0,123.0){\rule[-0.200pt]{0.400pt}{4.818pt}}
\put(574.0,133.0){\rule[-0.200pt]{0.400pt}{2.409pt}}
\put(792.0,133.0){\rule[-0.200pt]{0.400pt}{2.409pt}}
\put(946.0,133.0){\rule[-0.200pt]{0.400pt}{2.409pt}}
\put(1066.0,133.0){\rule[-0.200pt]{0.400pt}{2.409pt}}
\put(1164.0,133.0){\rule[-0.200pt]{0.400pt}{2.409pt}}
\put(1247.0,133.0){\rule[-0.200pt]{0.400pt}{2.409pt}}
\put(1319.0,133.0){\rule[-0.200pt]{0.400pt}{2.409pt}}
\put(1382.0,133.0){\rule[-0.200pt]{0.400pt}{2.409pt}}
\put(1439,82){\makebox(0,0){100}}
\put(1439.0,123.0){\rule[-0.200pt]{0.400pt}{4.818pt}}
\put(201.0,143.0){\rule[-0.200pt]{298.234pt}{0.400pt}}
\put(1439.0,143.0){\rule[-0.200pt]{0.400pt}{172.725pt}}
\put(201.0,860.0){\rule[-0.200pt]{298.234pt}{0.400pt}}
\put(40,501){\makebox(0,0){\hspace{-2cm}$M(\Lambda)/M(10)$}}
\put(820,21){\makebox(0,0){$\Lambda$(GeV)}}
\put(201.0,143.0){\rule[-0.200pt]{0.400pt}{172.725pt}}
\put(561,820){\makebox(0,0)[r]{conf$+$Coul}}
\put(581.0,820.0){\rule[-0.200pt]{24.090pt}{0.400pt}}
\put(201,143){\usebox{\plotpoint}}
\multiput(201.00,143.58)(0.585,0.500){635}{\rule{0.568pt}{0.120pt}}
\multiput(201.00,142.17)(371.822,319.000){2}{\rule{0.284pt}{0.400pt}}
\multiput(574.00,462.58)(0.808,0.499){267}{\rule{0.746pt}{0.120pt}}
\multiput(574.00,461.17)(216.452,135.000){2}{\rule{0.373pt}{0.400pt}}
\multiput(792.00,597.58)(1.441,0.500){363}{\rule{1.252pt}{0.120pt}}
\multiput(792.00,596.17)(524.402,183.000){2}{\rule{0.626pt}{0.400pt}}
\multiput(1319.00,780.58)(2.529,0.496){45}{\rule{2.100pt}{0.120pt}}
\multiput(1319.00,779.17)(115.641,24.000){2}{\rule{1.050pt}{0.400pt}}
\put(201,143){\makebox(0,0){$\times$}}
\put(574,462){\makebox(0,0){$\times$}}
\put(792,597){\makebox(0,0){$\times$}}
\put(1319,780){\makebox(0,0){$\times$}}
\put(1439,804){\makebox(0,0){$\times$}}
\put(631,820){\makebox(0,0){$\times$}}
\put(561,779){\makebox(0,0)[r]{conf$+$Coul$+$gen}}
\put(581.0,779.0){\rule[-0.200pt]{24.090pt}{0.400pt}}
\put(201,143){\usebox{\plotpoint}}
\multiput(201.00,143.58)(0.616,0.500){603}{\rule{0.592pt}{0.120pt}}
\multiput(201.00,142.17)(371.770,303.000){2}{\rule{0.296pt}{0.400pt}}
\multiput(574.00,446.58)(1.149,0.499){187}{\rule{1.018pt}{0.120pt}}
\multiput(574.00,445.17)(215.887,95.000){2}{\rule{0.509pt}{0.400pt}}
\multiput(792.00,539.92)(9.526,-0.497){53}{\rule{7.629pt}{0.120pt}}
\multiput(792.00,540.17)(511.167,-28.000){2}{\rule{3.814pt}{0.400pt}}
\multiput(1319.00,511.92)(0.660,-0.499){179}{\rule{0.627pt}{0.120pt}}
\multiput(1319.00,512.17)(118.698,-91.000){2}{\rule{0.314pt}{0.400pt}}
\put(201,143){\raisebox{-.8pt}{\makebox(0,0){$\Box$}}}
\put(574,446){\raisebox{-.8pt}{\makebox(0,0){$\Box$}}}
\put(792,541){\raisebox{-.8pt}{\makebox(0,0){$\Box$}}}
\put(1319,513){\raisebox{-.8pt}{\makebox(0,0){$\Box$}}}
\put(1439,422){\raisebox{-.8pt}{\makebox(0,0){$\Box$}}}
\put(631,779){\raisebox{-.8pt}{\makebox(0,0){$\Box$}}}
\end{picture}
\end{center}
\caption{Cut-off dependence of the pion mass
in the chiral limit $M(\Lambda)$, normalized to the pion mass
at $\Lambda=10GeV$, $M(10)$, (same parameters as in the Table 
\ref{tab:3.6}).
Line with crosses represents RPA solution with confinement$+$Coulomb,
the line with boxes stays for RPA solution when generated
potentials are added.}
\label{fig:3.11}
\end{figure}

Results for the pion and $\rho$-meson masses in RPA
and TDA approaches are presented in Tables \ref{tab:3.6}-\ref{tab:3.9}. 
In the chiral limit RPA gives ground state pion masses 
which are significantly lower than those obtained using TDA. 
Including the dynamical interaction terms reduces ground state pion 
mass even more, encreasing the mass splitting between the pion and $\rho$ 
meson (Tables \ref{tab:3.6} and \ref{tab:3.8}). In the chiral limit we get
\begin{eqnarray}
&& M_{\pi}=180\,MeV\,,\, M_{\rho}=700\,MeV 
\nonumber\\
&& M_{\rho}-M_{\pi}=520\,MeV
\,.\label{eq:3.62}\end{eqnarray} 
The $\pi-\rho$ mass splitting of $520\,MeV$ in the chiral limit
is close enough to the lattice data splitting of $600\,MeV$.
However, we are unable to get zero mass pion either in the BCS or
adding the leading order flow equations corrections. One of the reasons
is break of the covariance in this model.
Flow equations improve the $\pi-\rho$ mass splitting by $32\%$ in the TDA 
and by $24\%$ in the RPA.

Numerically obtained dependence of the $\pi-\rho$ mass splitting,
$M_{\rho}-M_{\pi}$, as a function of the bare mass of one of the quarks 
is shown in Figure \ref{fig:3.10}. We find $1/m_{const}$ behavior, where
$m_{const}$ is a constituent quark mass, which is valid for heavy quarks
and continues to be valid for lighter constituent quarks.
This fall of is more rapid in the RPA than in the TDA approach. 
From the RPA fit function, a constituent quark mass can be approximated,
uniformly for heavy and light quarks, as $m_{const}=m+20\,(MeV)$,
with the bare quark mass $m$. The $1/m_{const}$ behavior
is characteristic for the hyperfine interaction.
We find in the chiral limit, that in the TDA roughly $30\%$ of the $\pi-\rho$ 
mass splitting is due to the presence of the hyperfine interaction and $70\%$ 
due to the chiral symmetry breaking. In the RPA this ratio is $40\%$ for 
the hyperfine and $60\%$ for the chiral symmetry breaking. However,
the numerical value of this ratio depends on the details of the confining
interaction.

The dependence of the pion mass in the chiral limit on the cut-off parameter,
$M(\Lambda)$, is shown in Fig. \ref{fig:3.11}. The RPA solution, with only 
confining$+$Coulomb potential included, grows unlimited (due to Coulomb),
while adding the generated term stabilizes $M(\Lambda)$, which saturates
roughly at $M(10)$. Stable result confirms that the TDA/RPA equations
are completely renormalized when the generated by flow equations terms
are included.

\begin{figure}[!htb]
\begin{center}
\setlength{\unitlength}{0.240900pt}
\ifx\plotpoint\undefined\newsavebox{\plotpoint}\fi
\sbox{\plotpoint}{\rule[-0.200pt]{0.400pt}{0.400pt}}%
\begin{picture}(1500,900)(0,0)
\font\gnuplot=cmr10 at 10pt
\gnuplot
\sbox{\plotpoint}{\rule[-0.200pt]{0.400pt}{0.400pt}}%
\put(181,143){\makebox(0,0)[r]{0}}
\put(201.0,143.0){\rule[-0.200pt]{4.818pt}{0.400pt}}
\put(181,273){\makebox(0,0)[r]{2000}}
\put(201.0,273.0){\rule[-0.200pt]{4.818pt}{0.400pt}}
\put(181,404){\makebox(0,0)[r]{4000}}
\put(201.0,404.0){\rule[-0.200pt]{4.818pt}{0.400pt}}
\put(181,534){\makebox(0,0)[r]{6000}}
\put(201.0,534.0){\rule[-0.200pt]{4.818pt}{0.400pt}}
\put(181,664){\makebox(0,0)[r]{8000}}
\put(201.0,664.0){\rule[-0.200pt]{4.818pt}{0.400pt}}
\put(181,795){\makebox(0,0)[r]{10000}}
\put(201.0,795.0){\rule[-0.200pt]{4.818pt}{0.400pt}}
\put(221,82){\makebox(0,0){0}}
\put(221.0,123.0){\rule[-0.200pt]{0.400pt}{4.818pt}}
\put(343,82){\makebox(0,0){0.05}}
\put(343.0,123.0){\rule[-0.200pt]{0.400pt}{4.818pt}}
\put(465,82){\makebox(0,0){0.1}}
\put(465.0,123.0){\rule[-0.200pt]{0.400pt}{4.818pt}}
\put(586,82){\makebox(0,0){0.15}}
\put(586.0,123.0){\rule[-0.200pt]{0.400pt}{4.818pt}}
\put(708,82){\makebox(0,0){0.2}}
\put(708.0,123.0){\rule[-0.200pt]{0.400pt}{4.818pt}}
\put(830,82){\makebox(0,0){0.25}}
\put(830.0,123.0){\rule[-0.200pt]{0.400pt}{4.818pt}}
\put(952,82){\makebox(0,0){0.3}}
\put(952.0,123.0){\rule[-0.200pt]{0.400pt}{4.818pt}}
\put(1074,82){\makebox(0,0){0.35}}
\put(1074.0,123.0){\rule[-0.200pt]{0.400pt}{4.818pt}}
\put(1195,82){\makebox(0,0){0.4}}
\put(1195.0,123.0){\rule[-0.200pt]{0.400pt}{4.818pt}}
\put(1317,82){\makebox(0,0){0.45}}
\put(1317.0,123.0){\rule[-0.200pt]{0.400pt}{4.818pt}}
\put(1439,82){\makebox(0,0){0.5}}
\put(1439.0,123.0){\rule[-0.200pt]{0.400pt}{4.818pt}}
\put(221.0,143.0){\rule[-0.200pt]{293.416pt}{0.400pt}}
\put(1439.0,143.0){\rule[-0.200pt]{0.400pt}{172.725pt}}
\put(221.0,860.0){\rule[-0.200pt]{293.416pt}{0.400pt}}
\put(40,501){\makebox(0,0){\hspace{-0.3cm}$\psi^{0}_{\pi}$}}
\put(830,21){\makebox(0,0){$k$(GeV)}}
\put(221.0,143.0){\rule[-0.200pt]{0.400pt}{172.725pt}}
\sbox{\plotpoint}{\rule[-0.500pt]{1.000pt}{1.000pt}}%
\put(1279,820){\makebox(0,0)[r]{$\psi_{TDA}$}}
\multiput(1299,820)(20.756,0.000){5}{\usebox{\plotpoint}}
\put(1399,820){\usebox{\plotpoint}}
\put(221,677){\usebox{\plotpoint}}
\put(221.00,677.00){\usebox{\plotpoint}}
\put(241.63,676.00){\usebox{\plotpoint}}
\put(261.85,671.47){\usebox{\plotpoint}}
\put(281.51,664.83){\usebox{\plotpoint}}
\put(300.00,655.50){\usebox{\plotpoint}}
\put(317.92,645.05){\usebox{\plotpoint}}
\put(334.88,633.09){\usebox{\plotpoint}}
\put(351.27,620.36){\usebox{\plotpoint}}
\put(367.49,607.41){\usebox{\plotpoint}}
\put(382.66,593.25){\usebox{\plotpoint}}
\put(397.38,578.62){\usebox{\plotpoint}}
\put(412.06,563.94){\usebox{\plotpoint}}
\put(426.51,549.05){\usebox{\plotpoint}}
\multiput(437,538)(13.962,-15.358){2}{\usebox{\plotpoint}}
\put(468.66,503.23){\usebox{\plotpoint}}
\multiput(478,493)(14.053,-15.275){2}{\usebox{\plotpoint}}
\multiput(501,468)(14.078,-15.251){2}{\usebox{\plotpoint}}
\put(539.26,427.17){\usebox{\plotpoint}}
\multiput(550,416)(14.397,-14.951){2}{\usebox{\plotpoint}}
\multiput(576,389)(15.209,-14.123){2}{\usebox{\plotpoint}}
\multiput(604,363)(15.945,-13.287){2}{\usebox{\plotpoint}}
\multiput(634,338)(16.412,-12.706){2}{\usebox{\plotpoint}}
\multiput(665,314)(17.270,-11.513){2}{\usebox{\plotpoint}}
\multiput(698,292)(18.021,-10.298){2}{\usebox{\plotpoint}}
\put(751.45,263.29){\usebox{\plotpoint}}
\multiput(769,255)(19.372,-7.451){2}{\usebox{\plotpoint}}
\multiput(808,240)(19.739,-6.415){2}{\usebox{\plotpoint}}
\multiput(848,227)(20.108,-5.144){3}{\usebox{\plotpoint}}
\multiput(891,216)(20.261,-4.503){2}{\usebox{\plotpoint}}
\multiput(936,206)(20.461,-3.483){2}{\usebox{\plotpoint}}
\multiput(983,198)(20.495,-3.279){3}{\usebox{\plotpoint}}
\multiput(1033,190)(20.570,-2.769){2}{\usebox{\plotpoint}}
\multiput(1085,183)(20.589,-2.620){3}{\usebox{\plotpoint}}
\multiput(1140,176)(20.645,-2.136){3}{\usebox{\plotpoint}}
\multiput(1198,170)(20.659,-1.999){3}{\usebox{\plotpoint}}
\multiput(1260,164)(20.692,-1.617){3}{\usebox{\plotpoint}}
\multiput(1324,159)(20.700,-1.522){3}{\usebox{\plotpoint}}
\multiput(1392,154)(20.737,-0.882){2}{\usebox{\plotpoint}}
\put(1439,152){\usebox{\plotpoint}}
\sbox{\plotpoint}{\rule[-0.200pt]{0.400pt}{0.400pt}}%
\put(1279,779){\makebox(0,0)[r]{$X_{RPA}$}}
\put(221,788){\makebox(0,0){$\times$}}
\put(222,788){\makebox(0,0){$\times$}}
\put(223,788){\makebox(0,0){$\times$}}
\put(225,788){\makebox(0,0){$\times$}}
\put(227,788){\makebox(0,0){$\times$}}
\put(230,788){\makebox(0,0){$\times$}}
\put(234,788){\makebox(0,0){$\times$}}
\put(238,787){\makebox(0,0){$\times$}}
\put(242,786){\makebox(0,0){$\times$}}
\put(247,785){\makebox(0,0){$\times$}}
\put(253,783){\makebox(0,0){$\times$}}
\put(260,781){\makebox(0,0){$\times$}}
\put(267,778){\makebox(0,0){$\times$}}
\put(275,774){\makebox(0,0){$\times$}}
\put(284,769){\makebox(0,0){$\times$}}
\put(293,763){\makebox(0,0){$\times$}}
\put(303,755){\makebox(0,0){$\times$}}
\put(315,746){\makebox(0,0){$\times$}}
\put(327,735){\makebox(0,0){$\times$}}
\put(339,722){\makebox(0,0){$\times$}}
\put(353,707){\makebox(0,0){$\times$}}
\put(368,690){\makebox(0,0){$\times$}}
\put(384,670){\makebox(0,0){$\times$}}
\put(400,648){\makebox(0,0){$\times$}}
\put(418,624){\makebox(0,0){$\times$}}
\put(437,597){\makebox(0,0){$\times$}}
\put(457,568){\makebox(0,0){$\times$}}
\put(478,537){\makebox(0,0){$\times$}}
\put(501,505){\makebox(0,0){$\times$}}
\put(525,471){\makebox(0,0){$\times$}}
\put(550,438){\makebox(0,0){$\times$}}
\put(576,405){\makebox(0,0){$\times$}}
\put(604,373){\makebox(0,0){$\times$}}
\put(634,343){\makebox(0,0){$\times$}}
\put(665,316){\makebox(0,0){$\times$}}
\put(698,291){\makebox(0,0){$\times$}}
\put(733,270){\makebox(0,0){$\times$}}
\put(769,252){\makebox(0,0){$\times$}}
\put(808,236){\makebox(0,0){$\times$}}
\put(848,224){\makebox(0,0){$\times$}}
\put(891,213){\makebox(0,0){$\times$}}
\put(936,204){\makebox(0,0){$\times$}}
\put(983,195){\makebox(0,0){$\times$}}
\put(1033,188){\makebox(0,0){$\times$}}
\put(1085,180){\makebox(0,0){$\times$}}
\put(1140,173){\makebox(0,0){$\times$}}
\put(1198,167){\makebox(0,0){$\times$}}
\put(1260,161){\makebox(0,0){$\times$}}
\put(1324,156){\makebox(0,0){$\times$}}
\put(1392,152){\makebox(0,0){$\times$}}
\put(1349,779){\makebox(0,0){$\times$}}
\sbox{\plotpoint}{\rule[-0.400pt]{0.800pt}{0.800pt}}%
\put(1279,738){\makebox(0,0)[r]{$-Y_{RPA}$}}
\put(221,395){\makebox(0,0){$\star$}}
\put(222,395){\makebox(0,0){$\star$}}
\put(223,395){\makebox(0,0){$\star$}}
\put(225,395){\makebox(0,0){$\star$}}
\put(227,395){\makebox(0,0){$\star$}}
\put(230,395){\makebox(0,0){$\star$}}
\put(234,395){\makebox(0,0){$\star$}}
\put(238,395){\makebox(0,0){$\star$}}
\put(242,394){\makebox(0,0){$\star$}}
\put(247,394){\makebox(0,0){$\star$}}
\put(253,393){\makebox(0,0){$\star$}}
\put(260,392){\makebox(0,0){$\star$}}
\put(267,391){\makebox(0,0){$\star$}}
\put(275,389){\makebox(0,0){$\star$}}
\put(284,387){\makebox(0,0){$\star$}}
\put(293,384){\makebox(0,0){$\star$}}
\put(303,381){\makebox(0,0){$\star$}}
\put(315,377){\makebox(0,0){$\star$}}
\put(327,372){\makebox(0,0){$\star$}}
\put(339,366){\makebox(0,0){$\star$}}
\put(353,359){\makebox(0,0){$\star$}}
\put(368,352){\makebox(0,0){$\star$}}
\put(384,343){\makebox(0,0){$\star$}}
\put(400,333){\makebox(0,0){$\star$}}
\put(418,323){\makebox(0,0){$\star$}}
\put(437,311){\makebox(0,0){$\star$}}
\put(457,298){\makebox(0,0){$\star$}}
\put(478,285){\makebox(0,0){$\star$}}
\put(501,271){\makebox(0,0){$\star$}}
\put(525,257){\makebox(0,0){$\star$}}
\put(550,243){\makebox(0,0){$\star$}}
\put(576,229){\makebox(0,0){$\star$}}
\put(604,216){\makebox(0,0){$\star$}}
\put(634,204){\makebox(0,0){$\star$}}
\put(665,193){\makebox(0,0){$\star$}}
\put(698,184){\makebox(0,0){$\star$}}
\put(733,176){\makebox(0,0){$\star$}}
\put(769,170){\makebox(0,0){$\star$}}
\put(808,165){\makebox(0,0){$\star$}}
\put(848,162){\makebox(0,0){$\star$}}
\put(891,159){\makebox(0,0){$\star$}}
\put(936,157){\makebox(0,0){$\star$}}
\put(983,156){\makebox(0,0){$\star$}}
\put(1033,154){\makebox(0,0){$\star$}}
\put(1085,153){\makebox(0,0){$\star$}}
\put(1140,151){\makebox(0,0){$\star$}}
\put(1198,150){\makebox(0,0){$\star$}}
\put(1260,148){\makebox(0,0){$\star$}}
\put(1324,147){\makebox(0,0){$\star$}}
\put(1392,146){\makebox(0,0){$\star$}}
\put(1349,738){\makebox(0,0){$\star$}}
\end{picture}
\end{center}
\caption{Ground state pion wave function in TDA and RPA approaches.
Chiral limit, $m=0$, and confining$+$Coulomb$+$generated interactions
are taken  
($\alpha_s=0.4, \sigma=0.18 GeV^2, \Lambda=10 GeV$).}
\label{fig:3.12}
\end{figure}
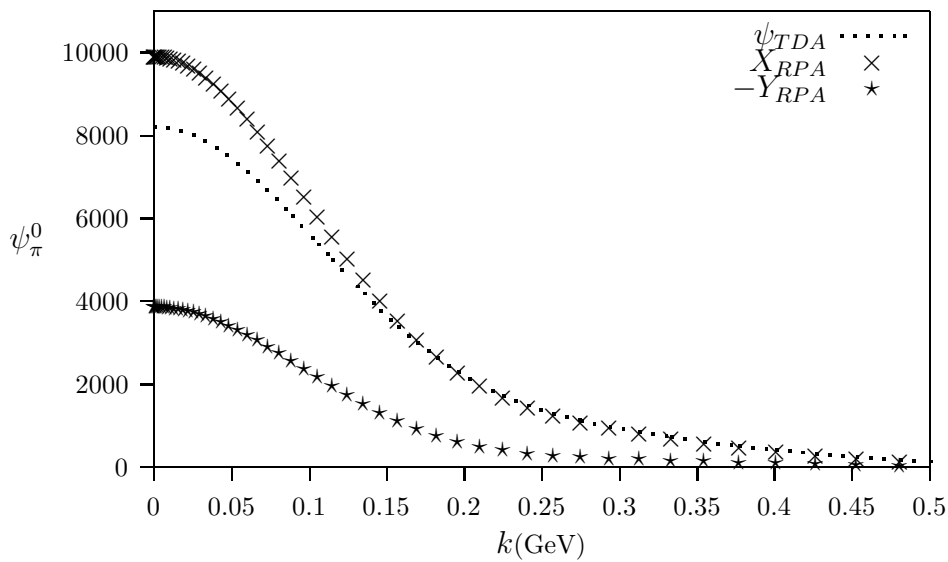

\begin{figure}[!htb]
\begin{center}
\setlength{\unitlength}{0.240900pt}
\ifx\plotpoint\undefined\newsavebox{\plotpoint}\fi
\sbox{\plotpoint}{\rule[-0.200pt]{0.400pt}{0.400pt}}%
\begin{picture}(1500,900)(0,0)
\font\gnuplot=cmr10 at 10pt
\gnuplot
\sbox{\plotpoint}{\rule[-0.200pt]{0.400pt}{0.400pt}}%
\put(181,143){\makebox(0,0)[r]{-4000}}
\put(201.0,143.0){\rule[-0.200pt]{4.818pt}{0.400pt}}
\put(181,245){\makebox(0,0)[r]{-2000}}
\put(201.0,245.0){\rule[-0.200pt]{4.818pt}{0.400pt}}
\put(181,348){\makebox(0,0)[r]{0}}
\put(201.0,348.0){\rule[-0.200pt]{4.818pt}{0.400pt}}
\put(181,450){\makebox(0,0)[r]{2000}}
\put(201.0,450.0){\rule[-0.200pt]{4.818pt}{0.400pt}}
\put(181,553){\makebox(0,0)[r]{4000}}
\put(201.0,553.0){\rule[-0.200pt]{4.818pt}{0.400pt}}
\put(181,655){\makebox(0,0)[r]{6000}}
\put(201.0,655.0){\rule[-0.200pt]{4.818pt}{0.400pt}}
\put(181,758){\makebox(0,0)[r]{8000}}
\put(201.0,758.0){\rule[-0.200pt]{4.818pt}{0.400pt}}
\put(181,860){\makebox(0,0)[r]{10000}}
\put(201.0,860.0){\rule[-0.200pt]{4.818pt}{0.400pt}}
\put(221,82){\makebox(0,0){0}}
\put(221.0,123.0){\rule[-0.200pt]{0.400pt}{4.818pt}}
\put(343,82){\makebox(0,0){0.05}}
\put(343.0,123.0){\rule[-0.200pt]{0.400pt}{4.818pt}}
\put(465,82){\makebox(0,0){0.1}}
\put(465.0,123.0){\rule[-0.200pt]{0.400pt}{4.818pt}}
\put(586,82){\makebox(0,0){0.15}}
\put(586.0,123.0){\rule[-0.200pt]{0.400pt}{4.818pt}}
\put(708,82){\makebox(0,0){0.2}}
\put(708.0,123.0){\rule[-0.200pt]{0.400pt}{4.818pt}}
\put(830,82){\makebox(0,0){0.25}}
\put(830.0,123.0){\rule[-0.200pt]{0.400pt}{4.818pt}}
\put(952,82){\makebox(0,0){0.3}}
\put(952.0,123.0){\rule[-0.200pt]{0.400pt}{4.818pt}}
\put(1074,82){\makebox(0,0){0.35}}
\put(1074.0,123.0){\rule[-0.200pt]{0.400pt}{4.818pt}}
\put(1195,82){\makebox(0,0){0.4}}
\put(1195.0,123.0){\rule[-0.200pt]{0.400pt}{4.818pt}}
\put(1317,82){\makebox(0,0){0.45}}
\put(1317.0,123.0){\rule[-0.200pt]{0.400pt}{4.818pt}}
\put(1439,82){\makebox(0,0){0.5}}
\put(1439.0,123.0){\rule[-0.200pt]{0.400pt}{4.818pt}}
\put(221.0,348.0){\rule[-0.200pt]{293.416pt}{0.400pt}}
\put(221.0,143.0){\rule[-0.200pt]{293.416pt}{0.400pt}}
\put(1439.0,143.0){\rule[-0.200pt]{0.400pt}{172.725pt}}
\put(221.0,860.0){\rule[-0.200pt]{293.416pt}{0.400pt}}
\put(40,501){\makebox(0,0){\hspace{-0.3cm}$\psi^{1}_{\pi}$}}
\put(830,21){\makebox(0,0){$k$(GeV)}}
\put(221.0,143.0){\rule[-0.200pt]{0.400pt}{172.725pt}}
\sbox{\plotpoint}{\rule[-0.500pt]{1.000pt}{1.000pt}}%
\put(1279,820){\makebox(0,0)[r]{$\psi_{TDA}$}}
\multiput(1299,820)(20.756,0.000){5}{\usebox{\plotpoint}}
\put(1399,820){\usebox{\plotpoint}}
\put(221,773){\usebox{\plotpoint}}
\put(221.00,773.00){\usebox{\plotpoint}}
\put(241.41,771.15){\usebox{\plotpoint}}
\put(261.55,766.33){\usebox{\plotpoint}}
\put(280.74,758.45){\usebox{\plotpoint}}
\put(298.91,748.45){\usebox{\plotpoint}}
\put(316.68,737.74){\usebox{\plotpoint}}
\put(333.03,724.97){\usebox{\plotpoint}}
\put(348.86,711.55){\usebox{\plotpoint}}
\put(364.19,697.56){\usebox{\plotpoint}}
\put(378.99,683.01){\usebox{\plotpoint}}
\put(392.81,667.54){\usebox{\plotpoint}}
\put(406.41,651.87){\usebox{\plotpoint}}
\multiput(418,639)(13.219,-16.002){2}{\usebox{\plotpoint}}
\put(446.44,604.20){\usebox{\plotpoint}}
\multiput(457,591)(12.453,-16.604){2}{\usebox{\plotpoint}}
\multiput(478,563)(13.174,-16.038){2}{\usebox{\plotpoint}}
\multiput(501,535)(12.706,-16.412){2}{\usebox{\plotpoint}}
\multiput(525,504)(13.029,-16.156){2}{\usebox{\plotpoint}}
\multiput(550,473)(13.338,-15.903){2}{\usebox{\plotpoint}}
\multiput(576,442)(13.912,-15.403){2}{\usebox{\plotpoint}}
\multiput(604,411)(14.676,-14.676){2}{\usebox{\plotpoint}}
\multiput(634,381)(15.157,-14.179){2}{\usebox{\plotpoint}}
\multiput(665,352)(16.544,-12.533){2}{\usebox{\plotpoint}}
\multiput(698,327)(17.572,-11.045){2}{\usebox{\plotpoint}}
\multiput(733,305)(18.768,-8.863){2}{\usebox{\plotpoint}}
\multiput(769,288)(19.535,-7.013){2}{\usebox{\plotpoint}}
\put(828.22,269.96){\usebox{\plotpoint}}
\multiput(848,266)(20.666,-1.922){3}{\usebox{\plotpoint}}
\multiput(891,262)(20.756,0.000){2}{\usebox{\plotpoint}}
\multiput(936,262)(20.681,1.760){2}{\usebox{\plotpoint}}
\multiput(983,266)(20.555,2.878){2}{\usebox{\plotpoint}}
\multiput(1033,273)(20.451,3.540){3}{\usebox{\plotpoint}}
\multiput(1085,282)(20.421,3.713){3}{\usebox{\plotpoint}}
\multiput(1140,292)(20.392,3.867){3}{\usebox{\plotpoint}}
\multiput(1198,303)(20.491,3.305){3}{\usebox{\plotpoint}}
\multiput(1260,313)(20.595,2.574){3}{\usebox{\plotpoint}}
\multiput(1324,321)(20.613,2.425){3}{\usebox{\plotpoint}}
\multiput(1392,329)(20.681,1.760){2}{\usebox{\plotpoint}}
\put(1439,333){\usebox{\plotpoint}}
\sbox{\plotpoint}{\rule[-0.200pt]{0.400pt}{0.400pt}}%
\put(1279,779){\makebox(0,0)[r]{$X_{RPA}$}}
\put(221,800){\makebox(0,0){$\times$}}
\put(222,800){\makebox(0,0){$\times$}}
\put(223,800){\makebox(0,0){$\times$}}
\put(225,800){\makebox(0,0){$\times$}}
\put(227,800){\makebox(0,0){$\times$}}
\put(230,800){\makebox(0,0){$\times$}}
\put(234,799){\makebox(0,0){$\times$}}
\put(238,799){\makebox(0,0){$\times$}}
\put(242,798){\makebox(0,0){$\times$}}
\put(247,797){\makebox(0,0){$\times$}}
\put(253,795){\makebox(0,0){$\times$}}
\put(260,793){\makebox(0,0){$\times$}}
\put(267,791){\makebox(0,0){$\times$}}
\put(275,787){\makebox(0,0){$\times$}}
\put(284,782){\makebox(0,0){$\times$}}
\put(293,777){\makebox(0,0){$\times$}}
\put(303,770){\makebox(0,0){$\times$}}
\put(315,761){\makebox(0,0){$\times$}}
\put(327,751){\makebox(0,0){$\times$}}
\put(339,739){\makebox(0,0){$\times$}}
\put(353,725){\makebox(0,0){$\times$}}
\put(368,708){\makebox(0,0){$\times$}}
\put(384,690){\makebox(0,0){$\times$}}
\put(400,669){\makebox(0,0){$\times$}}
\put(418,645){\makebox(0,0){$\times$}}
\put(437,619){\makebox(0,0){$\times$}}
\put(457,590){\makebox(0,0){$\times$}}
\put(478,560){\makebox(0,0){$\times$}}
\put(501,527){\makebox(0,0){$\times$}}
\put(525,494){\makebox(0,0){$\times$}}
\put(550,460){\makebox(0,0){$\times$}}
\put(576,426){\makebox(0,0){$\times$}}
\put(604,393){\makebox(0,0){$\times$}}
\put(634,361){\makebox(0,0){$\times$}}
\put(665,332){\makebox(0,0){$\times$}}
\put(698,307){\makebox(0,0){$\times$}}
\put(733,287){\makebox(0,0){$\times$}}
\put(769,271){\makebox(0,0){$\times$}}
\put(808,260){\makebox(0,0){$\times$}}
\put(848,254){\makebox(0,0){$\times$}}
\put(891,253){\makebox(0,0){$\times$}}
\put(936,257){\makebox(0,0){$\times$}}
\put(983,264){\makebox(0,0){$\times$}}
\put(1033,273){\makebox(0,0){$\times$}}
\put(1085,285){\makebox(0,0){$\times$}}
\put(1140,296){\makebox(0,0){$\times$}}
\put(1198,307){\makebox(0,0){$\times$}}
\put(1260,317){\makebox(0,0){$\times$}}
\put(1324,326){\makebox(0,0){$\times$}}
\put(1392,333){\makebox(0,0){$\times$}}
\put(1349,779){\makebox(0,0){$\times$}}
\sbox{\plotpoint}{\rule[-0.400pt]{0.800pt}{0.800pt}}%
\put(1279,738){\makebox(0,0)[r]{$-Y_{RPA}$}}
\put(221,441){\makebox(0,0){$\star$}}
\put(222,441){\makebox(0,0){$\star$}}
\put(223,441){\makebox(0,0){$\star$}}
\put(225,441){\makebox(0,0){$\star$}}
\put(227,441){\makebox(0,0){$\star$}}
\put(230,441){\makebox(0,0){$\star$}}
\put(234,441){\makebox(0,0){$\star$}}
\put(238,440){\makebox(0,0){$\star$}}
\put(242,440){\makebox(0,0){$\star$}}
\put(247,440){\makebox(0,0){$\star$}}
\put(253,440){\makebox(0,0){$\star$}}
\put(260,439){\makebox(0,0){$\star$}}
\put(267,439){\makebox(0,0){$\star$}}
\put(275,438){\makebox(0,0){$\star$}}
\put(284,437){\makebox(0,0){$\star$}}
\put(293,435){\makebox(0,0){$\star$}}
\put(303,434){\makebox(0,0){$\star$}}
\put(315,432){\makebox(0,0){$\star$}}
\put(327,430){\makebox(0,0){$\star$}}
\put(339,427){\makebox(0,0){$\star$}}
\put(353,424){\makebox(0,0){$\star$}}
\put(368,420){\makebox(0,0){$\star$}}
\put(384,416){\makebox(0,0){$\star$}}
\put(400,412){\makebox(0,0){$\star$}}
\put(418,407){\makebox(0,0){$\star$}}
\put(437,401){\makebox(0,0){$\star$}}
\put(457,396){\makebox(0,0){$\star$}}
\put(478,389){\makebox(0,0){$\star$}}
\put(501,383){\makebox(0,0){$\star$}}
\put(525,376){\makebox(0,0){$\star$}}
\put(550,370){\makebox(0,0){$\star$}}
\put(576,364){\makebox(0,0){$\star$}}
\put(604,358){\makebox(0,0){$\star$}}
\put(634,353){\makebox(0,0){$\star$}}
\put(665,348){\makebox(0,0){$\star$}}
\put(698,345){\makebox(0,0){$\star$}}
\put(733,342){\makebox(0,0){$\star$}}
\put(769,340){\makebox(0,0){$\star$}}
\put(808,339){\makebox(0,0){$\star$}}
\put(848,339){\makebox(0,0){$\star$}}
\put(891,339){\makebox(0,0){$\star$}}
\put(936,340){\makebox(0,0){$\star$}}
\put(983,341){\makebox(0,0){$\star$}}
\put(1033,342){\makebox(0,0){$\star$}}
\put(1085,344){\makebox(0,0){$\star$}}
\put(1140,345){\makebox(0,0){$\star$}}
\put(1198,345){\makebox(0,0){$\star$}}
\put(1260,346){\makebox(0,0){$\star$}}
\put(1324,347){\makebox(0,0){$\star$}}
\put(1392,347){\makebox(0,0){$\star$}}
\put(1349,738){\makebox(0,0){$\star$}}
\end{picture}
\end{center}
\caption{Pion wave function for the first exited state in TDA and RPA 
approaches. Chiral limit, $m=0$, and confining$+$Coulomb$+$generated 
interactions are taken into account (same parameters as in 
Fig. \ref{fig:3.10}).}
\label{fig:3.13}
\end{figure}
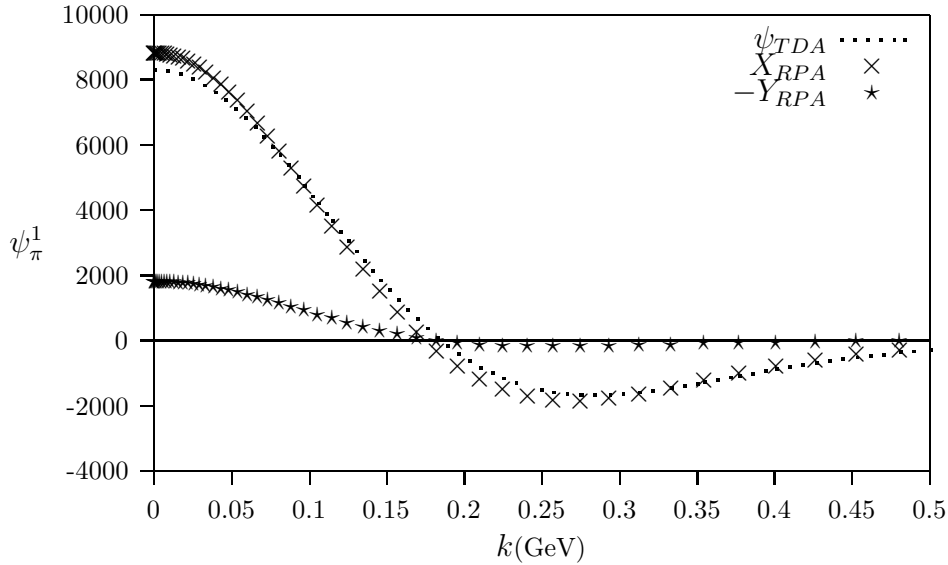

Pion wave functions for the ground and first excited states
are depicted at Figures \ref{fig:3.12} and \ref{fig:3.13}. 
Using the TDA and RPA wave functions, 
we obtained the pion decay constants. Though we obtained low value of quark 
condensate $-(155\,MeV)^3$, we obtain a realistic pion decay constant,
$92\, MeV$, in the RPA.

{\bf Summary III}

1) For the first time, due to the dynamical interactions, the UV and IR
finite equations are obtained in many-body technique. In the chiral limit,
no additional UV renormalization is required (no momentum dependent 
counterterms).

2) For the first time, the instantaneous as well as dynamical 
interactions are included (In this way we have utilized the success of the 
BCS model, where nonperturbative features such as dynamical chiral symmetry 
breaking and massive quasiparticle modes are explicitly present, and included 
perturbatively dynamical interactions in the BCS framework). 
In the RPA roughly $40\%$ of the $\pi-\rho$ mass splitting is due to the presence 
of the hyperfine interaction and $60\%$ due to the chiral symmetry breaking
(link between different model calcualtions). 

3) Due to the dynamical interactions, closed Ward-Takashi identities 
are obtained. Though we work in the gauge theory, there are no ambiguities,
since all fields in the effective Hamiltonian are physical.

\section{Conclusions}
\label{sec:4}

In this work we have outlined a strategy to derive an effective 
renormalized Hamiltonian by means of flow equations. Application of the
flow equations with the condition, that particle number conserving terms
are considered diagonal and those changing the particle number off-diagonal
led as in other cases to useful effective Hamiltonians.

The main advantage of flow equations as compared to other
many-body approaches is, that states of
different particle number are completely decoupled, since
the particle number violating contributions are eliminated down to 
$\lambda=0$. Thus one is able to truncate the Fock space and 
the bound state problem reduces to an eigenstate equation 
in the lowest Fock sector of quasiparticles.
Positronium (meson) problem is approximated as a state of two valence
electrons (constituent quark and antiquark), described by a Bethe-Salpeter
equation with an effective Hamiltonian in two-body sector.
Energy/mass of a quasiparticle is found from the gap 
equation with an effective Hamiltonian in one-body sector.
Coupled gap equation and Bethe-Salpeter equation with effective sector 
Hamilonians are analyzed further analytically and 
numerically for a one-particle mass gap and bound state spectrum,
respectively.

We applied flow equations to the light-front Hamiltonians of QED and QCD
and to the Hamiltonians of gluodynamics and QCD (with dynamical quarks 
added) in the Coulomb gauge. We found that dynamical new terms 
generated by flow equations are crucial in order to obtain an agreement with
covariant calculations and the experimental results.
In the light-front positronium problem, due to dynamical terms triplet states
are degenerate and thus rotational invariance (that means covariance 
in this case) is maintained. Using flow equations, we obtain the singlet-triplet 
splitting in the light-front framework which agrees with the experimental data.

In the Coulomb gauge QCD, due to the dynamical interactions we obtain 
for the first time the gap equation and the Bethe-Salpeter equation which are 
finite both in the UV and IR regions. We find that no additional 
renormalization is required in the chiral limit. For the first time,
we include dynamical interactions in the framework of many-body 
technique where confinement and chiral symmetry breaking explicitly 
present. In this way we have studied an overlap 
of hadron physics and high energy QCD.
We took into account the hyperfine interaction as well as chiral 
symmetry breaking and obtained the $\pi-\rho$ mass splitting caused
by the instantaneous and dynamical interactions.
We find in the RPA, that roughly $40\%$ of the $\pi-\rho$ mass splitting is due 
to the presence of the hyperfine interaction and $60\%$ due 
to the chiral symmetry breaking. 
Flow equations improve the gluon and quark condensates as compared
with other Hamiltonian methods (we obtain 
$\langle\alpha/\pi F_{\mu\nu}F_{\mu\nu}\rangle=1.3\cdot10^{-2}\,GeV^{4}$
agrees with the sum rules
and $\langle\bar{\psi}\psi\rangle=-(155\,MeV)^3$ is still low).

We find that for glueball states confinement plays the dominant role
and dynamical terms are necessary to insure the correct (finite)
UV behavior. However, in meson systems, dynamical terms are equally
important with the instantaneous terms responcible 
for chiral symmetry breaking.  
Using flow equations, we obtain
masses of low lying glueballs and mesons which agree with 
lattice data and the experimental results.

\newpage
  
{\bf The main publications on this topic are:}

\end{document}